\documentclass[11pt,hyper,letterpaper]{JHEP3}
\usepackage{graphicx,amssymb,amsmath,bbm,amsfonts,bm,psfrag,subfigure}
\def\R{\mathcal{R}}
\def\N{{\mathcal{N}}}

\def\<{\langle}

\def\>{\rangle}
\def\({\left (}
\def\){\right )}
\def\[{\left[}
\def\]{\right]}
\def\det{\rm{det}}
\def\beq{\begin{equation}}
\def\eeq{\end{equation}}

\def\a{\alpha}
\def\s{\sigma}

\newcommand{\bea}{\begin{eqnarray}}
\newcommand{\eea}{\end{eqnarray}}

\def\Op{{\cal{O}}}

\def\ra{\rightarrow}

\def\D{{\displaystyle{\not} D}}
\def\ds{{\displaystyle{\not} \partial}}
\def\ks{{\displaystyle{\not} k}}
\def\ra{\rightarrow}
\def\nn{\nonumber}

\def\a{\alpha}
\def\b{\beta}

\def\e{\epsilon}           
  
\def\g{\gamma}

\def\k{\kappa}             

\def\m{\mu}
\def\n{\nu}
  
\def\s{\sigma}                                   

\def\G{\Gamma}
\def\J{\Psi}

\title{ \LARGE Fermionic Operator Mixing in Holographic p-wave Superfluids}

\author{Martin Ammon,\!$^1$\footnotemark[1]\,
Johanna Erdmenger,\!$^1$\footnotemark[2]\,
Matthias Kaminski,\!$^2$\footnotemark[3]\,
and Andy O'Bannon\!$^1$\footnotemark[4]
\\
$^1$Max-Planck-Institut f\"{u}r Physik (Werner-Heisenberg-Institut) \\ F\"{o}hringer Ring 6, 80805 M\"{u}nchen, Germany
\\
$^2$Department of Physics, Princeton University \\ Jadwin Hall, Princeton, NJ 08544, USA}

\footnotetext[1]{E-mail address: \email{ammon@mppmu.mpg.de}}
\footnotetext[2]{E-mail address: \email{jke@mppmu.mpg.de}}
\footnotetext[3]{E-mail address: \email{mkaminsk@princeton.edu}}
\footnotetext[4]{E-mail address: \email{ahob@mppmu.mpg.de}}

\date{\today}

\abstract{We use gauge-gravity duality to compute spectral functions of fermionic operators in a strongly-coupled defect field theory in p-wave superfluid states. The field theory is (3+1)-dimensional $\N=4$ supersymmetric $SU(N_c)$ Yang-Mills theory, in the 't Hooft limit and with large coupling, coupled to two massless flavors of (2+1)-dimensional $\N=4$ supersymmetric matter. We show that a sufficiently large chemical potential for a $U(1)$ subgroup of the global $SU(2)$ isospin symmetry triggers a phase transition to a p-wave superfluid state, and in that state we compute spectral functions for the fermionic superpartners of mesons valued in the adjoint of $SU(2)$ isospin. In the spectral functions we see the breaking of rotational symmetry and the emergence of a Fermi surface comprised of isolated points as we cool the system through the superfluid phase transition. The dual gravitational description is two coincident probe D5-branes in $AdS_5 \times S^5$ with non-trivial worldvolume $SU(2)$ gauge fields. We extract spectral functions from solutions of the linearized equations of motion for the D5-branes' worldvolume fermions, which couple to one another through the worldvolume gauge field. We develop an efficient method to compute retarded Green's functions from a system of coupled bulk fermions. We also perform the holographic renormalization of free bulk fermions in any asymptotically Euclidean AdS space.
}

\keywords{AdS/CFT correspondence, Gauge/gravity correspondence}
\preprint{MPP-2010-28 \\ PUPT-2331}

\begin{document}

\section{Introduction and Summary}

The Anti-de Sitter/Conformal Field Theory correspondence (AdS/CFT) \cite{Maldacena:1997re, Witten:1998qj, Gubser:1998bc}, and more generally gauge-gravity duality, is a holographic duality between a weakly-coupled theory of gravity in some spacetime and a strongly-coupled field theory living on the boundary of that spacetime. Gauge-gravity duality thus provides a powerful new tool for studying strongly-coupled, scale-invariant field theories in states with finite charge density, and hence may be useful in condensed matter physics, for instance in understanding low-temperature systems near quantum criticality \cite{Hartnoll:2009sz,Herzog:2009xv, McGreevy:2009xe, Sachdev:2008ba}. In particular, many special properties of certain high-$T_c$ superconducting materials may be due to an underlying quantum critical point \cite{Hartnoll:2009sz,Herzog:2009xv, Sachdev:2008ba}. Gauge-gravity duality may provide valuable insight into the physics of such materials.

Of central importance for potential condensed matter applications is the holographic description of a Fermi surface\footnote{For an alternative approach, see ref. \cite{Rey:2008zz}.} \cite{Liu:2009dm,Cubrovic:2009ye,Faulkner:2009wj, Lee:2008xf}. On the field theory side, the minimal ingredients are some strongly-coupled theory with a global $U(1)$ symmetry, in a zero-temperature state with a finite $U(1)$ chemical potential, and some fermionic operator charged under the $U(1)$. Holographic calculations of the fermionic spectral function, as a function of frequency and momentum, reveal a pole at zero frequency but finite momentum, which defines the Fermi momentum. The pole represents an excitation about a Fermi surface. 

On the gravity side, the minimal ingredients are gravity and a $U(1)$ gauge field, plus some bulk Dirac fermion charged under the $U(1)$. The bulk geometry is a Reissner-Nordstr\"om black hole. The bulk fermion is dual to the fermionic operator, and the spectral function of the operator is extracted from solutions of the linearized bulk equation of motion, the Dirac equation. These Fermi liquids are, generically, not Landau Fermi liquids, although the exact properties depend on the mass and charge of the bulk fermion.

The bulk theory can also describe a phase transition to s-wave superfluid states, if a scalar charged under the $U(1)$ is present \cite{Gubser:2008px,Hartnoll:2008vx,Hartnoll:2008kx}. On the gravity side, the Reissner-Nordstr\"om black hole grows scalar hair at low temperature, that is, a solution with a non-trivial scalar becomes thermodynamically preferred to Reissner-Nordstr\"om. In the dual field theory, the thermodynamically-preferred state includes a nonzero expectation value for a scalar operator charged under the global $U(1)$, which we will refer to as the operator ``condensing.'' The phase transition is second order with mean-field exponents \cite{Hartnoll:2008vx,Hartnoll:2008kx}.

Gauge-gravity duality can also describe p-wave superfluids, that is, superfluids in which the condensing operator is a vector charged under the $U(1)$, thus breaking not only the $U(1)$ but also rotational symmetry (to some subgroup) \cite{Gubser:2008wv}. On the gravity side, the minimal ingredients are gravity and non-Abelian gauge fields. The simplest case is an AdS geometry and $SU(2)$ gauge fields, $A_M^a$, with Lorentz index $M$ and $a=1,2,3$ labels the $SU(2)$ generators $\tau_a$. Here the $U(1)$ is a subgroup of $SU(2)$, for example the $U(1)$ in the $\tau_3$ direction, which we will call $U(1)_3$. At high temperature the thermodynamically preferred geometry is Reissner-Nordstr\"om with nonzero $A_t^3$. At low temperature, the charged black hole grows vector hair: the preferred solution has non-trivial $A_x^1$. The dual field theory has three conserved currents, $J^{\mu}_a$, dual to the gauge fields. A chemical potential, producing a finite density $\langle J^t_3 \rangle$, explicitly breaks $SU(2)$ to $U(1)_3$, and the transition occurs at large chemical potential, where the thermodynamically preferred state has nonzero $\langle J^x_1\rangle$.

In bulk calculations for both the s- and p-wave, a major technical simplification is the so-called probe limit, in which the charge of the bulk scalar, or the $SU(2)$ Yang-Mills coupling, is sent to infinity, so that the scalar or Yang-Mills stress-energy tensor on the right-hand side of Einstein's equation becomes negligible. The bulk calculation then reduces to solving the scalar or Yang-Mills equation of motion in a fixed Reissner-Nordstr\"om background. The probe limit is sufficient to detect the transitions and determine that they are second order. In either case, however, if we cool the system then, as shown in refs. \cite{Hartnoll:2008kx}, the matter fields' stress-energy tensor grows and we can no longer trust the probe limit. Reaching zero temperature requires solving the fully-coupled equations, as done in refs. \cite{Hartnoll:2008kx, Ammon:2009xh, Gubser:2010dm}.

The zero-temperature limits of the bulk hairy black hole solutions generically involve a domain wall interpolating between two regions, one near the boundary and one deep in the interior of the spacetime. For example, the geometry may interpolate between a near-boundary AdS space and an interior AdS space with a different radius of curvature and speed of light \cite{Gubser:2008wz,Gubser:2009gp,Gubser:2009cg,Horowitz:2009ij,Basu:2009vv}. In field theory language, the interior AdS space represents an emergent conformal symmetry at low temperature and finite charge density. In other words, the emergent AdS represents a quantum critical point.

Holographic calculations of fermionic spectral functions in zero-temperature s-wave superfluid states \cite{Chen:2009pt,Faulkner:2009am,Gubser:2009dt,Basu:2010ak} exhibit the so-called `peak-dip-hump' structure \cite{Chen:2009pt}, expected to be relevant in high-$T_c$ superconductors \cite{Damascelli}, as well as, for suitable mass and charge of the bulk fermion, continuous bands of poles \cite{Gubser:2009dt} and, for suitable coupling to the bulk scalar, a gap, \textit{i.e.} poles in the spectral function at nonzero momentum \textit{and} nonzero frequency \cite{Faulkner:2009am}.

Generally, the bulk actions used in holographic constructions of superfluids and Fermi surfaces are not derived from any particular string theory construction. In other words, they are basically \textit{ad hoc} models built from the minimal ingredients needed to capture the essential physics. Simple models have one big advantage (besides simplicity!), namely a kind of universality: the results may be the same for many different theories, regardless of the details of their dynamics.

On the other hand, knowing the detailed dynamics of a specific dual theory, meaning the fundamental fields and Lagrangian of some microscopic, weak-coupling description, also has advantages. For example, the holographic results may tell us that a superfluid phase transition occurs, but may not tell us why. Is a nonzero $\langle J^x_1 \rangle$ the result of some pairing mechanism? If so, is the pairing mechanism the same in every dual theory? Knowing an exact dual theory may help to answer such questions, for example by providing some \textit{weak}-coupling intuition.\footnote{A good recent example is ref. \cite{Athanasiou:2010pv}.} Finding a dual Lagrangian means ``embedding'' the bulk theory into a full string or supergravity construction, built for example from D-branes (for which we know the worldvolume theories). String (and M-) theory embeddings of holographic s-wave superfluids appear in refs. \cite{Gubser:2009gp,Gubser:2009qm,Gauntlett:2009dn}.

A string theory embedding of holographic p-wave superfluids, in the probe limit, appears in refs. \cite{Ammon:2008fc,Basu:2008bh,Ammon:2009fe,Peeters:2009sr}. Here we begin with $N_c$ Dq-branes and $N_f$ Dp-branes. Taking the usual decoupling limit for the Dq-branes, which in particular means $N_c \rightarrow \infty$, we obtain supergravity in the near-horizon geometry of the Dq-branes. Non-extremal Dq-branes produce a black hole geometry. The probe limit consists of keeping $N_f$ fixed as $N_c \rightarrow \infty$, such that $N_f / N_c \rightarrow 0$. The dynamics of $N_f$ coincident Dp-branes is then described by the non-Abelian Born-Infeld action (plus Wess-Zumino terms) in the near-horizon Dq-brane geometry. Truncating that action to leading order in the field strength, we obtain a Yang-Mills action in a black hole geometry.

We then know precisely what the dual field theory is: the Dq-brane worldvolume theory, with gauge group $SU(N_c)$, in the large-$N_c$ and strong coupling limits, coupled to a number $N_f$ of fields in the fundamental representation of the gauge group, \textit{i.e.} flavor fields. We will call this the Dq/Dp theory. If the Dp-branes do not overlap with all $q$ spatial Dq-brane directions, then the flavor fields will only propagate along a defect. The probe limit consists of neglecting quantum effects due to the flavor fields, such as the running of the coupling, because these are suppressed by powers of $N_f / N_c$. These theories generically have bound states similar to mesons in Quantum Chromodynamics (QCD). The $U(N_f)$ gauge invariance of the Dp-branes is dual to a global $U(N_f)$ analogous to the isospin symmetry of QCD. In such systems the p-wave transition occurs when a sufficiently large isospin chemical potential triggers vector meson condensation (as we review in section \ref{pwave}).

Our goal is to use such a string theory system to compute fermionic spectral functions in the p-wave phase.

We choose our Dq-branes to be D3-branes. The dual field theory is then a CFT, $\N=4$ supersymmetric Yang-Mills (SYM) theory with gauge group $SU(N_c)$ in the 't Hooft limit of $N_c \ra \infty$ with large 't Hooft coupling, $\lambda \equiv g_{YM}^2 N_c \ra \infty$. The near-horizon gravity solution is (4+1)-dimensional AdS times a five-sphere, $AdS_5 \times S^5$, with $N_c$ units of Ramond-Ramond (RR) five-form flux on the $S^5$. At finite temperature AdS becomes AdS-Schwarzschild.

We consider supersymetric probe Dp-branes extended along $AdS_P \times S^Q$, where supersymmetry requires $|P-Q| = 2$ \cite{Skenderis:2002vf}. We focus on $P\geq3$, since only in those cases is a vector condensate $\langle J^x_1 \rangle$ possible.\footnote{One exception is a D5-brane along $AdS_2 \times S^4$, which we study in section \ref{eom1reduction} (but not in p-wave states).} We will study only trivial embeddings of such Dp-branes, that is, we will only study solutions in which all Dp-brane worldvolume scalars are trivial.

Our bulk fermions will be the Dp-branes' worldvolume fermions. These fermions are in a supermultiplet with the worldvolume scalars and gauge field, hence they are in the adjoint of the worldvolume $U(N_f)$, and couple to the gauge field via the gauge-covariant derivative.\footnote{Like all worldvolume fields, they are not charged under the diagonal $U(1) \subset U(N_f)$.} In other words, supersymmetry determines the charges of the fermions. For example, we will use $N_f = 2$, where we find three fermions with charges $+1$, $-1$ and $0$ under $U(1)_3$.

To compute fermionic spectral functions we need the linearized equations of motion, the Dirac equation, for these fermions. Fortunately, the fermionic part of the D-brane action, for D-branes in arbitrary backgrounds (including RR fields) is known to quadratic order \cite{Marolf:2003ye, Marolf:2003vf,Martucci:2005rb}. The form of the action is determined by supersymmetry and T-duality \cite{Martucci:2005rb}, as we review in section \ref{worldvolumefermions}. For our Dp-branes extended along $AdS_P \times S^Q$, we perform a reduction on the $S^Q$ to obtain a Dirac equation in $AdS_P$, following ref. \cite{Kirsch:2006he} very closely. The spectrum of $AdS_P$ fermion masses are fixed by $P$, $Q$ and the coupling to the background RR five-form.

We emphasize a major difference between our systems and the models of refs. \cite{Liu:2009dm,Cubrovic:2009ye,Faulkner:2009wj}: in our embedding of the Dirac equation into string theory, the mass and charge of the fermions are fixed by supersymmetry and T-duality. We are not free to dial the values of the mass and charge, unlike refs. \cite{Liu:2009dm,Cubrovic:2009ye,Faulkner:2009wj}.

Much of our analysis will be valid for any supersymmetric Dp-brane extended along $AdS_P \times S^Q$, with $P \geq 3$, but one particular Dp-brane is attractive for a number of reasons, namely the D5-brane extended along $AdS_4 \times S^2$ ($P=4$ and $Q=2$). From the bulk point of view, this D5-brane is the only Dp-brane with a \textit{massless} worldvolume fermion, as we show in section \ref{eom1reduction}. That makes both our numerical analysis, and comparison to refs. \cite{Liu:2009dm,Faulkner:2009wj} (in which the fermions were massless), much easier.

With two coincident $P=4$, $Q=2$ D5-branes, the dual field theory is (3+1)-dimensional $\N=4$ SYM coupled to $N_f = 2$ massless (2+1)-dimensional $\N=4$ supersymmetric flavor fields. The classical Lagrangian of the theory, with couplings that preserve the $SO(3,2)$ conformal symmetry of the (2+1)-dimensional defect, appears explicitly in refs. \cite{DeWolfe:2001pq, Erdmenger:2002ex}. We write the explicit form of the fermionic operators dual to the D5-branes' worldvolume fermions in section \ref{dualoperators}, following refs. \cite{Kirsch:2006he,DeWolfe:2001pq} very closely. These fermionic operators are mesinos, the supersymmetric partners of mesons.

The $P=4$, $Q=2$ D5-brane is also attractive for potential condensed matter applications. As mentioned in ref. \cite{Wapler:2009rf}, many real condensed matter systems are effectively (2+1)-dimensional degrees of freedom interacting with ambient (3+1)-dimensional degrees of freedom. The D3/D5 theory also exhibits a rich phase structure, explored in detail in refs. \cite{Wapler:2009rf,Myers:2008me,Evans:2008nf,Filev:2009xp,Wapler:2009tr,Filev:2009ai,Benincasa:2009be,Wapler:2010nq}, including for example a Berezinskii-Kosterlitz-Thouless transition (with finite charge density and magnetic field for the diagonal $U(1) \subset U(2)$) \cite{Jensen:2010ga}. Here we focus on the D3/D5 theory's phase diagram with finite isospin chemical potential.

As always in the probe limit, we cannot access the $T=0$, finite chemical potential ground state. The $P=4$, $Q=2$ D5-brane is again attractive, however, because we know that, unlike many Dp-branes, with zero temperature and \textit{zero} chemical potential, fully back-reacted solutions appear to preserve an AdS factor in the geometry, namely an $AdS_4$ \cite{D'Hoker:2007xz,Gomis:2006cu}. That suggests that the field theory retains $SO(3,2)$ conformal invariance even including quantum effects due to the flavor, which was indeed proven in ref. \cite{Erdmenger:2002ex}. Whether some scale invariance emerges with zero temperature and \textit{finite} chemical potential is unclear.

On a technical level, our goal is to solve the Dirac equation for a massless fermion in the adjoint of $SU(2)$ confined to an $AdS_P$ submanifold of (4+1)-dimensional AdS-Schwarzschild. For any Dp-brane, the three worldvolume fermions decouple in the normal (non-superfluid) phase, where $A_x^1$ is zero, but couple to one another in the superfluid phase, where $A_x^1$ is nonzero. These couplings indicate that, in the field theory, the dual fermionic operators experience operator mixing under renormalization group flow \cite{Amado:2009ts,Kaminski:2009dh}. In the field theory, the retarded Green's function, and hence the spectral function, becomes a matrix with off-diagonal entries.

We thus develop a method to compute the retarded Green's function for bulk fermions coupled to one another. Our method is essentially a combination of the method of ref. \cite{Amado:2009ts,Kaminski:2009dh}, for coupled bosonic fields, with the method of ref. \cite{Cubrovic:2009ye,Contino:2004vy}, for free fermions. Our method is actually very general, \textit{i.e.} applicable to any system of coupled bulk fermions, not just to fermions on the worldvolume of probe Dp-branes, and is especially convenient for numerical analysis. We thus explain our method first, in section \ref{mixing}.

As an added bonus, we also perform, to our knowledge for the first time, holographic renormalization for fermions in AdS.\footnote{For the holographic renormalization of fermions in Schr\"odinger spacetime, see ref. \cite{Leigh:2009ck}.} More precisely, we study a single free fermion in any space that asymptotically approaches Euclidean-signature AdS and determine the counterterms needed to render the on-shell action finite without spoiling the stationarity of the action. Our results rigorously justify many of the \textit{ad hoc} prescriptions used in the literature, where divergences of the on-shell action were simply discarded.

For the $P=4$, $Q=2$ D5-brane, using our method for coupled bulk fermions, we numerically compute spectral functions for mesinos as we cool the system through the p-wave superfluid phase transition. Due to the operator mixing, or equivalently the coupling of the fermions in the bulk, we see that the spectral function of even a \textit{neutral} fermion develops a nontrivial feature, a peak, as the system enters the p-wave phase.

Furthermore, as we lower the temperature, the zero-frequency spectral measure\footnote{As mentioned above, the retarded Green's function, and hence the spectral function, is generically a matrix. The spectral measure is simply the trace of the spectral function.} is clearly no longer rotationally invariant, and in fact at the lowest temperatures we can reliably access in the probe limit, the main features of the spectral measure are five largely isolated peaks in the $(k_x,k_y)$ plane, two on the $k_x$ axis, two on the $k_y$ axis, and one at the origin. These results are very similar to the $T=0$ results of ref. \cite{Gubser:2010dm}, where the bulk theory was gravity and $SU(2)$ gauge fields in (3+1)-dimensions, in the $T=0$ vector-hairy black hole geometry. In that case, for a fermion in the fundamental representation of $SU(2)$, the spectral measure consisted of two points on the $k_x$ axis, located symmetrically about the origin. The prediction of ref. \cite{Gubser:2010dm} for fermions in the adjoint representation would be three points on the $k_x$ axis, one at the origin and two at finite $k_x$, positioned symmetrically about the origin. At finite temperature we see five points emerging, but we strongly suspect that, if we could access the $T=0$ limit, we would indeed see only three points, as we discuss in section \ref{numerical}.

We cannot resist drawing an analogy between our system and certain experimentally-realized p-wave superconductors (see also ref. \cite{Kaminski:2010zu}).\footnote{We thank Ronny Thomale for many useful conversations about real p-wave superconductors.} In that context, a``reduction of the Fermi surface'' to certain points in momentum space has been proposed for the ruthenate compound $Sr_2 Ru O_4$~\cite{Maeno:2001cm}: the p-wave state is supported by ferromagnetic fluctuations that increase the propensity for electrons to form spin triplet Cooper pairs, with an odd (p-wave) Cooper pair wave function.\footnote{This is rather particular, bearing in mind that a large number of generic spin interactions, for example induced by superexchange processes, favor antiferromagnetic fluctuations.} Scattering channels with momentum transfer $Q=(0,0)$, as is the case in a ferromagnet, should be enhanced in the system, as opposed to scattering channels of $Q=(\pi,\pi)$, which is the case in an anti-ferromagnet. Small momentum transfer is best accomplished by a strongly peaked density of states at the Fermi level, as occurs for example with van Hove singularities, where the density of states diverges. This lies at the heart of the strong suspicion that a Fermi surface localized to certain points with a high density of states may account for a suitable setup to support p-wave pairing.

The paper is organised as follows. In section \ref{mixing}, we describe our method for computing retarded Green's functions for coupled bulk fermions. In section \ref{pwave} we review general features of Dq/Dp holographic p-wave superfluids and demonstrate a p-wave transition using the $P=4$, $Q=2$ D5-brane. In section \ref{worldvolumefermions}, we write the fermionic part of the Dp-brane action, perform the reduction of the worldvolume Dirac equation to $AdS_P$, and, for the $P=4$, $Q=2$ D5-brane, match bulk fermions to dual field theory operators. In section \ref{numerical} we present our numerical results for the fermionic retarded Green's functions using the $P=4$, $Q=2$ D5-brane. We conclude with suggestions for future research in section \ref{conclusions}. The holographic renormalization of fermions in AdS appears in the appendix.

Section \ref{mixing} and the appendix are technical, and may be read independently of the rest of the paper. Readers who only want to understand our system and our numerical results should read at least sections \ref{probedpbranesinads}, \ref{gaugecouplings} and \ref{numerical}.

\section{Holographic Fermionic Operator Mixing}
\label{mixing}

\subsection{Review: Free Fermions}
\label{freefermions}

We begin by studying a single free fermion in AdS space. In particular we will review how to extract the field theory fermionic two-point function from a solution for a bulk Dirac fermion, following refs. \cite{Liu:2009dm,Faulkner:2009wj,Iqbal:2009fd}.

In this section we will work mainly with Euclidean-signature AdS space, with the metric written in Fefferman-Graham form,\footnote{Capital Latin letters $A,B, \ldots$ will always denote all the $AdS_{d+1}$ directions, including the radial direction $u$, while lower-case Latin letters will denote field theory directions: $i,j = 1,\dots d$.}
\beq
\label{eq:fgeucads}
ds^2 = g_{AB} \, dx^A dx^B = \frac{du^2}{u^2} + \frac{1}{u^2} \delta_{ij} \, dx^i dx^j.
\eeq
The boundary is at $u=0$. Notice that throughout the paper we use units in which the radius of $AdS_{d+1}$ is equal to one.

We will study a bulk Dirac spinor $\Psi$. The Dirac action (plus boundary terms) is
\beq
\label{eq:diracaction}
S = \int d^{d+1} x \, \sqrt{g} \, \left ( \bar{\Psi} \, \D \Psi - m \, \bar{\Psi} \Psi \right) + S_{bdy},
\eeq
where, picking one of the spatial directions to be ``time,'' with corresponding $\g^t$, we define $\bar{\Psi} = \Psi^{\dagger} \g^t$. We write the $AdS_{d+1}$ Dirac operator $\D$ below. Here $S_{bdy}$ includes boundary terms that do not affect the equation of motion.

The AdS/CFT correspondence is the statement that a theory of dynamical gravity on $AdS_{d+1}$ is equivalent to a $d$ (spacetime) dimensional CFT that ``lives'' on the boundary of $AdS_{d+1}$. Every bulk field is dual to some operator in the boundary CFT. The precise statement of the correspondence equates the on-shell bulk action with the generating functional of connected CFT correlation functions. The bulk field $\Psi$ is dual to some fermionic operator $\Op$ in the dual $d$-dimensional field theory. The on-shell bulk action, $S$, acts as the generating functional for correlators involving $\Op$. In other words, to compute renormalized correlators of $\Op$, we take functional derivatives of $S$ with respect to some source.

Generically, however, both the on-shell bulk action and the CFT generating functional diverge. On the bulk side, the divergences arise from the infinite volume of $AdS_{d+1}$, \textit{i.e.} they are long-distance or infrared (IR) divergences. In the field theory the divergences are short-distance, ultraviolet (UV) divergences. To make the AdS/CFT correspondence meaningful we must regulate and renormalize these divergences.

Holographic renormalization proceeds as follows (see ref. \cite{Skenderis:2002wp} and references therein). We first regulate the on-shell bulk action by introducing a cutoff on the integration in the radial direction: we integrate not to $u=0$ but to some $u=\epsilon$. We then add counterterms on the $u=\epsilon$ surface to cancel any terms that diverge as we remove the regulator by taking $\epsilon \rightarrow 0$. Generically, the form of the counterterms is fixed by symmetries, and the coefficients of the counterterms are adjusted to cancel the divergences. Once the counterterms are known, we can proceed to compute functional derivatives of the on-shell bulk action, always taking $\epsilon \rightarrow 0$ in the end, thus obtaining renormalized CFT correlation functions in a way that is manifestly covariant and preserves all symmetries.

As first observed in ref. \cite{Henningson:1998cd}, when we evaluate the Dirac action on a solution, the bulk term obviously vanishes. The nonzero contribution to the on-shell action comes from $S_{bdy}$, which involves terms localized on the $u=\epsilon$ surface. As observed in ref. \cite{Henneaux:1998ch}, the form of $S_{bdy}$ is fixed by demanding a well-defined variational principle for the Dirac action. Formally, $S_{var}$ thus includes two types of terms,
\beq
S_{bdy} = S_{var} + S_{CT},
\eeq
where $S_{var}$ are the terms required for the variation of the action to be well-defined \cite{Henneaux:1998ch}, while $S_{CT}$ are the counterterms, which do not affect the variation of the action. 

In the appendix we perform the holographic renormalization of the Dirac action. In particular, we determine the counterterms in $S_{CT}$. The details of holographic renormalization are well-known for various species of bulk fields, for example for the metric \cite{deHaro:2000xn}, scalar fields \cite{deHaro:2000xn}, and gauge fields \cite{Bianchi:2001kw}. To our knowledge, the only detailed analysis of holographic renormalization for fermions was in the (more complicated) context of \textit{non}-relativistic gauge-gravity duality, in ref. \cite{Leigh:2009ck}. As shown in the appendix, however, in the relativistic case the holographic renormalization procedure for fermions very closely parallels the procedure for scalars. 

As shown in the appendix, the details of the holographic renormalization depend on the value of $m$. Some values of $m$ are special, for example when $m$ is half-integer (in units of the $AdS_{d+1}$ radius), counterterms logarithmic in $\epsilon$ (rather than just polynomial in $\epsilon$) are needed. For simplicity, in this section we will restrict to values of $m$ that are positive and not half-integer. Our arguments are easy to generalize to any value of $m$.

In this section we will also restrict to four- and five-dimensional AdS spaces, which we will collectively denote as $AdS_{d+1}$ with $d=3,4$, primarily for pedagogical reasons: in these cases the bulk Dirac spinor has four complex components, and we can write explicit $4 \times 4$ bulk Dirac $\G$-matrices. Additionally, we note that $AdS_{d+1}$ spaces with $d \leq 4$ are the cases most relevant for condensed matter applications (as opposed to, say, $AdS_7$). The generalization to other dimensions is straightforward. In the appendix we work with arbitrary $d$.

In later sections we will be interested in computing finite-temperature, real-time correlation functions, in particular the retarded Green's functions, in which case the bulk geometry will be Lorentzian-signature AdS-Schwarzschild. We review the prescription for obtaining the retarded Green's function in such cases at the end of this subsection.

Varying the above action we obtain the bulk equation of motion, the Dirac equation,
\beq
e^M_{~A} \,  \g^A \, D_M \Psi - m \, \Psi = 0,
\eeq
where $e^M_{~A} = u \, \delta^M_{~A}$ are the inverse vielbeins associated with the metric in eq. (\ref{eq:fgeucads}).\footnote{Recall that for inverse vielbeins, the upper index is general coordinate and the lower index is local Lorentz. The $\g^A$ obey the usual algebra $\{\g^A,\g^B\} = 2 \, \delta^{AB}$.} The curved-space covariant derivative is
\beq
D_M = \partial_M + \frac{1}{4} \left( \omega_M \right)_{AB} \left [ \g^A, \g^B \right],
\eeq
where $\left(\omega_M\right)_{AB}$ is the spin connection associated with the metric in eq. (\ref{eq:fgeucads}). The only nonzero components of the spin connection are $\left( \omega_i \right)_{uj} = \frac{1}{u} \, \delta_{ij}$, so that $D_u = \partial_u$ and the other components of $D_M$ are
\beq
D_i = \partial_i + \frac{1}{4} \frac{1}{u} \, \left [ \g^u, \g^i \right].
\eeq
We can now simplify the Dirac equation,
\bea
0 & = & e^M_{~A} \,  \g^A \, D_M \Psi - m \, \Psi \nonumber \\ & = & u \, \g^M \partial_M \Psi + \frac{1}{4} \g^i \left [ \g^u, \g^i \right] \Psi - m \Psi \nonumber \\ & = & \left [ u \, \g^M \partial_M - \frac{d}{2} \,\g^u - m \right ] \Psi.
\eea
We will work with a single Fourier mode, so we let $\Psi \rightarrow e^{ikx}\,\Psi$, where, without loss of generality, we have chosen the momentum to point in the $\hat{x}$ direction.\footnote{\label{pwavefootnote1} In a p-wave superfluid phase rotational symmetry is broken, so there, to study the most general case, we must use a momentum with nonzero components in different directions, as we will discuss in section \ref{worldvolumefermions}.} The Dirac equation is then
\beq
\label{eq:ddimsimplifieddiraceq}
\left [u \g^u \partial_u + i k\, u \, \g^x - \frac{d}{2} \g^u - m \right] \Psi = 0.
\eeq

We will now choose an explicit basis for the $\G$-matrices. We will use a basis in which all the $\G$-matrices are Hermitian,
\beq
\g^u = \left( \begin{array}{cc} - \s_3 & 0 \\ 0 & -\s_3 \end{array} \right), \qquad \g^t = \left( \begin{array}{cc} \s_1 & 0 \\ 0 & \s_1 \end{array} \right), \qquad \g^x = \left( \begin{array}{cc} -\s_2 & 0 \\ 0 & \s_2 \end{array} \right),
\eeq
where $\sigma_1$, $\sigma_2$ and $\sigma_3$ are the usual Pauli matrices,
\beq
\s_1 = \left ( \begin{array}{cc} 0 & 1 \\ 1 & 0 \end{array} \right), \qquad \s_2 = \left ( \begin{array}{cc} 0 & -i \\ i & 0 \end{array} \right), \qquad \s_3 = \left ( \begin{array}{cc} 1 & 0 \\ 0 & -1 \end{array} \right).
\eeq
Next we will define two sets of projectors. The first set is
\beq
\Pi_+ = \frac{1}{2}  \left( 1 + \g^u\right) = \begin{pmatrix} 0 & & & \\ & 1 & & \\ & & 0 & \\ & & & 1\end{pmatrix}, \qquad \Pi_- = \frac{1}{2}  \left( 1 - \g^u\right) = \begin{pmatrix} 1 & & & \\ & 0 & & \\ & & 1 & \\ & & & 0\end{pmatrix}.
\eeq
We use these to define $\Psi_{\pm} = \frac{1}{2} \left( 1 \pm \g^u\right) \Psi$ so that $\g^u \Psi_{\pm} = \pm \Psi_{\pm}$. The second set of projectors was used for example in refs. \cite{Liu:2009dm,Faulkner:2009wj},
\beq
\label{eq:defpionetwoprojectors}
\Pi_1 = \frac{1}{2} \left( 1 + i \g^u \g^t \g^x \right) = \begin{pmatrix} 0 & & & \\ & 0 & & \\ & & 1 & \\ & & & 1\end{pmatrix}, \quad \Pi_2 = \frac{1}{2} \left( 1 - i \g^u \g^t \g^x \right) = \begin{pmatrix} 1 & & & \\ & 1 & & \\ & & 0 & \\ & & & 0\end{pmatrix}.
\eeq
To make converting between $\Psi_{\pm}$ and $\Psi_{1,2}$ easy, we explicitly write $\Psi$ first as $\Psi_+ + \Psi_-$ and then as $\Psi_1 + \Psi_2$,
\beq
\label{eq:psiexplicit}
\Psi = \begin{pmatrix} 0 \\ \Psi_{+u} \\ 0 \\ \Psi_{+d} \end{pmatrix} + \begin{pmatrix} \Psi_{-u} \\ 0 \\ \Psi_{-d} \\ 0 \end{pmatrix} = \begin{pmatrix} 0 \\ 0 \\ \Psi_{1u} \\ \Psi_{1d} \end{pmatrix} + \begin{pmatrix} \Psi_{2u} \\ \Psi_{2d} \\ 0 \\ 0 \end{pmatrix},
\eeq
where the subscripts $u$ and $d$ indicate the ``up'' and ``down'' components of the effectively two-component $\Psi_{\pm}$ and $\Psi_{1,2}$. Identifications such as $\Psi_{+u} = \Psi_{2d}$ are then obvious.

We have a choice of whether to use $\Psi_{\pm}$ or $\Psi_{1,2}$, although of course, we can easily translate between the two options using eq. (\ref{eq:psiexplicit}). We will choose whatever is most convenient for a given question.

For example, the projectors $\Pi_{1,2}$ commute with the operator in eq. (\ref{eq:ddimsimplifieddiraceq}), which tells us that, for a free fermion, the equations for $\Psi_{1,2}$ decouple. That makes $\Psi_{1,2}$ especially attractive for numerical analysis, hence we employ them in sections \ref{worldvolumefermions} and \ref{numerical}.\footnote{As mentioned in footnote \ref{pwavefootnote1}, in the p-wave superfluid phase, the most general momentum has nonzero components in multiple directions. That means $\Psi_1$ and $\Psi_2$ will no longer decouple because other $\G$-matrices, such as $\g^y$, will appear in the equation of motion, and these do not commute with $\Pi_{1,2}$. Nevertheless, when studying the p-wave superfluid phase we use $\Psi_{1,2}$ to make the comparison with the rotationally-symmetric case easier.} Explicitly, the equations for $\Psi_{1,2}$ are
\bea
\label{eq:diraceqpi12projectors}
\left[ u \, \partial_u - \frac{d}{2} + m \, \s_3  - k u \right] \Psi_1 & = & 0, \\ \label{eq:1storderdiracb} \left[ u \, \partial_u - \frac{d}{2} + m \, \s_3  + k u \right] \Psi_2 & = & 0.
\eea

On the other hand, the asymptotic behavior of $\Psi$ is most succinctly described using $\Psi_{\pm}$, hence we use these frequently below, especially in the appendix. In terms of $\Psi_{\pm}$, the equation of motion becomes 
\bea
\label{eq:1storderdiraca}
\left( u \, \partial_u - \frac{d}{2} - m \right) \Psi_+ + k u \, \s_3 \Psi_- & = & 0, \\ \label{eq:1storderdiracb} \left( u \, \partial_u - \frac{d}{2} + m \right) \Psi_- + k u \, \s_3 \Psi_+ & = & 0.
\eea
These first-order equations give rise to the second-order equations
\beq
\label{eq:2ndorderdiraceq}
\left [ \partial_u^2 - \frac{d}{u} \partial_u + \frac{1}{u^2} \left( - m^2 \pm m + \frac{d^2}{4} + \frac{d}{2} \right) - k^2 \right]\Psi_{\pm} = 0.
\eeq
The leading asymptotic behaviors of $\Psi_{\pm}$ are
\beq
\label{eq:leadingudependence}
\Psi_{\pm} = c_{\pm}(k) \, u^{\frac{d}{2} \pm m} + O\left(u^{\frac{d}{2} +1 \pm m}\right).
\eeq
where $c_{\pm}(k)$ are spinors that obey $\Pi_{\pm} c_{\pm}(k) = \pm c_{\pm}(k)$, and which may depend on $k$, as indicated.

As reviewed above, to compute renormalized correlators of the dual operator $\Op$, we take functional derivatives of $S$ with respect to some source. We identify the source for $\Op$ as the coefficient of the dominant term in $\Psi$'s near-boundary expansion (the term that grows most quickly as $u\ra0$). From eq. (\ref{eq:leadingudependence}), we see that the dominant term is the $u^{\frac{d}{2}-m}$ term, hence we 
identify $c_-(k)$ as the source for $\Op$. More formally, we equate
\beq
\label{eq:generatingdef}
e^{-S_{ren}[c_-,\bar{c}_-]} = \left \< \mbox{exp} \left [ \int d^dx \, \left(\bar{c}_- \, \Op + \bar{\Op} \, c_- \right)\right]  \right\>,
\eeq
where the left-hand-side is the exponential of minus the action in eq. (\ref{eq:diracaction}), evaluated on a solution and properly renormalized (hence the subscript), and the right-hand-side is the generating functional of the dual field theory, with $c_-(k)$ acting as the source for the operator $\Op$.\footnote{As we review in the appendix, for a bulk fermion with mass $m$, in a standard quantization the dimension $\Delta$ of $\Op$ is $\Delta = \frac{d}{2}+|m|$ \cite{Henningson:1998cd,Henneaux:1998ch}. In the appendix we also discuss the chirality of $\Op$ (when $d$ is even).} Upon taking minus the logarithm of both sides, we find that the on-shell bulk action is the generator of connected correlators.

For bulk bosonic fields, we must solve a straightforward Dirichlet problem: we fix the leading asymptotic value of the field, allow the field to vary, and then impose a regularity condition in the interior of the space to fix the entire solution. This procedure is dual to the statement that once we choose a source, the dynamics of the theory determines the expectation values of the dual operator.

The story for fermions is more subtle, because $\Psi_+(u,k)$ and $\Psi_-(u,k)$ are not independent \cite{Henningson:1998cd,Henneaux:1998ch}. Each one determines the canonical momentum associated with the other (see for example ref. \cite{Iqbal:2009fd}). In the bulk Dirichlet problem, then, we cannot fix their asymptotic values $c_{\pm}(k)$ simultaneously, but can fix only one, the coefficient of the dominant term, $c_-(k)$, and then vary the field. As shown in refs. \cite{Henneaux:1998ch}, for the action to remain stationary under such variations, we must add a boundary term to the action,
\beq
\label{eq:svardef}
S_{var} = \int d^dx \, \sqrt{\g} \, \bar{\Psi}_+ \Psi_-,
\eeq
where the integration is over the $u=\epsilon$ hypersurface, $\sqrt{\g} = \epsilon^{-d}$ is the square root of the determinant of the induced metric at $u=\epsilon$, and $\Psi_{\pm}$ are evaluated at $u=\epsilon$.

Indeed, since the bulk action is first-order in derivatives, the only nonzero contribution to the on-shell action comes from the boundary terms $S_{bdy} = S_{var} + S_{CT}$. Generically, when evaluated on a solution, divergent terms appear in $S_{var}$, which are canceled by the counterterms in $S_{CT}$. Notice that, to preserve stationarity of the action, $S_{CT}$ must involve only $\Psi_-(\epsilon,k)$, since that is held fixed under variations. We write the counterterms explicitly in the appendix.

The principal result of the appendix is the renormalized on-shell action: we evaluate $S_{bdy}$ on a solution and take $\epsilon \rightarrow 0$ to obtain (for positive, non-half-integer $m$)
\beq
S_{ren} = \int d^dx \, \bar{c}_+ \, c_-,
\eeq
We can now easily compute the renormalized connected correlators of $\Op$ and $\bar{\Op}$ by taking functional derivatives of $S_{ren}$. For example, the renormalized one-point function of $\bar{\Op}$ is
\beq
\left \< \bar{\Op} \right \>_{ren} = - \frac{\delta S_{ren}}{\delta c_-} = - \bar{c}_+.
\eeq
If we use the fact that the on-shell bulk action must be Hermitian, $S=S^{\dagger}$, then we also have
\beq
S_{ren} =  S^{\dagger}_{ren} = \int d^dx \, \left [ \bar{c}_+ c_- \right]^{\dagger} = \int d^dx \, \bar{c}_- c_+, \nonumber
\eeq
hence we also find, as we should,
\beq
\left \< \Op \right \>_{ren} =  - \frac{\delta S_{ren}}{\delta \bar{c}_-} = - c_+.
\eeq
We can obtain two-point functions via second functional derivatives, for example
\beq
\label{eq:secondderivativeofonshellaction}
\left \< \Op \, \bar{\Op} \right \>_{ren} = - \, \frac{\delta^2 S_{ren}}{\delta c_- \delta \bar{c}_-} = - \frac{\delta c_+}{\delta c_-}.
\eeq
The equation of motion plus some regularity condition in the interior of the spacetime will relate $c_+$ and $c_-$ (recalling that we fix $c_-$ and vary $c_+$). The equation is linear, hence the relation will be linear: $c_+ = - G(k) \, \g^t \, c_-$, for some matrix $G(k)$ which will turn out to be the Euclidean Green's function. We include a factor of $\g^t$ because, as discussed in refs. \cite{Liu:2009dm,Faulkner:2009wj}, the Euclidean Green's function is actually $\left \< \Op \, \Op^{\dagger} \right \>_{ren}$, which differs from $\left \< \Op \, \bar{\Op} \right \>_{ren}$ by a factor of $\g^t$. We indeed find
\beq
\label{eq:euclideangreensfunctionfromonshellaction}
\left \< \Op \, \bar{\Op} \right \>_{ren} = G(k) \, \g^t, \qquad \left \< \Op \, \Op^{\dagger} \right \>_{ren} = G(k).
\eeq
In general, we must extract $G(k)\, \g^t$ from a solution by imposing some regularity condition in the bulk of the spacetime (in our coordinates, the $u \rightarrow \infty$ region), which fixes $c_+$ in terms of $c_-$. We review that procedure for Euclidean $AdS_{d+1}$ in the appendix and for Lorentzian-signature AdS-Schwarzschild below.

We can also reproduce the formulas used in refs. \cite{Liu:2009dm,Faulkner:2009wj} by switching to $\Psi_{1,2}$. In that case, the equations for $\Psi_1$ and $\Psi_2$ decouple, hence in the Green's function the $\Pi_1$ and $\Pi_2$ subspaces will not mix. Writing $c_+ = - G(k) \, \g^t \, c_-$ explicitly, we will have (suppressing the $k$ dependence of $c_{\pm}(k)$)
\beq
\begin{pmatrix} 0 \\ c_{+u} \\ 0 \\ c_{+d} \end{pmatrix} = - \begin{pmatrix} G_{22}(k) \, \mathbf{1}_2 & \\ & G_{11}(k) \, \mathbf{1}_2 \end{pmatrix} \begin{pmatrix} 0 & 1 & & \\ 1 & 0\ & & \\ & & 0 & 1 \\ & & 1 & 0 \end{pmatrix} \begin{pmatrix} c_{-u} \\ 0 \\ c_{-d} \\ 0 \end{pmatrix} = - \begin{pmatrix} G_{22}(k) \, \mathbf{1}_2 & \\ & G_{11}(k) \, \mathbf{1}_2 \end{pmatrix} \begin{pmatrix} 0 \\ c_{-u} \\ 0 \\ c_{-d} \end{pmatrix}, \nonumber
\eeq
where blank entries represent zero, $\mathbf{1}_2$ is the $2 \times 2$ identity matrix, and $G_{11}$ and $G_{22}$ represent the components of the Green's function in the $\Pi_1$ and $\Pi_2$ subspaces, respectively. Given a bulk solution for $\Psi$, we obtain the Green's functions simply by reading off the asymptotic values of $c_+(k)$ and $c_-(k)$ and then constructing
\beq
G_{22}(k) = - \frac{c_{+u}}{c_{-u}}, \qquad G_{11}(k) = - \frac{c_{+d}}{c_{-d}}.
\eeq

Finally, we review the prescription of ref. \cite{Iqbal:2009fd} to compute the retarded two-point function in the finite-temperature, Lorentzian-signature case. Here the geometry is AdS-Schwarzschild, with a horizon at some position $u_h$. To obtain the retarded two-point function, we require that, near the horizon, the bulk solution for $\Psi$ has the form of wave traveling into the horizon (out of the spacetime), \textit{i.e.} an in-going wave. The asymptotic form for $\Psi$ near the boundary is the same as in eq. (\ref{eq:leadingudependence}) (for positive, non-half-integer $m$). Following ref. \cite{Iqbal:2009fd}, in the regime of linear response, we have
\beq
\label{eq:lorentziancplus}
c_+(\omega,k) = - i G^R(\omega,k) \, \g^t \, c_-(\omega,k),
\eeq
where $G^R(\omega,k)$ is the retarded Green's function. Notice that here we distinguish the frequency $\omega$ from the momentum $k$, and $\g^t$ is now anti-Hermitian,
\beq
\g^t = \begin{pmatrix} i\sigma_1 & 0 \\ 0 & i \sigma_1 \end{pmatrix}.
\eeq
Eq. (\ref{eq:lorentziancplus}) is essentially just an analytic continuation from the Euclidean case: $\g^t \rightarrow i \g^t$. For a free fermion, we obtain (see also eq. (A17) of ref. \cite{Faulkner:2009wj})
\beq
\label{eq:freelorentziangreens}
G^R_{22}(\omega,k) = \frac{c_{+u}}{c_{-u}}, \qquad G^R_{11}(\omega,k) = \frac{c_{+d}}{c_{-d}}.
\eeq

\subsection{Coupled Fermions}
\label{coupledfermions}

We now consider multiple bulk fermions, say $N$ of them, $\Psi_a$ with $a=1,\ldots,N$, coupled to one another. The fact that the linearized fluctuation of the $\Psi_a$ couple in the bulk is dual to the statement that the fermionic operators in the field theory mix with one another under renormalization group flow.

We will work in Lorentzian signature, and finite temperature, so that the bulk geometry is AdS-Schwazrschild, with a horizon at some position $u_h$. We consider fermions with quadratic couplings of the form (with implicit summation over $a,b$)
\beq
\label{eq:coupleddiracaction}
S = i \int d^{d+1} x \, \sqrt{g} \, \left ( \bar{\Psi}_a \, \D \Psi_a - \bar{\Psi}_a \Lambda_{ab} \Psi_b \right) + S_{bdy},
\eeq
for some matrix $\Lambda_{ab}$ that need not be diagonal in either the $a,b$ indices or the spinor indices. As a concrete example, in later sections we will introduce a bulk $SU(2)$ gauge field $A_M$ and a bulk fermion valued in the adjoint of $SU(2)$. The indices $a,b$ are then $SU(2)$ indices, hence we will have three bulk fermions (for $\tau_1$, $\tau_2$, and $\tau_3$) with a coupling, coming from the gauge-covariant derivative, of the form $\epsilon_{abc} \bar{\Psi}_a \, e^{M}_{~A} \g^A  \left(A_M\right)_b \Psi_c$, which is obviously not diagonal in either $SU(2)$ indices or in spinor indices (because of the $\g^A$).

For the following arguments, we do not need to know any details about the equations of motion. We will only exploit one important feature. Using the $\Pi_{\pm}$ projectors, we will always obtain equations similar to eqs. (\ref{eq:1storderdiraca}) and (\ref{eq:1storderdiracb}). We will then always be able to write these equations in the form
\beq
\label{eq:generalcoupledeom}
\nabla_{ab\pm} \Psi_{b\pm} = M_{ac\pm} \Psi_{c\mp},
\eeq
where $\nabla_{ab\pm}$ is some differential operator, involving in particular $\partial_u$, and $M_{ac\pm}$ is a matrix representing the couplings among not only the $\Psi_a$, which come from $\Lambda_{ab}$, but also the terms from $\D \Psi_a$ that produce couplings between $\Psi_{a+}$ and $\Psi_{a-}$, for example the terms proportional to the momentum $k$ in eqs. (\ref{eq:1storderdiraca}) and (\ref{eq:1storderdiracb}). The key feature is that only the $\Psi_{a\pm}$ are on the left-hand-side, while only the $\Psi_{a\mp}$ are on the right-hand-side.

In practical terms, the total number of complex functions for which we must solve is $4 \times N$, since each $\Psi_a$ has four complex components. In other words, we need to decompose the $\Psi_{a}$ not only into $\Psi_{a+}$ and $\Psi_{a-}$, but also into the up and down components, $\Psi_{a + u}$, $\Psi_{a+d}$, $\Psi_{a-u}$, and $\Psi_{a-d}$. When convenient, we may sometimes think of eq. (\ref{eq:generalcoupledeom}) as equations describing these $4 \times N$ coupled functions, which we may sometimes refer to as ``fields.''

Clearly, if we solve for all the $\Psi_a$, insert the solutions into the bulk action, and take functional derivatives, we will obtain field theory retarded Green's functions that are matrices, $G^R_{ab}\left(\omega,k\right)$. In principle, we may be able to diagonalize the equations of motion and obtain decoupled equations, in which case the Green's function will be diagonal. Given the bulk solutions for the $\Psi_a$, we then extract the elements of $G^R_{ab}\left(\omega,k\right)$ using eq. (\ref{eq:freelorentziangreens}). In some cases, however, diagonalizing the equations of motion may be prohibitively difficult, \textit{i.e.} practically impossible. We can always resort to numerics to find solutions, but we will then be forced to compute elements of the \textit{un}-diagonalized $G^R_{ab}\left(\omega,k\right)$. We thus need to know what combinations of the asymptotic values $c_{a+}$ and $c_{a-}$ give an arbitrary element $G^R_{ab}(\omega,k)$.

We will describe a prescription to obtain the matrix $G^R_{ab}(\omega,k)$, assuming we have bulk solutions for the $\Psi_a$. The method is a hybrid of the methods in refs. \cite{Amado:2009ts,Kaminski:2009dh} and refs. \cite{Cubrovic:2009ye,Contino:2004vy}. Refs. \cite{Amado:2009ts,Kaminski:2009dh} described a general method to construct a retarded Green's function for coupled bulk scalar and gauge fields, while refs. \cite{Cubrovic:2009ye,Contino:2004vy} described general methods for computing Green's functions from fermions in the bulk.

The first observation is that we can construct second-order equations for the bulk fields, the $\Psi_{a\pm}$, that will be similar to eq. (\ref{eq:2ndorderdiraceq}). We actually don't care about  the exact form of these equations. We only need to know that such equations exist. We thus have a system of $2 N$ second-order linear equations, for which we expect $2 \times 2 N$ linearly-independent solutions. We must therefore fix two boundary conditions for each field to specify a solution for the entire system. Following refs. \cite{Kaminski:2009dh}, we fix these boundary conditions near the horizon $u_h$. For example, the $\Psi_{a + u}$ will have the near-horizon form
\beq
\label{eq:fermioningoingwave}
\Psi_{a + u} = n_{a + u} \left( u - u_h \right)^{i \alpha} + \ldots.
\eeq
where $n_{a + u}$ and $\a$ are constants (independent of $u$) and $\ldots$ represents terms that decay faster, as $u \rightarrow u_h$, than the terms shown. The two constants $n_{a + u}$ and $\a$ are the two degrees of freedom we have to specify the solution. Generically, the equation of motion will only be satisfied for two values of $\a$, one describing an in-going wave and the other describing an out-going wave. As is well-known, to obtain the retarded Green's function, we must use an in-going wave. We still need to choose the normalization $n_{a + u}$. As shown in refs. \cite{Liu:2009dm,Faulkner:2009wj}, for fermions, once we choose an in-going wave solution, if we use the projectors $\Pi_{1,2}$, then when we fix the normalization of the up component $\Psi_{a1u}$ to be $n_{a1u}$, the equation of motion fixes the down component $\Psi_{a1d}$ to have normalization $i$ times $n_{a1u}$. The same applies to the up and down components of $\Psi_{a2}$. Switching to the $\Pi_{\pm}$ projectors (recall eq. (\ref{eq:psiexplicit})), the statement is that once we fix the normalization of $\Psi_{a-d}$ to be $n_{a-d}$, then $\Psi_{a+d}$ must have normalization $i$ times $n_{a-d}$. The same statement applies to $\Psi_{a-u}$ and $\Psi_{a+u}$.

We thus need only fix $2N$ normalizations, for the up and down components of the $\Psi_{a-}$. Let us arrange these normalizations into a row vector $\vec{n}$
\beq
\vec{n} = \left( n_{1-u}, n_{1-d}, n_{2-u}, n_{2-d},\ldots, n_{N-u},n_{N-d}\right).
\eeq
Following refs. \cite{Kaminski:2009dh}, we use these horizon normalizations to construct a basis of solutions as follows. We solve the equations of motion $2N$ times, each time with a different choice of $\vec{n}$. The first time we use $\vec{n} = (+1,+1,+1,\ldots,+1,+1)$, the second time we use $\vec{n} = (+1,-1,+1,\ldots,+1,+1)$, the third time we use $\vec{n} = (+1,+1,-1,\ldots,+1,+1)$, and so on. We label these choices $\vec{n}^{(i)}$, with $i = 1, \ldots, 2N$. For each choice of normalizations, we obtain solutions $\Psi^{(i)}_{a\pm}$. We now have a basis of solutions, so we can write any particular solution as a linear combination of these. To do so, we construct matrices that we will call $\tilde{P}^{\pm}_{aj}(u,\omega,k)$ from the basis solutions, where each row corresponds to a field and each column corresponds to a choice of normalization (the $i$ index). For example, (suppressing the $\Psi_{a-}$'s dependence on all variables)
\beq
\label{eq:ptildedef}
\left [ \tilde{P}^-_{aj}(u,\omega,k) \right ] = \begin{pmatrix} \Psi^{(1)}_{1-} & \Psi^{(2)}_{1-} & \ldots & \Psi^{(2N)}_{1-} \\ \Psi_{2-}^{(1)} & \Psi_{2-}^{(2)} & \ldots & \Psi_{2-}^{(2N)} \\ \ldots & \ldots & & \ldots \\ \Psi_{N-}^{(1)} & \Psi_{N-}^{(2)} & \ldots & \Psi_{N-}^{(2N)} \end{pmatrix},
\eeq
with $\tilde{P}^+_{aj}(u,\omega,k)$ defined similarly. The $\tilde{P}^{\pm}_{aj}(u,\omega,k)$ are $2N \times 2N$ matrices. For later convenience, we will factor out the leading asymptotic behavior of the solutions, defining new matrices $P^{\pm}_{aj}(u,\omega,k)$,
\beq
\label{eq:pmatricesdef}
\tilde{P}^{\pm}_{aj}(u,\omega,k) \equiv u^{\frac{d}{2} \pm m} \, P^{\pm}_{aj}(u,\omega,k).
\eeq
We can now write any solution as a linear combination of the basis solutions:
\bea
\label{eq:psilinearcombinations}
\Psi_{a+}(u,\omega,k) = u^{\frac{d}{2} + m} \, P^+_{aj}(u,\omega,k) \, \left(P^+(\epsilon,\omega,k)^{-1}\right)_{jb} \, c_{b+}(\omega,k), \nonumber \\ \Psi_{a-}(u,\omega,k) = u^{\frac{d}{2} - m} \, P^-_{aj}(u,\omega,k) \, \left(P^-(\epsilon,\omega,k)^{-1}\right)_{jb} \, c_{b-}(\omega,k),
\eea
with a summation over the $j$ index. Notice that we take the solutions $\Psi_{a\pm}$ to be linear in the ``sources,'' $c_{a\pm}$. As emphasized in refs. \cite{Kaminski:2009dh}, eq. (\ref{eq:psilinearcombinations}) is simply saying that the sources $c_{a\pm}$ will source various linear combinations of fields in the bulk, and that we can write those linear combinations as linear combinations of our basis solutions. Notice that when we evaluate the solutions at $u=\epsilon$, we reproduce the leading asymptotic form, $\Psi_{a\pm} \sim c_{a\pm} \, u^{\frac{d}{2} \pm m}$.

Now we arrive at the main difference between bulk fermions and bulk bosons: $c_{a+}$ and $c_{a-}$ are not independent. The equation of motion relates them \cite{Henningson:1998cd,Henneaux:1998ch}. Indeed, we saw above that only the $c_{a-}$ are sources, while the $c_{a+}$ give one-point functions (roughly speaking). To relate them, we follow refs. \cite{Cubrovic:2009ye,Contino:2004vy}. We return to the equation of motion as written in eq. (\ref{eq:generalcoupledeom}). We focus only on the equation with $\Psi_{a+}$ on the left-hand-side, and simply insert solutions as written in eq. (\ref{eq:psilinearcombinations}) (suppressing all $\omega$ and $k$ dependence)
\beq
\label{eq:pminusrhs}
\nabla_{ab +} \, u^{\frac{d}{2} + m} P^+_{bj}(u) \, \left( \left( P^+ (\epsilon)^{-1} \right)_{jd} \, c_{d+} \right) = M_{ae+} \, u^{\frac{d}{2} - m} P^-_{ej}(u) \left( \left( P^-(\epsilon)^{-1}\right)_{jf} c_{f-}\right),
\eeq
where the parentheses separate $u$-dependent factors from $u$-independent factors. We now observe that the matrices $P^{\pm}_{aj}$ also solve the equation of motion, by construction, since they are built from  solutions. We thus have
\beq
\nabla_{ab +} \, u^{\frac{d}{2} + m} P^+_{bj}(u)= M_{ac+} \, u^{\frac{d}{2} - m} P^-_{cj}(u).
\eeq
Here we have a free $j$ index, so we actually have $2N$ such equations. (Recall that the index $j$ labels the choice of normalization vector $\vec{n}$.) The above equation is just the statement that one column of the $P^{\pm}_{aj}$ matrices solves the equation of motion. We are free to act on the right with the vector $\left( P^+ (\epsilon)^{-1} \right)_{jd} \, c_{d+}$, so that we obtain
\beq
\label{eq:pplusrhs}
\nabla_{ab +} \, u^{\frac{d}{2} + m} P^+_{bj}(u) \left( \left( P^+ (\epsilon)^{-1} \right)_{jd} \, c_{d+} \right) = M_{ac+} \, u^{\frac{d}{2} - m} P^-_{cj}(u) \left( \left( P^+ (\epsilon)^{-1} \right)_{jd} \, c_{d+} \right).
\eeq
We now simply compare eqs. (\ref{eq:pminusrhs}) and (\ref{eq:pplusrhs}). The left-hand sides are identical, so we may equate the right-hand sides. Acting on the left with some inverse matrices, we obtain the desired relation between the $c_{a+}$ and $c_{a-}$,
\beq
\label{eq:coupledcpluseq}
c_{a+} = P^+(\epsilon)_{aj} \, \left( P^-(\epsilon)^{-1} \right)_{jb} c_{b-}.
\eeq
Invoking eq. (\ref{eq:lorentziancplus}), we now just need to perform two operations to extract the retarded two-point function $G^R_{ab}(\omega,k)$ from $P^+(\epsilon)_{aj} \, \left( P^-(\epsilon)^{-1} \right)_{jb}$: we take $\epsilon \rightarrow 0$ and then act on the right with $-i \g^t$. 

The effect of taking $\epsilon \rightarrow 0$ is easy to understand. From the definition of the $\tilde{P}^{\pm}_{aj}(u)$ in eq. (\ref{eq:ptildedef}) and the definition of the $P^{\pm}_{aj}(u)$ in eq. (\ref{eq:pmatricesdef}), we can identify the $\epsilon \rightarrow 0$ limit of the $P^{\pm}_{aj}(u)$ as
\beq
\label{eq:identifypminuswithcminus}
\lim_{\epsilon \rightarrow 0} \left [ P^-_{aj}(\epsilon)  \right ] =  \begin{pmatrix} c^{(1)}_{1-} & c^{(2)}_{1-} & \ldots & c^{(2N)}_{1-} \\ c_{2-}^{(1)} & c_{2-}^{(2)} & \ldots & c_{2-}^{(2N)} \\ \ldots & \ldots & & \ldots \\ c_{N-}^{(1)} & c_{N-}^{(2)} & \ldots & c_{N-}^{(2N)}  \end{pmatrix},
\eeq
and similarly for $\lim_{\epsilon \rightarrow 0} P^+_{aj}(\epsilon)$. In short, the matrices $P^{\pm}_{aj}$, when evaluated at the boundary, are simply matrices of the $c_{a+}$ and $c_{a-}$.

Notice that the $P^-(\epsilon)^{-1}$ matrix will introduce a factor of ${\det} P^-(\epsilon)$ in the denominator of the Green's function. Generically, then, if ${\det} P^-(\epsilon)$ has a zero, the Green's function will have a pole, which means a quasi-normal mode appears in the bulk spectrum, as in the bosonic cases of ref. \cite{Kaminski:2009dh}. Given the identification in eq. (\ref{eq:identifypminuswithcminus}), then, to identify quasi-normal modes we need only identify the zeroes of the matrix of $c_-$'s.

Understanding how $\g^t$ acts on $P^+(\epsilon)_{aj} \, \left( P^-(\epsilon)^{-1} \right)_{jb}$ is a little tricky. Luckily, the way we have written $P^+(\epsilon)_{aj} \, \left( P^-(\epsilon)^{-1} \right)_{jb}$ means that $-i\g^t$ acts trivially. To see that, notice that eq. (\ref{eq:coupledcpluseq}) is written in a two-component form: here $c_{a\pm}$ are two component spinors. To restore them to four-component form, we take a direct product,
\beq
c_{a+} = \begin{pmatrix} c_{a+u} \\ c_{a+d} \end{pmatrix} \rightarrow c_{a+} \otimes \begin{pmatrix} 0 \\ 1 \end{pmatrix} = \begin{pmatrix} 0 \\ c_{a+u} \\ 0 \\ c_{a+d} \end{pmatrix}, \quad c_{a-} = \begin{pmatrix} c_{a-u} \\ c_{a-d} \end{pmatrix} \rightarrow c_{a-} \otimes \begin{pmatrix} 1 \\ 0 \end{pmatrix} = \begin{pmatrix} c_{a-u} \\ 0 \\ c_{a-d} \\ 0 \end{pmatrix}.
\eeq
To restore the $P^{\pm}_{aj}$ matrices to the same four-component form, we recall recall eq. (\ref{eq:ptildedef}), which shows that we should perform exactly the same direct products (suppressing the dependence on all variables):
\beq
P^{+}_{aj} \rightarrow P^+_{aj} \otimes \begin{pmatrix} 0 \\ 1 \end{pmatrix}, \quad P^{-}_{aj} \rightarrow P^-_{aj} \otimes \begin{pmatrix} 1 \\ 0 \end{pmatrix},
\eeq
which implies $\left(P^-\right)^{-1}_{ja} \rightarrow \left(P^-\right)^{-1}_{ja} \otimes \begin{pmatrix} 1 & 0 \end{pmatrix}$. Eq. (\ref{eq:coupledcpluseq}) thus becomes
\beq
c_{a+} \otimes \begin{pmatrix} 0 \\ 1 \end{pmatrix} = \left \{ \left [ P^+(\epsilon)_{aj} \, \left( P^-(\epsilon)^{-1} \right)_{jb} \right] \otimes \begin{pmatrix} 1 & 0 \\ 0 & 0 \end{pmatrix} \right \} \, \left [ c_{b-} \otimes \begin{pmatrix} 1 \\ 0 \end{pmatrix} \right ].
\eeq
We now simply observe that, in such a representation, $-i\g^t = \mathbf{1}_N \otimes \s_1$. In the $N \times N$ subspace we want, $-i\g^t$ merely acts as the identity.

In summary, the retarded Green's function for coupled bulk fermions is
\beq
G^R_{ab}(\omega,k) = \lim_{\epsilon \rightarrow 0} \left ( P^+(\epsilon)_{aj} \, P^-(\epsilon)^{-1}_{jb} \right),
\eeq
with the matrices $P^{\pm}_{aj}$ defined in eq. \eqref{eq:pmatricesdef}.

Finally, as an important check, let us use our prescription to reproduce the result for free fermions, eq. (\ref{eq:freelorentziangreens}). For illustration, we consider $N=2$, so we have two bulk fermions, which we will call $\Psi_a$ and $\Psi_b$. We return to the equation of motion as written in eq. (\ref{eq:generalcoupledeom}), and assume the equations for $\Psi_a$ and $\Psi_b$ decouple, so that $\nabla_{ab \pm}$ and $M_{ab\pm}$ become diagonal in the $a$ and $b$ indices. We can further decouple the equations of motion by using the projectors $\Pi_{1,2}$. Acting with these, we obtain equations similar to eq. (\ref{eq:diraceqpi12projectors}). We thus find four decoupled equations, for $\Psi_{a1}$, $\Psi_{a2}$, $\Psi_{b1}$ and $\Psi_{b2}$.

We now solve the equations $2N = 4$ times, each time with a different normalization vector $\vec{n}$ for the $\Psi_{a-}$ and $\Psi_{b-}$ fields. In the first solution, all four fields have normalizations $\vec{n} = (n_{a-u},n_{a-d},n_{b-u},n_{b-d}) = (+1,+1,+1,+1)$. In the second solution, we use $\vec{n} = (+1,-1,+1,+1)$. The key observation is that the field $\Psi_{a-d}$ whose normalization we change is $\Psi_{a-d} = \Psi_{a1u}$ (recall eq. (\ref{eq:psiexplicit})), and hence couples only to $\Psi_{a1d} = \Psi_{a+d}$. The change in normalization thus leaves the other three fields, $\Psi_{a-u}$, $\Psi_{b-u}$, and $\Psi_{b-d}$ unchanged. The solutions for these fields will thus be identical to what they were using the original $+1$ normalizations. The $P^-_{aj}$ matrix thus takes the form (here we must write the up and down components explicitly)
\beq
\left [ \tilde{P}^-_{aj}(u,\omega,k) \right ] = \begin{pmatrix} \Psi^{(1)}_{1-u} & \Psi^{(1)}_{1-u} & \Psi^{(1)}_{1-u} & \Psi^{(1)}_{1-u} \\ \Psi^{(1)}_{1-d} & \Psi^{(2)}_{1-d} & \Psi^{(1)}_{1-d} & \Psi^{(1)}_{1-d} \\  \Psi^{(1)}_{2-u} & \Psi^{(1)}_{2-u} & \Psi^{(3)}_{2-u} & \Psi^{(1)}_{2-u} \\ \Psi^{(1)}_{2-d} & \Psi^{(1)}_{2-d} & \Psi^{(1)}_{2-d} & \Psi^{(4)}_{2-d} \end{pmatrix}, 
\eeq
with $\tilde{P}^+_{aj}(u,\omega,k)$ being identical except all $-$ subscripts become $+$. The main feature here is that all the superscripts are the same, except on the diagonal. A straightforward exercise (especially simple for $2 \times 2$ matrices) then shows that taking the inverse $\left(P^-\right)^{-1}_{ja}$ and then contracting with $P^+_{aj}$, and taking $\epsilon \rightarrow 0$, reproduces exactly the purely diagonal $c_{+u}/c_{-u}$ and  $c_{+d}/c_{-d}$ form of eq. (\ref{eq:freelorentziangreens}).

In summary: by combining the methods of refs. \cite{Amado:2009ts,Kaminski:2009dh} and \cite{Cubrovic:2009ye,Contino:2004vy}, we have provided a relatively simple prescription to compute the matrix-valued retarded two-point function from bulk solutions for coupled fermions. We simply solve the equations of motion (typically numerically) $2N$ times, using a different normalization vector $\vec{n}$ each time, use those solutions to construct the matrices $P^{\pm}_{aj}(\epsilon)$, and then take $\lim_{\epsilon \rightarrow 0} P^+(\epsilon)_{aj} \left( P^-(\epsilon)^{-1} \right)_{jb}$.

\section{Probe Branes and Holographic p-wave Superfluids}
\label{pwave}

In this section we review how to obtain a holographic p-wave phase transitions from simple string theory constructions of intersecting Dq-branes and Dp-branes \cite{Ammon:2008fc,Basu:2008bh,Ammon:2009fe,Peeters:2009sr} (see also ref. \cite{Roberts:2008ns}). We also present some new results for the particular D3/D5 system we subsequently explore in later sections.

\subsection{p-waves, Probe Branes, and Vector Meson Condensation}

The minimal ingredients for a holographic p-wave phase transition are gravity in a black hole spacetime with holographic variable $u$ (and some dual field theory), plus non-Abelian bulk gauge fields. We will consider the simple example of $SU(2)$ gauge fields $A_{M}^a$, where $a=1,2,3$ labels the generators $\tau_a$ of $SU(2)$, although other non-Abelian groups besides $SU(2)$ work just as well \cite{Gubser:2010dm,Manvelyan:2008sv}. As described in the introduction, the p-wave superfluid transition appears in the bulk as a charged black hole growing vector hair at low temperature.

As observed in refs. \cite{Ammon:2008fc,Basu:2008bh,Ammon:2009fe,Peeters:2009sr}, we can easily obtain holographic p-wave superfluids using well-known intersections of $N_c$ coincident Dq-branes with $N_f$ coincident Dp-branes in type II string theory, which we will refer to as Dq/Dp systems. The idea is to take the usual decoupling limit for the Dq-branes, which in particular means $N_c \rightarrow \infty$, to obtain type II supergravity in the near-horizon geometry of the Dq-branes. Starting with non-extremal Dq-branes produces a black hole spacetime.

If we keep $N_f$ fixed as $N_c \rightarrow \infty$, so that $N_f \ll N_c$, then we may neglect the effect of the Dp-branes on the supergravity fields. The Dp-branes are then probes of the background geometry, and their dynamics is described by the non-Abelian Born-Infeld action (possibly plus Wess-Zumino terms) with gauge group $U(N_f)$. If we introduce exactly $N_f = 2$ probe branes, then we have $U(2)$ gauge fields on the worldvolume of the Dp-branes. The $SU(2)$ subgroup gives us the $SU(2)$ gauge fields we want. Given that the non-Abelian Born-Infeld action is not known to all orders in the field strength, we are typically limited to working at the leading non-trivial order, which is the Yang-Mills term. We have thus obtained $SU(2)$ gauge fields in a black hole spacetime.

Crucially, notice that these Dq/Dp systems give rise to \textit{probe} gauge fields, rather than gauge fields coming from the supergravity sector. The probe limit is sufficient to study many properties of the p-wave phase transition \cite{Gubser:2008wv, Ammon:2008fc, Basu:2008bh,Ammon:2009fe,Peeters:2009sr}, however, the probe limit is known to fail at low temperatures, because the solutions with nonzero $A_x^1(u)$ have a field strength that increases as we cool the system, so that we can no longer neglect the back-reaction (or trust the Yang-Mills approximation to the non-Abelian Born-Infeld action) \cite{Ammon:2008fc,Ammon:2009fe}. To reach zero temperature, we must solve the fully coupled equations of motion, which to date has only been done in \textit{ad hoc} models \cite{Ammon:2009xh,Gubser:2010dm,Basu:2009vv,Gubser:2008zu}.

The benefit of the Dq/Dp construction is that we can identify the dual theory, which we will call the Dq/Dp theory. The $N_c$ Dq-branes generically give rise to an $SU(N_c)$ gauge theory with fields only in the adjoint representation of $SU(N_c)$. Open strings from the $N_f$ Dp-branes to the Dq-branes give rise to fields in the fundamental representation of $SU(N_c)$, \textit{i.e.} flavor fields. In analogy with (supersymmetric) QCD, we will call any flavor fermions or scalars ``quarks'' or ``squarks,'' respectively. If the Dp-branes do not overlap with all $q$ spatial directions of the Dq-branes, then the flavor fields will be confined to propagate along some defect of nonzero codimension.

In the field theory, the probe limit consists of neglecting quantum effects due to the flavors, such as the effect on the running of the coupling, because such effects are parametrically suppressed by $N_f / N_c$. In the language of perturbation theory, the probe limit consists of discarding all diagrams involving quark or squark loops.

If we separate the D-branes in an overall transverse direction, we may give the Dq-Dp strings a finite length and hence the flavor fields a finite mass, although in this paper we consider only massless flavor fields (unless stated otherwise).

The $U(N_f)$ gauge invariance on the Dp-branes' worldvolume is dual to a $U(N_f)$ flavor symmetry, analogous to the vector symmetry of QCD. The overall $U(1)$ we identify as baryon (or really quark) number, and the $SU(2)$ subgroup we identify as isospin.

We can thus easily see what the bulk transition looks like in the field theory. We have a strongly-coupled, large-$N_c$ non-Abelian gauge theory coupled to $N_f = 2$ species of massless flavor fields, which may be confined to a defect. We study thermal equilibrium states with temperature $T$, and introduce an isospin chemical potential $\mu$ for $U(1)_3$. For sufficiently large $\mu$, the system develops a nonzero $\langle J^x_1 \rangle$. The operator $J^x_1$ is a gauge-invariant bilinear in the flavor fields, valued in the adjoint of $SU(N_f)$. For example, in two-flavor massless QCD, with up and down quarks $u$ and $d$, $J^x_1 \sim \bar{u} \g^x d$.

Such an operator is precisely what we would call a vector meson, and the phase transition appears to be vector meson condensation. To be precise, the spectrum of the Dq/Dp theory includes gauge-invariant bound states of flavor fields. We will refer to such bosonic or fermionic bound states as ``mesons'' or ``mesinos,'' respectively. For massive flavor fields, these mesons/mesinos are typically the lightest flavor degrees of freedom in the theory \cite{Kruczenski:2003be,Myers:2006qr}. We may thus imagine writing an effective theory for these degrees of freedom, analogous to the chiral Lagrangian of QCD. An isospin chemical potential $\mu$ will act as a negative mass-squared for any mesons/mesinos charged under $U(1)_3$. If we make $\mu$ sufficiently large, then we expect Bose-Einstein condensation of mesons. In QCD, we expect the lightest charged mesons, the pions, to condense first producing a scalar condensate (and hence an s-wave superfluid), while the heavier vector mesons may condense at higher $\mu$ \cite{Son:2000by,Sannino:2003fj}. Which mesons condense first in Dq/Dp systems depends on the details of the system. In Dq/Dp holographic models of QCD, such as the Sakai-Sugimoto model \cite{Sakai:2004cn}, holographic calculations suggest that indeed the pions condense first and the vector mesons second, as we increase $\mu$ \cite{Aharony:2007uu,Rebhan:2008ur}. The general lesson from these Dq/Dp systems is that the p-wave superfluid phase transition appears to be vector meson condensation, which is in line with our weak-couling intuition.

Moreover, thinking of the p-wave states as a Bose-Einstein condensate makes many potentially confusing features of the p-wave state transparent. For example, the p-wave transition appears to involve the spontaneous generation of a persistent current $\langle J^x_1 \rangle$, that is, at high density charges begin moving without experiencing dissipation.\footnote{Crucially, however, no net momentum is flowing. In holographic calculations, in both the probe and fully back-reacted cases \cite{Ammon:2009xh,Gubser:2010dm,Basu:2009vv}, the Yang-Mills stress-energy tensor and the metric are diagonal, and indeed the bulk spacetime is static, which indicates that the expectation value of the field theory stress-energy tensor is strictly diagonal. The system thus has zero net momentum. If charges are moving, they must be doing so in pairs that move in opposite directions. A static bulk spacetime also indicates that the energy density of the field theory is not changing in time: the system is not heating up, consistent with the fact that the moving charges experience no dissipation (no frictional forces).} While not impossible, such a scenario naturally raises some questions. Why do charges start moving? How does that lower the free energy? Vector meson condensation neatly accounts for all of the physics: we merely see Bose-Einstein condensation, \textit{i.e.} bosons populating a zero-momentum state, the main novelty being that the bosons are vectors, not scalars.

We are interested in condensed matter applications, and in particular quantum critical theories, which are scale-invariant, hence our Dq-branes will be D3-branes, whose near-horizon geometry is $AdS_5 \times S^5$. The dual theory is then a CFT, namely (3+1)-dimensional $\N=4$ SYM with large $N_c$ and and large 't Hooft coupling. Our Dp-branes will preserve half the supersymmetry of the background, which means Dp-branes extended along $AdS_P \times S^Q$ with $P+Q = p+1$, where supersymmetry requires $|P-Q|=2$ \cite{Skenderis:2002vf}. Well-known examples include D7-branes extended along $AdS_5 \times S^3$ \cite{Karch:2002sh,Erdmenger:2007cm} or $AdS_3 \times S^5$ \cite{Harvey:2007ab,Buchbinder:2007ar,Harvey:2008zz}, D5-branes extended along $AdS_4 \times S^2$ \cite{DeWolfe:2001pq,Erdmenger:2002ex}, or D3-branes along $AdS_3 \times S^1$ \cite{Constable:2002xt}.\footnote{These are the cases in which the flavor fields propagate in at least one spatial dimension, and hence a nonzero $\langle J^x_1 \rangle$ is possible.} For two coincident D7-branes with $P=5$ and $Q=3$, holographic calculations have shown that a p-wave transition occurs precisely when peaks in the Green's function, namely those corresponding to vector mesons charged under $U(1)_3$ in the vacuum state, cross into the upper-half of the complex frequency plane, indicating an instability toward Bose-Einstein condensation.

With an eye toward condensed matter applications, and for technical reasons we explain in section \ref{worldvolumefermions}, we will work with (two coincident) D5-branes with $P=4$ and $Q=2$. The dual theory is thus $\N=4$ SYM coupled to $N_f=2$ flavor fields that propagate only in 2+1 dimensions (a codimension one defect) and preserve (2+1)-dimensional $\N=4$ supersymmetry (eight real supercharges). The field content and Lagrangian of the D3/D5 theory were determined in refs. \cite{DeWolfe:2001pq, Erdmenger:2002ex}, which we review in section \ref{dualoperators}. The D3/D5 system exhibits rich thermodynamics, studied in detail (holographically) in \cite{Wapler:2009rf,Myers:2008me,Evans:2008nf,Filev:2009xp,Wapler:2009tr,Filev:2009ai,Wapler:2010nq,Jensen:2010ga}. Here we initiate the study of the D3/D5 theory with a finite isospin chemical potential and finite temperature. As expected, we will find a p-wave phase transition.

Our ultimate goal is to compute, holographically, fermionic retarded Green's functions in p-wave superfluid states. We discuss fermionic excitations on the worldvolume of the D5-branes, and the dual mesinos, in section \ref{worldvolumefermions}. We will be able to compare our numerical results with previous studies, however, we cannot exploit the analytic results for the form of fermionic Green's functions derived in ref. \cite{Faulkner:2009wj}. The analysis of ref. \cite{Faulkner:2009wj} involved fermions in an extremal Reissner-Nordstr\"om background, and in particular made great use of the emergent near-horizon $AdS_2$ factor. Without having access to the $T=0$ finite-density state, we do not know for sure whether the geometry exhibits an emergent $AdS_2$ or something similar. Nevertheless, the equations of motion for our fermions are formally similar to those of refs. \cite{Liu:2009dm,Faulkner:2009wj}, and hence we can recover similar finite-temperature results.

Despite the bad news that we cannot reach $T=0$ in the probe limit, we do have good news: we can study fermionic response near the p-wave transition. The high-temperature normal phase is rotationally symmetric, but the p-wave phase is of course not. We will see the breaking occur explicitly in fermionic spectral functions in section \ref{numerical}.

\subsection{Probe Dp-branes in $AdS_5 \times S^5$}
\label{probedpbranesinads}

We now want to study a p-wave superfluid transition for (2+1)-dimensional flavor fields described holographically by two coincident D5-branes with $P=4$ and $Q=2$. Although our (numerical) analysis will be for the D5-brane, in the interest of generality, and to connect to our discussion in section \ref{worldvolumefermions}, we will write formulas for an arbitrary Dp-brane with $P \geq 3$.

The background supergravity solution includes a metric and Ramond-Ramond (RR) five-form. The five-form will be important in section \ref{worldvolumefermions}. The spacetime is (4+1)-dimensional AdS-Schwarzschild times $S^5$, with metric
\beq
ds^2 = \frac{1}{u^2} \left( \frac{du^2}{f(u)} -f(u) dt^2 + d\vec{x}^2 \right) + ds^2_{S^5},
\eeq
with
\beq
f(u) = 1 - \frac{u^4}{u_h^4}, \qquad u_h = \frac{1}{\pi T}.
\eeq
In our units, where the AdS radius is one, we can convert between string theory and field theory quantities using $\alpha'^{-2} = 4\pi g_s N_c = 2 g_{YM}^2 N_c = 2 \lambda$.

As we will be studying fermions in section \ref{worldvolumefermions}, we will need the vielbeins and spin connection associated with the metric above. We record these here for later use. The nonzero vielbeins of (4+1)-dimensional AdS-Schwarzschild are (recall that upper index is local Lorentz, and the lower index is general coordinate),
\beq
\label{eq:adsbhvielbeinsdef}
e^u_{~u} = \frac{1}{u\, \sqrt{f}}, \qquad e^t_{~t} = \frac{\sqrt{f}}{u}, \qquad e^i_{~j} = \frac{1}{u} \, \delta^i_{~j}.
\eeq
The spin connection $\omega$ of (4+1)-dimensional AdS-Schwarzschild then has the nonzero components
\beq
\omega_{tu} = \left( \frac{f(u)}{u} - \frac{f'(u)}{2} \right) dt, \qquad \omega_{\vec{x} u} = - \frac{\sqrt{f(u)}}{u} \, d\vec{x},
\eeq
where $\omega_{\vec{x} u}$ indicates the three components $\omega_{xu}$, $\omega_{yu}$, and $\omega_{zu}$.

Next we introduce two coincident probe Dp-branes extended along $AdS_P \times S^Q.$ We will only consider the trivial embedding of the Dp-branes, that is, we consider solutions in which all the Dp-branes' worldvolume scalars (including scalars in $AdS_5$ directions) are zero. The dual flavor fields are then massless. The induced metric on the Dp-branes
\beq
ds^2_{Dp} = \frac{1}{u^2} \left( \frac{du^2}{f(u)} -f(u) dt^2 + d\vec{x}^2 \right) + ds^2_{S^Q},
\eeq
where now $d\vec{x}^2$ represents the appropriate-dimensional Euclidean metric.

We want non-trivial worldvolume $SU(2)$ gauge fields. The action for the gauge fields, to leading non-trivial order, is
\beq
S_{Dp} = - T_{Dp} N_f \int d^{p+1} \xi \, \sqrt{-g_{Dp}} \, \left [ 1 + (2\pi\alpha')^2 \frac{1}{2} Tr \left( F_{\m\n} F^{\m\n} \right) \right ], \nonumber \\
\eeq
where $T_{Dp} = \left(2\pi\right)^{-p} g_s^{-1} \left(\alpha'\right)^{-\frac{p+1}{2}}$ is the tension of the Dp-brane, $N_f = 2$, the integral is over the worldvolume coordinates $\xi^{\mu}$, $g_{Dp}$ is the determinant of the induced metric, and the trace is over gauge indices. We use $SU(2)$ generators $\tau_a = \frac{1}{2} \s_a$ such that, with $\epsilon_{123} = +1$
\beq
\left [ \tau_a, \tau_b \right] = i \, \epsilon_{abc} \, \tau_c.
\eeq
The field strength $F_{\mu \nu} = F_{\mu \nu}^a \, \tau_a$.

The equation of motion for the gauge field is simply the Yang-Mills equation,
\beq
\label{ymeq}
\nabla_{\mu} F^{\mu \nu}_a + f_{abc} \, (A_{\mu})_b \, F^{\mu \nu}_c = 0.
\eeq
For probe Dp-branes wrapping $AdS_P \subseteq AdS_5$ with $P\geq 3$, we will consider solutions of eq. (\ref{ymeq}) of the form
\beq
\label{gaugefieldansatzpwave}
A = A^1_x(u) \, \tau_1 \, dx + A^3_t(u) \, \tau_3 \, dt,
\eeq
in which case the Yang-Mills equation becomes
\begin{subequations}
\label{eoms}
\beq
\label{Ateom}
\left(A_t^3\right)^{''} + \frac{4-P}{u} \left(A_t^3\right)' - \frac{1}{f(u)} A_t^3 \left(A_x^1\right)^2 =0,
\eeq
\beq
\label{Azeom}
\left(A_x^1\right)^{''} + \left(\frac{4-P}{u} + \frac{f'(u)}{f(u)}\right) \left(A_x^1\right)' + \frac{1}{f(u)^2}\left(A_t^3\right)^2 A_x^1 = 0,
\eeq
\end{subequations}
where primes denote $\partial_u$. The equations of motion determine the asymptotic forms of the solutions,\footnote{Notice that $A_x^1(u)$ has no leading constant the way $A^t_3(u)$ does, so no source for $J^x_1$ is present in the field theory: $U(1)_3$ will be broken spontaneously.}
\beq
A_t^3(u) = \mu - d^3_t  \, u^{P-3} + \ldots, \qquad A_x^1(u) = d^1_x \, u^{P-3} + \ldots,
\eeq
where $\ldots$ represent terms that decay faster than $u^{P-3}$ as $u \ra 0$. Here the constant $d_t^3$ is related to $\langle J^t_3 \rangle$ as
\beq
\label{eq:densityconstantproportionality}
\langle J^t_3 \rangle = N_f T_{Dp} \left( 2\pi\alpha'\right)^2 \, d_t^3 = \left( 2\pi\right)^{-p+3} \, 2^{\frac{p-3}{4}} \, N_f N_c \, \lambda^{\frac{p-7}{4}} \, d_t^3,
\eeq
where in the second equality we converted to field theory quantities. Similarly, $\langle J^x_1 \rangle = N_f T_{Dp} \left( 2\pi\alpha'\right)^2 \, d_x^1$. Notice in particular that both $\langle J^t_3 \rangle$ and $\langle J^x_1 \rangle$ are proportional to $N_f N_c$.

One solution of eqs. (\ref{eoms}) has $A_x^1(u)=0$ and
\beq
\label{gaugefieldansatznormal}
A^3_t(u) = \mu \left(1-\frac{u^{P-3}}{u_h^{P-3}}\right).
\eeq
Such a solution corresponds in the field theory to the normal phase, in which the chemical potential $\mu$ explicitly breaks the $SU(2)$ isospin symmetry down to the $U(1)_3$, but no spontaneous symmetry breaking occurs. These solutions exist for all values of $\mu$.

Notice that these D3/Dp theories in the probe limit are scale-invariant\footnote{In the probe limit we neglect the quantum effects that would cause the $\N=4$ SYM coupling to run, and we are working with massless flavor fields. The theory is thus in a limit where no intrinsic scale appears.}, so the only meaningful physical quantity is $\mu/T$, so fixing $T$ and increasing $\mu$ is equivalent to fixing $\mu$ and reducing $T$. We will think in terms of the latter. Any transition must occur at a temperature $T_c$ set by the chemical potential, $T_c \propto \mu$.

For sufficiently low $T$ (or large $\mu$), other solutions of eq. (\ref{ymeq}) exists in which $A_x^1(u)$ is nonzero. For the D7-brane with $P=5$ and $Q=3$, such solutions were found numerically in refs. \cite{Ammon:2008fc,Basu:2008bh,Ammon:2009fe}.  These solutions correspond in the field theory to superfluid states, with nonzero $\langle J_x^1 \rangle$, so $U(1)_3$ is spontaneously broken. For $P \geq 4,$ the field theory's spatial rotational symmetry\footnote{When $P=3$ the flavor fields are confined to a (1+1)-dimensional defect and hence have no spatial rotational symmetry. Notice that in the $P=3,4$ cases the large $N_c$ limit is what permits spontaneous symmetry breaking to occur, by suppressing the fluctuations that would destroy long-range order.} is also broken from $SO(P-2)$ down to $SO(P-3)$.

For sufficiently low $T$, we have two solutions, so we need to determine which is thermodynamically preferred. As shown in refs. \cite{Ammon:2008fc,Basu:2008bh,Ammon:2009fe} for the D7-brane with $P=5$ and $Q=3$, the superfluid phase is thermodynamically preferred relative to the normal phase for all $T/\mu$ where the solutions with nonzero $A^x_1(u)$ exist. The transition between the phases is second order, with mean-field exponents. In particular, near the transition, the condensate has mean field exponent $1/2$: $\langle J^x_1 \rangle \propto \left( 1 - T/T_c\right)^{1/2}$.

\begin{figure}
\begin{center}
\psfrag{d1x}[r]{$\frac{2^{3/2} \lambda^{1/2}}{N_f N_c T^2} \langle J_x^1 \rangle$}
\psfrag{T}{$T/T_c$}
\includegraphics{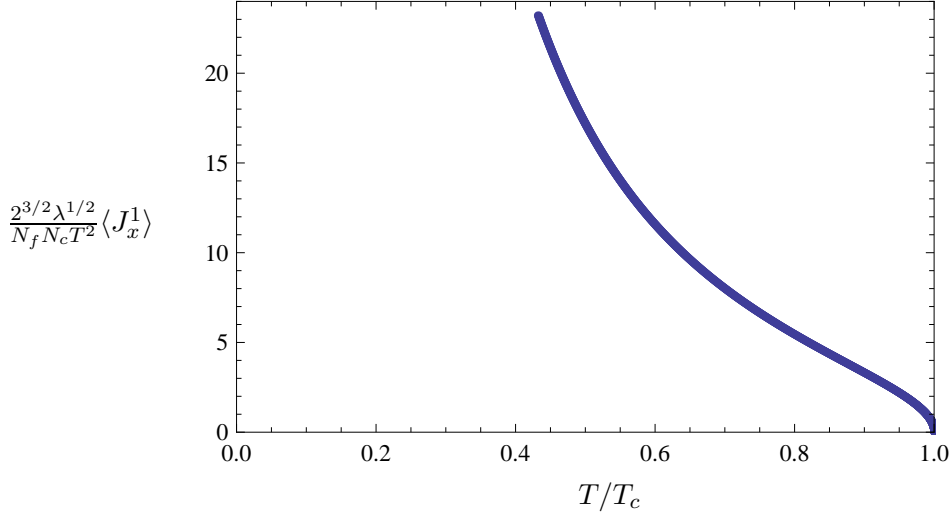}
\caption{The condensate $\langle J_x^1 \rangle$, times $\frac{2^{3/2} \lambda^{1/2}}{N_f N_c T^2}$, versus the rescaled temperature $T / T_c.$}
\label{superconducting}
\end{center}
\end{figure}

For two coincident D5 branes with $P=4$ and $Q=2$, the story is qualitatively the same. For $\mu \geq 3.81 \times \left( \pi T\right)$ the state with nonzero $\langle J_x^1 \rangle$ has lower free energy. In other words, $T_c = \frac{\mu}{3.81 \times \pi}$. In figure (\ref{superconducting}) we plot the constant $d_x^1$, which is proportional to $\langle J^x_1 \rangle$, versus the rescaled temperature $T/T_c$. Near the transition, $\langle J^x_1 \rangle$ appears to have a mean-field exponent of $1/2$ as in the D7-brane case \cite{Ammon:2008fc,Basu:2008bh,Ammon:2009fe}.

Notice that as $T$ decreases, the condensate grows, which, as explained in refs. \cite{Hartnoll:2008vx,Hartnoll:2008kx}, suggests that we are leaving the probe limit. We will only present results in the p-wave phase for $T/T_c \gtrsim 0.4$ (as plotted in figure (\ref{superconducting})), where we have some hope that the probe approximation captures the essential physics faithfully.

\section{The Worldvolume Fermions}
\label{worldvolumefermions}

We want to study fluctuations of fermionic operators of the D3/D5 theory with finite temperature and isospin chemical potential, in the two phases described above, the normal (non-superfluid) phase and the superfluid phase. On the field theory side, we will study mesino operators valued in the adjoint representation of the $SU(2)$ isospin symmetry. As we will see below, we will thus have three mesinos, two with equal and opposite charges under the $U(1)_3$, and one that is neutral. We discuss the form of the mesino operators in section \ref{dualoperators}.

To be specific, we will compute holographically the retarded two-point function of mesinos as a function of frequency and momentum. On the gravity side, that means studying fermionic fluctuations of the Dp-branes, and in particular solving their linearized equation of motion in the background we found in the last section, where the geometry is (4+1)-dimensional AdS-Schwarzschild and the D5-branes have non-trivial worldvolume gauge fields. As for all supersymmetric Dp-branes, the worldvolume fermions are in a supermultiplet with the worldvolume gauge field and scalars, and hence are in the adjoint of the worldvolume $SU(2)$ gauge group, which is dual to the statement that the mesinos are in the adjoint of the isospin symmetry.\footnote{Obviously, the mesinos carry no baryon number, which is dual to the statement that the worldvolume fermions, like all of the worldvolume fields, do not couple via a gauge-covariant derivative to the diagonal $U(1)$ part of the worldvolume gauge field.}

As we will see, in the normal phase the three bulk fermions decouple, which is dual to the statement that the retarded Green's function is diagonal. In the superfluid phase, however, where $A_x^1(u)$ is nonzero, the three fermions couple, indicating that the dual operators mix under renormalization group flow in the field theory. The Green's function is then a $6 \times 6$ matrix, where the $6$ is the number of fermions times the two components of the fermions (two, using the $\Pi_{1,2}$ projectors). We thus have a perfect testing ground for the method we developed in section \ref{mixing} for computing Green's functions from coupled bulk fermions.

The fermionic part of general Dp-brane actions, to quadratic order in the fermionic fields and in backgrounds with non-trivial RR forms, was determined in refs. \cite{Marolf:2003ye, Marolf:2003vf,Martucci:2005rb}. The general couplings were derived by starting with the action for a supermembrane in M-theory, written in a superspace formalism, expanding the action to second order in the Grassmann variables, reducing to type IIA supergravity, and then performing a T-duality to type IIB. The form of the quadratic fermionic action on the Dp-brane worldvolume is thus determined completely by supersymmetry and T-duality.

Using the worldvolume fermion actions of refs. \cite{Marolf:2003ye, Marolf:2003vf,Martucci:2005rb}, the spectra of mesinos in the D3/D7 theory (for the D7-brane with $P=5$ and $Q=3$) and in the Sakai-Sugimoto model were determined in refs. \cite{Kirsch:2006he, Heise:2007rp}. We will very closely follow the D7-brane analysis of ref. \cite{Kirsch:2006he}, which in turn was the fermionic generalization of the analysis of ref. \cite{Kruczenski:2003be} for mesons. For a Dp-brane extended along $AdS_P \times S^Q$, we consider a worldvolume spinor that is a spherical harmonic on the $S^Q$. We reduce the worldvolume spinor on the $S^Q$, obtaining an effective Dirac action in $AdS_P$. This procedure fixes the masses of the bulk fermionic excitations, which allows us to identify the dimensions of the dual mesinos, and more generally to map bulk fluctuations to mesino operators. As emphasized in ref. \cite{Kirsch:2006he}, the coupling to the RR five-form is crucial to obtain the correct bulk masses.\footnote{The authors of ref. \cite{Hung:2009qk} appear to omit the coupling to the five-form when they study fermionic fluctuations of the probe D3-brane extended along $AdS_3 \times S^1$.}

One of our main points is: because we work with a particular string theory system, we do not have the freedom to change the mass or the charge of our bulk fermion, in stark contrast to the models of refs. \cite{Liu:2009dm,Cubrovic:2009ye,Faulkner:2009wj}. Both the masses and charges are, ultimately, fixed by supersymmetry and T-duality, as explained above.

\subsection{Equation of Motion I: Reduction to AdS}
\label{eom1reduction}

We will now repeat the analysis of ref. \cite{Kirsch:2006he}, in which the fermionic action of a D7-brane extended along $AdS_5 \times S^3$ was reduced to an effective Dirac action in $AdS_5$, but now for more general Dp-branes extended along $AdS_P \times S^Q$, with emphasis on D5-branes with $P=4$ and $Q=2$.\footnote{In section \ref{pwave} we were interested in p-wave states and hence required $P \geq 3$. In this subsection we relax that constraint. Our results will thus also apply for D5-branes extended along $AdS_2 \times S^4$, which were used in ref. \cite{Kachru:2009xf} to construct a holographic model of fermions at lattice sites (that can pair to form dimers).}

The quadratic action for fermionic fluctuations of the Dp-branes is (refs. \cite{Martucci:2005rb,Kirsch:2006he})
\bea
\label{eq:dpbranefermionicaction}
S_{Dp} = N_f T_{Dp} \int d^{p+1} \xi \, \sqrt{-g_{Dp}} \, \frac{1}{2} \, Tr \left [ \hat{\bar{\J}} P_- \G^{\hat{A}} \left( D_{\hat{A}} + \frac{1}{8} \frac{i}{2 * 5!} F_{\hat{N}\hat{P}\hat{Q}\hat{R}\hat{S}} \G^{\hat{N}\hat{P}\hat{Q}\hat{R}\hat{S}} \G_{\hat{A}} \right) \hat{\J} \right ], \nonumber
\eea
Here $\hat{\J}$ is a ten-dimensional positive-chirality Majorana-Weyl spinor of type IIB supergravity, the $\G_{\hat{A}}$ are the pullback of the ten-dimensional $\G$-matrices to the Dp-brane worldvolume, $\G_{\hat{A}} = \G_{\hat{M}} \partial_{\hat{A}} x^{\hat{M}}$ (we use a trivial embedding, so the pullback is trivial), $P_-$ is a $\k$-symmetry projector that ensures $\k$-symmetry invariance of the action, $D_A$ is a (gauge and curved-space) covariant derivative, and $F_{\hat{N}\hat{P}\hat{Q}\hat{R}\hat{S}}$ is the five-form of the background. Notice that here $F_{\hat{N}\hat{P}\hat{Q}\hat{R}\hat{S}}$ is \textit{not} the pullback of the five-form to the Dp-brane worldvolume, rather, it is the five-form evaluated on the submanifold spanned by the Dp-brane. Indeed, no part of the expression $F_{\hat{N}\hat{P}\hat{Q}\hat{R}\hat{S}} \G^{\hat{N}\hat{P}\hat{Q}\hat{R}\hat{S}}$ involves a pullback. Here $\hat{A}, \hat{B}, \ldots$ denote all worldvolume indices, while below $A,B,\ldots$ denote AdS-Schwarzschild coordinates which are wrapped by the probe brane. Moreover, the indices of the coordinates on the sphere $S^Q$ are labelled by $a,b,\ldots.$ Notice that the fermion $\hat{\J}$ is in the adjoint representation of $SU(2)$.

The equation of motion for the fermion is (for now we suppress gauge indices)
\beq
\label{eq:dpbranefermiondiracaction}
\left [ \G^{\hat{A}} D_{\hat{A}} + \frac{1}{8} \frac{i}{2 * 5!} \G^{\hat{A}} F_{\hat{N}\hat{P}\hat{Q}\hat{R}\hat{S}} \G^{\hat{N}\hat{P}\hat{Q}\hat{R}\hat{S}} \G_{\hat{A}} \right] \hat{\J} = 0
\eeq
We will reduce the equation of motion for the fermion to a Dirac equation in $AdS_P$, following ref. \cite{Kirsch:2006he} very closely. First we decompose every ten-dimensional spinor and $\G$-matrix into parts associated with $AdS_5$ and $S^5$. In a local Lorentz frame, the $\G$-matrices decompose as
\beq \label{gammamatrices}
\G^M = \s_2 \otimes \mathbf{1}_4 \otimes \g^M, \qquad \G^m = \s_1 \otimes \g^m \otimes \mathbf{1}_4,
\eeq
where $\mathbf{1}_4$ is the $4\times4$ identity matrix, the index $M$ runs over $AdS_5$ directions (which we will generically call $01234$), and the index $m$ runs over $S^5$ directions (which we will call $56789$). The $\g$-matrices are five-dimensional, obeying the usual relations
\beq
\left \{ \g^M, \g^N \right\} = 2 \eta^{MN}, \qquad \left \{ \g^m, \g^n \right \} = 2 \delta^{mn}.
\eeq
Given the above decompositions, we then have
\beq
\G^{01234} = i \s_2 \otimes \textbf{1}_4 \otimes \textbf{1}_4, \qquad \G^{56789} = \s_1 \otimes \textbf{1}_4 \otimes \textbf{1}_4
\eeq
\beq
\G^{11} = \G^{0123456789} = \s_3  \otimes \textbf{1}_4 \otimes \textbf{1}_4
\eeq
The ten-dimensional spinor $\hat{\J}$ has positive chirality, $\G^{11} \hat{\J} = \hat{\J}$, and decomposes as
\beq \label{typeIIBspinor}
\hat{\J} = \uparrow \otimes \chi \otimes \J,
\eeq
where $\uparrow = \left( \begin{array}{c} 1 \\ 0 \end{array} \right)$, and $\chi$ and $\J$ are four-component spinors of $SO(5)$ and $SO(4,1)$, which act on the tangent spaces of $S^5$ and $AdS_5$, respectively. The spinor $\chi$ further decomposes as $\chi = \chi_{\parallel} \otimes \chi_{\perp}$, where $\chi_{\parallel}$ is a spinor associated with the $S^Q$ that the Dp-brane wraps and $\chi_{\perp}$ is associated with the $S^5$ directions transverse to the $S^Q$.

We parameterize the five-form in terms of the volume forms of $AdS_5$ and $S^5$, which we denote as $\Omega_{AdS_5}$ and $\Omega_{S^5}$,
\beq
F_{NPQRS} = 4 \, \left(\Omega_{AdS_5}\right)_{NPQRS}, \qquad F_{npqrs} = 4 \, \left( \Omega_{S^5}\right)_{npqrs}. \nonumber
\eeq
Using the decomposition of the type IIB spinor in eq. \eqref{typeIIBspinor} and of the $\g$-matrices in eq. \eqref{gammamatrices} we obtain
\bea \label{gammamatrixidentity}
\G^{a} \left( \Gamma^{01234} + \Gamma^{56789} \right) \G_{a} \left( \uparrow \otimes \chi \otimes \J \right) &=& 2 \downarrow \otimes \chi \otimes \J \nonumber \, , \\
\G^{A} \left( \Gamma^{01234} + \Gamma^{56789} \right) \G_{A} \left( \uparrow \otimes \chi \otimes \J \right) &=& -2 \downarrow \otimes \chi \otimes \J \, ,
\eea
In eq. \eqref{gammamatrixidentity} we have not summed over $a$ or $A.$ Using this result we can simplify the coupling of the spinor to the five-form,
\beq
\frac{1}{8} \frac{i}{2 * 5!} \G^{\hat{A}} F_{\hat{N}\hat{P}\hat{Q}\hat{R}\hat{S}} \G^{\hat{N}\hat{P}\hat{Q}\hat{R}\hat{S}} \G_{\hat{A}} \hat{\J} = \frac{i}{4} \G^{\hat{A}} \left( \left( \s_1 + i \s_2 \right) \otimes \mathbf{1}_4 \otimes \mathbf{1}_4 \right) \G_{\hat{A}} \hat{\J} = - \frac{i}{2} (P-Q) \left( \downarrow \otimes \chi \otimes \J \right)
\eeq
where here we do sum over $\hat{A}$. We can also extract the $S^Q$ and the $AdS_P$ part of the derivative terms as
\bea
\G^{\hat{A}} D_{\hat{A}} \hat{\J} & = & \G^A D_A \hat{\J} + \G^a D_a \hat{\J} \nonumber \\ & = & \left( \left( \s_2 \otimes \mathbf{1}_4 \otimes \g^A D_A \right) + \left( \s_1 \otimes \g^a D_a \otimes \mathbf{1}_4 \right) \right) \left( \uparrow \otimes \chi \otimes \J \right) \nonumber \\ & = & \left( i \left( \mathbf{1}_2 \otimes \mathbf{1}_4 \otimes \g^A D_A \right) + \left(\mathbf{1}_2 \otimes \g^a D_a \otimes \mathbf{1}_4 \right) \right) \left( \downarrow \otimes \chi \otimes \J \right) \nonumber \\ & \equiv & \left( i \D_{AdS_P} + \D_{S^Q} \right) \left( \downarrow \otimes \chi \otimes \J \right),
\eea
where $\D_{AdS_P}$ and $\D_{S^Q}$ are the Dirac operators of $AdS_P$ and $S^Q$, respectively. The Dirac operator on a sphere $S^Q$ has spinor spherical harmonics $\chi_{\ell}^{\pm}$ that obey
\beq
\label{eq:diracsphericalharmonicspectrum}
\D_{S^Q} \chi_{\ell}^{\pm} = \mp \frac{i}{R_Q} \left( \ell + \frac{Q}{2} \right) \chi_{\ell}^{\pm},
\eeq
where $\ell \geq 0$ and $R_Q$ is the radius of the $S^Q$. In our units, $R_Q=1$. For $Q=3$, relevant for the D7-brane along $AdS_5 \times S^3$, the spinors $\chi_{\ell}^+$ are in the $\left( \frac{\ell+1}{2}, \frac{\ell}{2}\right)$ representation of the $SO(4)$ that acts on $S^3$, while the spinors $\chi_{\ell}^-$ are in the  $\left( \frac{\ell}{2}, \frac{\ell+1}{2}\right)$ representation. For $Q=2$, relevant for the D5-brane along $AdS_4 \times S^2$, the spinors $\chi_{\ell}^{\pm}$ are in the $\left(\ell + \frac{1}{2} \right)$ of the $SU(2) \simeq SO(3)$ that acts on $S^2$.

Inserting everything into eq. \eqref{eq:dpbranefermiondiracaction}, we find
\beq
\label{eq:dpbraneworldvolumefermionsdiracequation}
\left( \D_{AdS_P} \mp \left( \ell + \frac{Q}{2} \right) - \frac 1 2 (P - Q) \right) \, \J^{\pm}_{\ell} = \left \{ \begin{array}{c} \left( \D_{AdS_P} - \left( \ell + \frac{1}{2} P \right) \right) \, \J^+_{\ell} \\ \left( \D_{AdS_P} + \left( \ell - \frac{1}{2} P + Q\right) \right) \, \J^-_{\ell} \end{array} \right \} = 0.
\eeq
The fermions\footnote{The $\pm$ superscript here refers to the sign of the eigenvalue of the Dirac operator in eq. \eqref{eq:diracsphericalharmonicspectrum}, not to the projectors $\Pi_{\pm}$ defined in section \ref{mixing}.} $\J^{\pm}$ thus have masses (in our units, where the radius of AdS is one)
\beq
\label{MassesOfDualOperators}
m^+_{\ell} = \ell + \frac{P}{2}, \qquad m^-_{\ell} = - \left( \ell + Q - \frac{1}{2} P \right).
\eeq
We collect the values of $m^{\pm}_{\ell}$ for our Dp-branes of interest in the table below.

\begin{table}[h!]
\begin{center}
\begin{tabular}{|c|c|c|c|c|c|c|c|c|}
\hline
Dp & $P$ & $Q$ & $[\psi]$ & $[q]$ & $m^+_{\ell} = \ell + P/2$ & $\Delta^+_{\ell}$ & $|m^-_{\ell}| = \ell + Q - P/2$ & $\Delta^-_{\ell}$ \\ \hline \hline
D7 &  5 & 3 & 3/2 & 1 & $\ell + 5/2$ & $\ell + 9/2$ & $\ell + 1/2$ & $\ell + 5/2$ \\ \hline
D7 & 3 & 5 & 1/2 & - & $\ell + 3/2$ & $\ell + 5/2$ & $\ell + 7/2$ & $\ell + 9/2$ \\ \hline
D5 & 4 & 2 &  1 & 1/2 & $\ell + 2$ & $\ell + 7/2$ & $\ell$ & $\ell + 3/2$ \\ \hline
D5 & 2 & 4 & 0 & - & $\ell + 1$ & $\ell + 3/2$ & $\ell +3$ & $\ell + 7/2$ \\ \hline
D3 & 3 & 1 & 1/2 & 0 & $\ell + 3/2$ & $\ell + 5/2$ & $\ell - 1/2$ & $\ell + 1/2$ \\ \hline
\end{tabular}
\end{center}
\caption{Masses of fermionic excitations on the worldvolume of a Dp-brane extended along $AdS_P \times S^Q$ inside $AdS_5 \times S^5$. We list Dp-branes that are known to preserve eight real supercharges (at zero temperature and density), in which case $|P-Q| = 2$. Here $\psi$ denotes a generic quark field and $q$ denotes a generic squark field. $\Delta^{\pm}_{\ell}$ denotes the dimension of the operator dual to the bulk fermion with mass $m^{\pm}_{\ell}$, with $\Delta^{\pm}_{\ell} = \frac{P-1}{2} + |m^{\pm}_{\ell}|$. For the D3-brane, the values of $m^-_{\ell}$ and $\Delta^-_{\ell}$ shown are for $\ell \geq 1$ only, whereas for $\ell = 0$, $|m^-_0|= 1/2$ and $\Delta^-_0 = 3/2$.}
\end{table}

Notice that since $P$ and $Q$ are integers, the $m_{\ell}^{\pm}$ will always be integer or half-integer. As we review in the appendix, a bulk fermion with an integer or half-integer mass $m$ is dual to a fermionic operator of dimension $\Delta = \frac{P-1}{2} + |m|$. We include the values of $\Delta^{\pm}_{\ell} = \frac{P-1}{2} + |m^{\pm}_{\ell}|$ in the table.

To get a rough idea of which operators correspond to which bulk fermion, we can do some dimension counting.  Let us denote a generic quark as $\psi$, a generic squark as $q$, a generic adjoint Majorana fermion as $\lambda$, and a generic adjoint real scalar as $X$. The dimensions of the fields are $[\psi] = \frac{P-2}{2}$, $[q]=\frac{P-3}{2}$, $[\lambda] = 3/2$, $[X] = 1$.

For the D7-brane extended along $AdS_5 \times S^3$, the D5-brane along $AdS_4 \times S^2$, and the D3-brane along $AdS_3 \times S^1$, all of which have $P-Q=2$, the dual flavor fields comprise a supermultiplet with both quarks $\psi$ and squarks $q$. In these cases, we can build a gauge-invariant mesino in two ways  \cite{Kirsch:2006he}. One way is to construct an operator of the form $\bar{\psi} \lambda \psi + q^{\dagger} X \lambda q$, with dimension $\Delta = P - 1/2$. We can additionally include some number $\ell$ of adjoint scalars\footnote{Notice these are not necessarily all the same scalar, \textit{i.e.} $X^{\ell}$ could represent $\ell$ distinct scalars. At the moment we are just counting dimensions, ignoring this subtlety.} as $\bar{\psi} \lambda X^{\ell} \psi + q^{\dagger} X \lambda X^{\ell} q$, so that the dimension is $\Delta = \ell + P - 1/2$. Inspecting the table, these are precisely the $\Delta^+_{\ell}$, so apparently these kinds of mesinos are dual to the bulk fermions with masses $m^+_{\ell}$. The other way to build a mesino is to construct $\bar{\psi} X^{\ell} q$ (plus the Hermitian conjugate), with dimension $\Delta = \ell + P - 5/2$. For Dp-branes with $P-Q=2$, these dimensions are precisely the $\Delta^-_{\ell}$, so apparently mesinos of this type are dual to the fermions with masses $m^-_{\ell}$.

For the D7-brane extended along $AdS_3 \times S^5$ and D5-brane along $AdS_2 \times S^4$, which have $P-Q=-2$, the dual flavor fields are quarks alone, with no squarks \cite{Harvey:2007ab,Buchbinder:2007ar,Harvey:2008zz,Gomis:2006sb}. The mesinos with dimensions $\Delta^+_{\ell}$ are of the same form, $\bar{\psi} \lambda X^{\ell} \psi$, but the mesinos with dimensions $\Delta^-_{\ell}$ must obviously have a different form. We leave a detailed study of these mesinos for the future.

Looking at the table, we immediately notice that the D5-branes are special: for these, the masses of the worldvolume fermions are integers. The reason is that the D5-branes wrap even-dimensional spheres, so the eigenvalue in eq. (\ref{eq:diracsphericalharmonicspectrum}) is $\pm i$ times an integer.

In the next subsection we focus on the D5-brane extended along $AdS_4 \times S^2$, explaining in more detail the symmetries of the theory and the form of the mesinos. In the subsequent sections, we focus on the single worldvolume fermion with mass $m^-_0 = 0$, for a number of reasons. First, of all the worldvolume fermions, these have the smallest mass, hence the dual operator will have the lowest dimension, $\Delta^-_{0} = 3/2$, and hence be the most relevant mesino. Second, a numerical analysis is simpler when the fermion's mass is zero. Third, with a massless bulk fermion we can directly compare to the results of refs. \cite{Liu:2009dm,Faulkner:2009wj}, where most of the analysis focused on massless bulk fermions. Fourth, as we show in the appendix, a massless fermion requires no counterterms.

\subsection{The Dual Operators}
\label{dualoperators}
In this section we focus on the D5-brane along $AdS_4 \times S^2$ and study in detail the operators in the D3/D5 theory dual to the fermionic fluctuations considered above. For the D7-brane along $AdS_5 \times S^3$, a similar analysis appears in ref. \cite{Kirsch:2006he}. We begin with $N_f=1$ and generalize to $N_f>1$ at the end.

The dual field theory is (3+1)-dimensional $\N=4$ SYM coupled to defect flavor fields preserving (2+1)-dimensional $\N=4$ supersymmetry (eight real supercharges). The couplings of the theory were determined in refs. \cite{DeWolfe:2001pq, Erdmenger:2002ex}. Coupling the defect fields to the ambient fields requires decomposing the (3+1)-dimensional $\N=4$ multiplet into two (2+1)-dimensional $\N=4$ multiplets, a vector multiplet and a hypermultiplet. The bosonic content of the (3+1)-dimensional $\N=4$ multiplet is the vector $A_\mu$ and six scalars\footnote{In the initial type IIB D3/D5 intersection, the D3-branes are extended along $0123$, and these scalars represent fluctuations of the D3-branes in the $456789$ directions, hence our notation. The D5-branes are extended along $012456$, so they break the $SO(6)$ rotational symmetry in $456789$ down to $SO(3) \times SO(3) \simeq SU(2) \times SU(2)$, one rotating $456$ and one rotating $789$.} $X^4,X^5,\ldots,X^9$. The bosonic content of the (2+1)-dimensional vector multiplet is the (2+1)-dimensional vector field $A_k$ and the three scalars $X_V = (X^7,X^8,X^9)$. The bosonic content of the (2+1)-dimensional hypermultiplet is the scalar $A_3$ and the three scalars $X_H = (X_4,X_5,X_6)$. The flavor fields form a (2+1)-dimensional hypermultiplet with two fermions (quarks) $\psi$ and two complex scalars (squarks) $q$.

The classical Lagrangian preserves (2+1)-dimensional $SO(3,2)$ conformal symmetry but breaks the $SO(6)$ R-symmetry down to a subgroup $SU(2)_H \times SU(2)_V$, under which the scalars in $X_H$ transform in the $(1,0)$ representation and the scalars in $X_V$ transform in the $(0,1)$. We use an upper index to denote these representations: $X^A_V$ and $X^I_H$. The adjoint fermions $\lambda^{im}$ transform in the $(1/2,1/2)$. Here $i$ is the $SU(2)_V$ index and $m$ is the $SU(2)_H$ index. The quarks $\psi^i$ transform in the $(1/2,0)$ and the squarks $q^m$ transform in the $(0,1/2)$. In table \ref{tab:fieldcontent} (borrowed from ref. \cite{DeWolfe:2001pq}), we summarize the field content and quantum numbers, including the conformal dimensions of the fields.

\begin{table}
\begin{center}
\begin{tabular}{|c|c|c|c|c|c|}\hline
Mode &  Spin & $SU(2)_H$ & $SU(2)_V$ & $SU(N_c)$ & $\Delta$ \\ \hline
$A_k$ & $1$ &  $0$ & $0$ & adj & $1$ \\
$X^A_V$ & $0$ &  $0$ & $1$ & adj & $1$ \\
$A_3$ & $0$ &  $0$ & $0$ & adj & $1$ \\
$X^I_H$ & $0$ &  $1$ & $0$ & adj & $1$ \\ 
$\lambda^{im}$ & $\frac 1 2$ & $\frac 1 2$ & $\frac 1 2$ & adj &
$\frac 3 2$ \\
$q^m$ & $0$ & $\frac 1 2$ & $0$ & {\bf N} & $\frac 1 2$ \\
$\psi^i$ & $\frac 1 2$ & $0$ & $\frac 1 2$ & {\bf N} & $1$ \\ \hline
\end{tabular}
\vspace{0.5cm}
\caption{The field content of the D3/D5 theory. (Adapted from ref. \cite{DeWolfe:2001pq}.) Here $A_k$, $X^A_V$, $A_3$, $X^I_H$ and $\lambda^{im}$ are the adjoint fields of (3+1)-dimensional $\N=4$ SYM decomposed into (2+1)-dimensional $\N=4$ multiplets. $A_k$ and $X^A_V$ are the bosons in a (2+1)-dimensional vector multiplet while $A_3$ and $X^I_H$ are the bosons in a (2+1)-dimensional hypermultiplet. $q^m$ and $\psi^i$ are the (2+1)-dimensional flavor fields, which are in an $\N=4$ hypermultiplet.}
\label{tab:fieldcontent}
\end{center}
\end{table}

Let us now match fluctuations of the D5-brane probe to dual field theory operators, building on the matching of bosonic fields and operators in refs. \cite{DeWolfe:2001pq,Myers:2006qr,Fraga:2008va}. These fluctuations correspond to mesonic operators in the dual theory \cite{Kruczenski:2003be,Karch:2002sh}, which can be arranged into a (2+1)-dimensional massive $\N=4$ supersymmetric multiplet.

First we consider the bosonic fluctuations of the D5-brane, as studied in refs. \cite{DeWolfe:2001pq,Myers:2006qr,Fraga:2008va}. The bosonic fluctuations consists of three real scalars, which in the notation of ref. \cite{DeWolfe:2001pq} are $\phi_{l}, \ (b+z)^{(-)}_{l}$ and $(b+z)^{(+)}_{l}$, as well as a vector $b^k_{l}.$ Here $\phi_{l}$ corresponds to fluctuations of the embedding in $S^5$ directions (transverse to the $S^2$), $(b+z)^{(-)}_{l}$ and $(b+z)^{(+)}_{l}$ are linear combinations of the fluctuations of the $S^2$ components of the worldvolume gauge field with the fluctuation of the embedding in $AdS_5$ transverse to $AdS_4$, and $b^k_{l}$ ($k=0,1,2$) corresponds to fluctuations of the worldvolume gauge field in the $AdS_4$ directions. We summarize the quantum numbers of these fluctuations in table \ref{tab:bosonicfluctuations} (borrowed from ref. \cite{DeWolfe:2001pq}). Notice that our definition of $l$ differs from that in ref. \cite{DeWolfe:2001pq}. In our notation, fluctuations with the same $l$ have the same mass. Later we will show that all operators with the same quantum number $l$ fit into a super multiplet. Note that $(b+z)^{(+)}_{l}$ is not present in the lowest multiplet with $l=0.$

\begin{table}
\begin{center}
\begin{tabular}{|c|c|c|c|c|c|} \hline
Mode &  $\Delta$ & $SU(2)_H$ & $SU(2)_V$ &Operator& Operator in
lowest multiplet\\[1mm] \hline 
$b^k_{l}$ &  $l+2$ & $l,$ \ \ \ \ \; $l \geq 0$ & $0$ &${\cal J}_{l}$&
$i \bar{q}^m \overleftrightarrow{D^k} \, q^m + \bar\psi^i \rho^k
\psi^i$\\[1mm] 
$\phi_{l}$ &  $l+2$  & $l,$ \ \ \ \ \; $l \geq 0$ & $1$ &$\mathcal{E}_{l}$&
$\bar\psi_i \sigma_{ij}^A \psi_j + 2  \bar{q}^m X_V^{Aa} T^a
q^m$ \\[1mm]
$(b+z)^{(-)}_{l+1}$ &  $l+1$  & $l+1,$\; $l \geq 0$ & $0$ &${\cal{C}}_{l}$&
$\bar{q}^m \sigma_{mn}^I q^n$ \\[1mm] 
$(b+z)^{(+)}_{l-1}$ &  $l+3$
& $l-1,$ \; $l \geq 1$ & $0$ &${\cal D}_{l}$ & --- \\ \hline
\end{tabular}
\end{center}
\vspace{0.5cm}
\caption{The bosonic fluctuations and their dual field theory operators for the D3/D5 system. (Adapted from ref. \cite{DeWolfe:2001pq}.) Here $\sigma$ are Pauli matrices, $T^a$ are the generators of $SU(2)_V$, and $\rho^k$ are the (2+1)-dimensional $\Gamma$-matrices.} 
\label{tab:bosonicfluctuations}
\end{table}

We studied the fermionic fluctuations\footnote{In the last subsection we used a subscript $\ell$, while here we use a subscript $l$. For $\Psi_{\ell}^-$ the two  are identical: $\ell = l$. For $\Psi_{\ell}^+$ we take $\ell = l -1$.} $\Psi_{l}^\pm$ of the D5-brane in the last section. The fluctuations $\Psi_{l}^-$ with $l \geq 0$ correspond to operators with dimensions $\Delta_{l}^- = l + 3/2$ that are in the $l+1/2$ representation of $SU(2)_H.$ The fluctuations $\Psi_{l-1}^+$ with $l \geq 1$ correspond to operators with dimensions $\Delta_{l-1}^+ = l + 5/2$ in the $l- 1/2$ representation of $SU(2)_H.$ Since both fluctuations are fermionic they transform in the $1/2$ representation $1/2$ of $SU(2)_V.$ We summarize the fermionic modes in table \ref{tab:fermionicfluctuations}.

\begin{table}
\begin{center}
\begin{tabular}{|c|c|c|c|c|} \hline
D5-brane Mode & $\Delta$  & $SU(2)_H$  & $SU(2)_V$ & Operator \\ \hline
$\Psi_{l}^-$ & $ l + 3/2$ & $l+ 1/2$, \; $l \geq 0$ & $1/2$& ${\cal F}_{l}$ \\
$\Psi_{l-1}^+$ & $l + 5/2$  & $l -1/2$, \; $l \geq 1$ & $1/2$ & ${\cal G}_{l}$ \\ \hline
\end{tabular}
\end{center}
\caption{Matching between fermionic fluctuations of the D5-brane and field theory operators. The explicit form of the fermionic operators ${\cal F}_{l}$ and ${\cal G}_{l}$ appear in the text below.} 
\label{tab:fermionicfluctuations}
\end{table}

We first review the lowest multiplet, \textit{i.e.} $l=0$, which appears already, including the fermionic operators, in ref.  \cite{DeWolfe:2001pq}. According to tables \ref{tab:bosonicfluctuations} and \ref{tab:fermionicfluctuations}, the D5-brane fluctuation corresponding to the lowest-dimension operator is $(b+z)^{(-)}_{l+1}$ with $l=0$. Only one operator exists on the field theory side with the same quantum numbers as $(b+z)^{(-)}_{1}$, so we can match $(b+z)^{(-)}_{1} $ with the operator $\mathcal{C}_{0}^{I} = q^{\dagger m} \sigma_{mn}^{I} q^n$, where $\sigma^I$ are the Pauli matrices of $SU(2)_H$. $\mathcal{C}_{0}$ transforms in the $(1,0)$ representation of $SU(2)_H \times SU(2)_V.$ Moreover $\mathcal{C}_{0}$ is the lowest chiral primary in the multiplet since all other operators dual to D5-brane fluctuations have larger conformal dimensions. We can thus construct all operators in the same multplet as ${\cal C}_{0}$ by applying supersymmetry generators to $\mathcal{C}_{0}.$ The supersymmetry generators form a $2 \times 2$ matrix of Majorana spinors $\eta^{im}$, which transforms like $\lambda^{im}$, \textit{i.e.} in the representation $(1/2, 1/2)$ of $SU(2)_H\times SU(2)_V.$ Applying the supersymmetry generators to $\mathcal{C}_0$ we obtain the fermionic operator ${\cal F}_{0}^{im}= \bar{\psi}^i q^m + q^{\dagger m} \psi^i$ with conformal dimension $\Delta = 3/2$ and $SU(2)_H \times SU(2)_V$ quantum numbers $(1/2, 1/2)$. ${\cal F}_0^{im}$ is dual to the fermionic D5-brane fluctuation $\Psi_{l=0}^-.$ Applying another supersymmetry generator to ${\cal F}_{0}^{im}$ we obtain either ${\cal J}_0$ or $\mathcal{E}_0$, the forms of which appear in table \ref{tab:bosonicfluctuations}. Both ${\cal J}_0$ and ${\mathcal{E}}_0$ have conformal dimension $\Delta =2$ and are singlets under $SU(2)_H$ but can be distinguished by their $SU(2)_V$ quantum number: ${\cal J}_0$ is a singlet whereas $\mathcal{E}_0$ is a triplet under $SU(2)_V$. 

Let us now discuss the general multiplet dual to the higher-$l$ fluctuations of the D5-brane. As in the $l=0$ case, we construct the multiplet by applying supersymmetry generators to the lowest chiral primary in the multiplet, $\mathcal{C}_l$, which is dual to $(b+z)^{(-)}_{l}.$ According to ref. \cite{DeWolfe:2001pq}, the lowest chiral primary is $\mathcal{C}_l^{I_0 I_1 ... I_l} = C_{0}^{( I_0}  \left(X_H^{l}\right)^{I_1 ... I_l )},$ where $(X_H^l)$ stands for the traceless symmetric product of $l$ copies of the field $X_H^I$. $\mathcal{C}_l$ has conformal dimension $\Delta = l+1$ and is in the $(l+1,0)$ representation of $SU(2)_H \times SU(2)_V$. Applying a supersymmetry generator to $\mathcal{C}_l$ we find the fermionic operator ${\cal F}_l$ with conformal dimension $\Delta = l + 3/2$, which is dual to the D5-brane fluctuation $\Psi_l^-$.  ${\cal F}_l$ is in the $(l+1/2,1/2)$ representation of $SU(2)_H \times SU(2)_V$. Explicitly, ${\cal F}_l$ is of the form
\begin{gather}
\label{eq:foperatorexplicit}
{\cal F}_l^{I_1 ... I_l \, im} = \bar{\psi}^i \left( X_H^l \right)^{I_1 ... I_l} q^m + q^{\dagger m}
\left( X_H^l \right)^{I_1 ... I_l} \psi^i \,.
\end{gather}
Applying another supersymmetry generator to ${\cal F}_l$ we obatin ${\cal J}_l$ or $\mathcal{E}_l$, which have the same conformal dimension $\Delta = l+2$, but differ in the $SU(2)_H \times SU(2)_V$ representation. ${\cal J}_l$ transforms in the $(l,0)$ representation whereas $\mathcal{E}_l$ has quantum numbers $(l,1).$ To obtain the precise form of ${\cal J}_l$ or $\mathcal{E}_l$ we insert the operator $X_H^l$ into the operator ${\cal J}_0$ or $\mathcal{E}_0$, respectively. 

In contrast to the $l=0$ multiplet, other operators also appear in the multiplet for $l \geq 1$, which we construct by applying three or four supersymmetry generators to $\mathcal{C}_l$: a fermionic operator ${\cal G}_l$ and a bosonic operator ${\cal D}_l$. ${\cal G}_l$ has conformal dimension $\Delta = l + 5/2 $ and $SU(2)_H \times SU(2)_V$ quantum numbers $(l - 1/2, 1/2)$. These are precisely the quantum numbers of the fermionic D5-brane fluctuation $\Psi_{l-1}^+.$ Explicitly, ${\cal G}_l$ has the form
\begin{gather}
{\cal G}_l^{I_1 ... I_{l-1} \, im} = \bar{\psi}^j \left( X_H^{l-1} \right)^{I_1 ... I_{l-1}} \lambda^{im} \, \psi_j +
q^{\dagger n} \left( X_H^{l-1} \right)^{I_1 ... I_{l-1}} \lambda^{im} \, X_{H,I}\, \sigma^I_{np} \, q^p \, .
\end{gather}
Finally ${\cal D}_l$ has conformal dimension $\Delta = l + 3 $ and $SU(2)_H \times SU(2)_V$ quantum numbers $(l - 1, 0)$ and therefore can be identified with the D5-brane fluctuation $(b+z)^{(+)}_{l-1}.$

We have constructed the supermultiplet for the cases $l=0$ and $l\geq 1$. For $l=0$ they multiplet consists of the bosonic operators $\mathcal{C}_0, {\cal J}_0$ and $\mathcal{E}_0$ and of the fermionic operator ${\cal F}_0.$ The multiplet includes are eight bosonic and eight fermionic degrees of freedom. The multiplet containing ${\cal C}_l, \ l\geq 1$ has $16 l +1$ fermionic and bosonic degrees of freedom:
\begin{itemize}
\item One real scalar ${\cal C}_l$ in the $(l+1,0)$ representation with $\Delta = l+1$
\item One spinor ${\cal F}_l$ in the $(l+1/2,1/2)$ representation with $\Delta = l+3/2,$
\item One massive vector ${\cal J}_l$ in the  $(l,0)$ representation with $\Delta = l+2,$
\item One real scalar ${\cal E}_l$ in the  $(l,1)$ representation with $\Delta = l+2,$
\item One spinor ${\cal G}_l$ in the $(l-1/2,1/2)$ representation with $\Delta = l+5/2,$
\item One real scalar ${\cal D}_l$ in the $(l-1,0)$ representation with $\Delta = l+3$.
\end{itemize}
Moreover we mapped the operators in the supermultiplet to the fluctuations of the probe brane summarized in tables \ref{tab:bosonicfluctuations} and \ref{tab:fermionicfluctuations}. 

Finally, we consider $N_f>1$ coincident probe D5-branes. The dual field theory then has $N_f$ massless flavors, with a global $U(N_f)$ flavor symmetry. The overall $U(1)$ we identify as baryon (more accurately quark) number, while the $SU(N_f)$ subgroup we identify as isospin. The mesino operators ${\cal F}_l$ and ${\cal G}_l$ of course have zero baryon number charge and are valued in the adjoint of $SU(N_f)$. For example, in our case with $N_f=2$ the mesinos acquire an $SU(2)$ isospin index, ${\cal F}_l^a$ and ${\cal G}_l^a$.

As explained at the end of the last subsection, for our numerical analysis we use the D5-brane fermions with zero mass, $\Psi_0^-$. The dual fermionic operator is ${\cal F}_0 \sim \bar{\psi} q + q^{\dagger} \psi$.

\subsection{Equation of Motion II: Gauge Couplings}
\label{gaugecouplings}
In this section we return to the equation of motion for the worldvolume fermions, eq. \eqref{eq:dpbraneworldvolumefermionsdiracequation}, and specialize to our case of interest, namely two coincident Dp-branes in (4+1)-dimensional AdS-Schwarzschild with trivial worldvolume scalars but non-trivial worldvolume gauge fields $A_t^3(u)$ and $A_x^1(u)$. More specifically, we will explicitly unpack the gauge- and curved-space covariant Dirac operator $\D_{AdS_P}$ for the $AdS_P$ submanifold of (4+1)-dimensional AdS-Schwarzschld and see how, when $A_x^1(u)$ is nonzero, the three worldvolume fermions couple to one another. In this subsection we assume $P \geq 3$.

The linearized equation of motion for the worldvolume fermions in eq. \eqref{eq:dpbraneworldvolumefermionsdiracequation} is\footnote{Here the $\pm$ does not refer to the projectors $\Pi_{\pm}$ of section \ref{mixing}, but rather to the $\pm$ sign labeling the eigenvalues of the Dirac operator of $S^Q$ in eq. \eqref{eq:diracsphericalharmonicspectrum}.}
\beq
\left( \D_{AdS_P} - m_l^\pm \right ) \, \J_l^\pm = 0 \, ,
\eeq
where the masses $m_l^\pm$ appear in eq. \eqref{MassesOfDualOperators}. To simplify the notation we write $m$ instead of $m_l^\pm$ as well as $\J$ instead of $\J_l^\pm.$ $\D_{AdS_P} = e^M_{~A} \g^A D_M$ is the gauge and curved-space covariant Dirac operator. The index $A$ runs over the worldvolume directions inside $AdS_5.$ Notice that here $e^M_{~A}$ are the \textit{inverse} vielbeins. The $\g^A$ are the $\G$-matrices of (4+1)-dimensional Minkowski space, which obey $\left \{ \g^A, \g^B \right\} = 2 \, \eta^{AB}$.

For the Dirac operator, we have (here $a$ is a gauge index)
\bea
\left [ \left(\D_{AdS_P} - m \right )\, \J \right]_a & = & \left( u \sqrt{f} \, \g^u \, \partial_u + \frac{u}{\sqrt{f}} \, \g^t \, \partial_t + u \, \g^i \, \partial_i + \left [ - \frac{P-1}{2} \sqrt{f} + \frac{1}{4} \, u \, \frac{f'}{\sqrt{f}} \, \right] \, \g^u \right) \J_a\nonumber \\ & + & e^M_{~A} \, i \, \g^A \left[ A_M \,, \J \right]_a - m \, \J_a,
\eea
where $f' = \partial_u f$. When $T=0$ and hence $f(u) = 1$, the operator in parentheses on the right-hand side in the first line is the Dirac operator of $AdS_P$. At finite temperature, where $f(u) = 1 - u^4 / u_h^4$, the operator is that of an $AdS_P$ submanifold of (4+1)-dimensional AdS-Schwarzschild.\footnote{For a general (d+1)-dimensional AdS-Schwarzschild space, $f(u) = 1-u^d/u_h^d$.} The coupling to the gauge field in the second line is of course fixed by gauge invariance.

We need an ansatz for $\J_a$. For the coordinate dependence, our ansatz will be similar to the one in refs. \cite{Liu:2009dm,Cubrovic:2009ye,Faulkner:2009wj},
\beq
\label{fermionansatz}
\J = \J_a \, \s_a = u^{(P-1)/2} \, f^{-\frac{1}{4}} \, e^{i k_{\mu} x^{\mu}} \, \psi_a(u) \, \tau_a,
\eeq
where $\mu$ runs over field theory directions, the $\psi(u)_a$ are three spinor functions for which we must solve, and we extract a factor of $u^{(P-1)/2} \, f^{-\frac{1}{4}}$ to make the Dirac equation look nice later (in the language of ref. \cite{Faulkner:2009am} these factors will ``remove the spin connection'' from the equation of motion).

Using our ansatz for the fermion in eq. (\ref{fermionansatz}) and the ansatz for the gauge field in eq. (\ref{gaugefieldansatzpwave}) (from which we can recover eq. (\ref{gaugefieldansatznormal}) simply by setting $A_x^1=0$), we find three Dirac equations,
\bea
\label{eq:diraceq1}
0 & = & \left ( \sqrt{f} \g^u \partial_u - \frac{i\,\omega}{\sqrt{f}} \, \g^t + i k_{i} \g^{i} - \frac{1}{u} m \right) \psi_1 + \, \frac{A_t^3(u)}{\sqrt{f}} \, \g^t \, \psi_2, \\
\label{eq:diraceq2}
0 & = & \left ( \sqrt{f} \g^u \partial_u - \frac{i\,\omega}{\sqrt{f}} \, \g^t + i k_{i} \g^{i} - \frac{1}{u} m \right) \psi_2 - \, \frac{A_t^3(u)}{\sqrt{f}} \, \g^t \, \psi_1 + A_x^1(u) \, \g^x \, \psi_3, \\
\label{eq:diraceq3}
0 & = & \left ( \sqrt{f} \g^u \partial_u - \frac{i\,\omega}{\sqrt{f}} \, \g^t + i k_{i} \g^{i} - \frac{1}{u} m \right) \psi_3 - \, A_x^1(u) \, \g^x \, \psi_2.
\eea

In what follows we use the Lorentzian-signature $\g^A$ from section \ref{mixing}, in which all the $\g^A$ are Hermitian except for $\g^t$, which will be anti-Hermitian:\footnote{We only need $\g^y$ when $P\geq4$.}
\beq
\label{eq:gbasis2}
\g^u = \left( \begin{array}{cc} - \s_3 & 0 \\ 0 & -\s_3 \end{array} \right), \qquad \g^t = \left( \begin{array}{cc} i\s_1 & 0 \\ 0 & i\s_1 \end{array} \right), \qquad \g^x = \left( \begin{array}{cc} -\s_2 & 0 \\ 0 & \s_2 \end{array} \right), \qquad \g^y = \left( \begin{array}{cc} 0 & \s_2 \\ \s_2 & 0 \end{array} \right).
\eeq
We will also use the projectors $\Pi_{1,2}$, which in Lorentzian signature are defined as
\beq
\label{eq:Pi}
\Pi_{\alpha} \equiv \frac{1}{2} \left (1-\left(-1\right)^{\alpha} \g^u \g^t \g^x \right),
\eeq
with $\alpha = 1,2$.

\subsubsection{Normal Phase}

First consider the normal phase, where $A_x^1(u)=0$. In that case, $\psi_3$ decouples from $\psi_1$ and $\psi_2$, and its equation of motion becomes that of a free neutral fermion, as expected. We can then simplify the remaining two equations by taking linear combinations of them. Defining\footnote{Here the $\pm$ index refers to linear combinations of the worldvolume fermions that diagonalize the equations of motion when $A_x^1(u)=0$. Nowhere in this subsection or the next do we use a $\pm$ index to refer to the projectors $\Pi_{\pm}$ of section \ref{mixing} or to the eigenvalues of the Dirac operator on $S^Q$ of eq. \eqref{eq:diracsphericalharmonicspectrum}.} $\psi_{\pm} \equiv \psi_2 \pm i \psi_1$, we find three decoupled equations,
\bea
\label{diraceqexplicit}
0 & = & \left ( \sqrt{f} \g^u \partial_u - \frac{i\,\omega}{\sqrt{f}} \, \g^t + i k_{i} \g^{i} - \frac{1}{u} m \right) \psi_+ \, + \, i \, \frac{A_t^3(u)}{\sqrt{f}} \, \g^t \, \psi_+,\\
0 & = & \left ( \sqrt{f} \g^u \partial_u - \frac{i\,\omega}{\sqrt{f}} \, \g^t + i k_{i} \g^{i} - \frac{1}{u} m \right) \psi_- \, - \, i \, \frac{A_t^3(u)}{\sqrt{f}} \, \g^t \, \psi_-,\\
0 & = & \left ( \sqrt{f} \g^u \partial_u - \frac{i\,\omega}{\sqrt{f}} \, \g^t + i k_{i} \g^{i} - \frac{1}{u} m \right) \psi_3
\eea
which are precisely the equations of motion for fermions (in the fundamental representation of the unbroken $U(1)_3 \subset SU(2)$) with charges $q=\mp1,0$. As in section \ref{worldvolumefermions}, we emphasize that, because we consider a particular embedding of the Dirac equation into string theory, the allowed values of the mass and charge of the fermions are constrained by supersymmetry and T-duality.

We will now follow appendix A of ref. \cite{Faulkner:2009wj} to simplify the equation of motion further. First we rewrite the equation as\footnote{Starting now, we will use the notation $\psi$ to refer to any of our three fermions, when we are making general statements.},
\beq
\label{diraceqexplicitsimplified}
\left ( \sqrt{f} \, \g^u \partial_u - \frac{1}{u} m \right) \psi + i K_{\m}(u) \g^{\mu} \psi = 0,
\eeq
\beq
\label{eq:K}
K_{\mu}(u) = \left( -v(u),k_i\right), \qquad v(u) = \frac{1}{\sqrt{f}} \left( \omega + q A_t(u) \right),
\eeq
where the index $i$ runs over spatial directions, $q=\mp 1$ for $\psi_{\pm}$ and $q=0$ for $\psi_3$. Notice that near the boundary, $v(u) \rightarrow \omega + q \mu$, so the frequency $\omega$ is measured relative to ($q$ times) the chemical potential.

The system is rotationally invariant, so without loss of generality we can take only $k_x$ to be nonzero. (Obviously, this will not be the case in the superfluid phase, where rotational symmetry is broken.) The fermion's equation of motion then depends only on $\g^u$, $\g^t$ and $\g^x$, so the projectors $\Pi_{\alpha}$ commute with the operator acting on the $\psi$ in eq. \eqref{diraceqexplicitsimplified}, hence the equations for $\phi_{\alpha} \equiv \Pi_{\alpha} \psi$ decouple from one another. In terms of the $\phi_{\alpha}$, the equation of motion becomes
\beq
\label{eq:normalphasesimplifiedeoms}
\left( \partial_u + \frac{1}{u \sqrt{f}} \, m \, \s_3 \right) \phi_{\alpha}  + \frac{1}{\sqrt{f}} \, \left( - i v(u) \s_2 - \left( -1\right)^{\alpha} k_x \s_1 \right) \phi_{\alpha} = 0.
\eeq
We thus obtain six decoupled equations, four for the $\phi_{\pm \, \alpha}$ and two for the $\phi_{3 \alpha}$.

Eq. \eqref{eq:normalphasesimplifiedeoms} is \textit{almost} identical to eq. (A14) of ref. \cite{Faulkner:2009wj}. The biggest difference is the function $f(u)$, which for us is the $f(u)$ of (4+1)-dimensional AdS-Schwarzschild and in ref. \cite{Faulkner:2009wj} was the $f(u)$ of (3+1)-dimensional AdS-Schwarzschild. Given that we will solve nearly identical equations of motion, we will obtain qualitatively similar finite-temperature results. As mentioned above, however, we cannot reach $T=0$ within the probe approximation, so we will not be able to reproduce the $T=0$ results of refs. \cite{Liu:2009dm,Faulkner:2009wj}, including in particular the influence of an emergent $AdS_2$.

\subsubsection{Superfluid Phase}
\label{sec:superfluidphase}

In the solution corresponding to the superfluid phase, where $A_x^1(u)$ is nonzero, we cannot write linear combinations of $\psi_1$, $\psi_2$ and $\psi_3$ to diagonalize the system and produce three decoupled equations. To make comparison to the normal phase easier, we will again work with $\psi_{\pm} = \psi_2 \pm i \psi_1$, so that eqs. (\ref{eq:diraceq1})-(\ref{eq:diraceq3}) become
\bea
0 & = & \left ( \sqrt{f} \g^u \partial_u - \frac{i\,\omega}{\sqrt{f}} \, \g^t + i k_{i} \g^{i} - \frac{1}{u} m \right) \psi_+ \, + \, i \, \frac{A_t^3(u)}{\sqrt{f}} \, \g^t \, \psi_+ + A_x^1(u) \, \g^x \, \psi_3,\\
0 & = & \left ( \sqrt{f} \g^u \partial_u - \frac{i\,\omega}{\sqrt{f}} \, \g^t + i k_{i} \g^{i} - \frac{1}{u} m \right) \psi_- \, - \, i \, \frac{A_t^3(u)}{\sqrt{f}} \, \g^t \, \psi_- + A_x^1(u) \, \g^x \, \psi_3, \\
0 & = & \left ( \sqrt{f} \g^u \partial_u - \frac{i\,\omega}{\sqrt{f}} \, \g^t + i k_{i} \g^{i} - \frac{1}{u} m \right) \psi_3 - \frac{1}{2} \, A_x^1(u) \, \g^x \, \left(\psi_+ + \psi_- \right).
\eea
Clearly the three fermions $\psi_{\pm}$ and $\psi_3$ couple to one another via a nonzero $A_x^1(u)$. Here we have a concrete example of the couplings described in section \ref{coupledfermions} (especially around eq. (\ref{eq:generalcoupledeom})). We simplify the $\psi_+$ and $\psi_-$ equations again by writing them as
\bea
0 & = & \left ( \sqrt{f} \, \g^u \partial_u - \frac{1}{u} m \right) \psi_+ + i K_{\m}(u) \g^{\mu} \psi_+ + A_x^1(u) \, \g^x \, \psi_3, \\
0 & = & \left ( \sqrt{f} \, \g^u \partial_u - \frac{1}{u} m \right) \psi_- + i K_{\m}(u) \g^{\mu} \psi_- + A_x^1(u) \, \g^x \, \psi_3, \\
0 & = & \left ( \sqrt{f} \, \g^u \partial_u - \frac{1}{u} m \right) \psi_3 + i K_{\m}(u) \g^{\mu} \psi_3 - \frac{1}{2} \, A_x^1(u) \, \g^x \, \left(\psi_+ + \psi_- \right),
\eea
where $K_{\mu}$ is defined the same way as in eq. (\ref{eq:K}). Recall that $\psi_+$ has charge $q=-1$, $\psi_-$ has charge $q=+1$ and $\psi_3$ has charge $q=0$.

Now we come to a big difference from the solution corresponding to the normal phase, at least for $Dp$-branes wrapping $AdS_P$ with $P \geq 4$. In the solution corresponding to the superfluid phase, rotational symmetry is broken from $SO(P-2)$ to $SO(P-3)$. Using the $SO(P-3)$ rotational symmetry, the most general momentum we can pick has nonzero $k_x$ \textit{and} nonzero $k_y$. The equations for $\psi_+$ and $\psi_-$ then depend on $\g^u$, $\g^t$, $\g^x$ and now also $\g^y$, hence the $\Pi_{\alpha}$ projectors no longer commute with the operators acting on the fermions in the equations of motion, and the equations for the $\phi_{\alpha} \equiv \Pi_{\alpha} \psi$ will not decouple from each other. Upon acting with the projectors $\Pi_{\alpha}$, the equations of motion become
\begin{subequations}
\label{eq:brokenphasesimplifiedeoms}
\begin{multline}
0 = \left[ \partial_u + \frac{m}{u \sqrt{f}} \, \s_3 + \frac{1}{\sqrt{f}} \, \left( - i v(u) \s_2 - \left( -1\right)^{\alpha} k_x \s_1 \right) \right] \phi_{+ \, \alpha} \\ - \frac{k_y}{\sqrt{f}} \s_1  \left( -1\right)^{\alpha+1} \epsilon^{\alpha \beta} \phi_{+ \, \b} - \frac{A_x^1}{\sqrt{f}} \left( -1\right)^{\alpha} i \s_1 \phi_{3 \, \alpha},
\end{multline}
\begin{multline}
0 = \left[ \partial_u + \frac{m}{u \sqrt{f}} \, \s_3  + \frac{1}{\sqrt{f}} \, \left( - i v(u) \s_2 - \left( -1\right)^{\alpha} k_x \s_1 \right) \right] \phi_{- \, \alpha} \\ - \frac{k_y}{\sqrt{f}} \s_1 \left(-1\right)^{\alpha+1}\epsilon^{\alpha \beta} \phi_{- \, \b} - \frac{A_x^1}{\sqrt{f}} \left( -1\right)^{\alpha} i \s_1 \phi_{3 \, \alpha},
\end{multline}
\begin{multline}
0 = \left[ \partial_u + \frac{m}{u \sqrt{f}} \, \s_3 + \frac{1}{\sqrt{f}} \, \left( - i v(u) \s_2 - \left( -1\right)^{\alpha} k_x \s_1 \right) \right] \phi_{3 \, \alpha} \\ - \frac{k_y}{\sqrt{f}} \s_1 \left(-1\right)^{\alpha +1} \epsilon^{\alpha \beta} \phi_{3 \, \b} + \frac{1}{2} \frac{A_x^1}{\sqrt{f}} \left( -1\right)^{\alpha} i \s_1 \left( \phi_{+ \, \alpha} + \phi_{- \, \alpha}\right).
\end{multline}
\end{subequations}
Here $\epsilon^{\alpha \beta}$ is antisymmetric with $\epsilon^{12}=+1$. Notice that when $k_y$ is nonzero, the $\phi_1$ and $\phi_2$ couple.

Eq. \eqref{eq:brokenphasesimplifiedeoms} is the result for any Dp-brane extended along $AdS_P \times S^Q$. What will change from one Dp-brane to another are the allowed values of $m$ and the solutions for $A_t^3(u)$ and $A_x^1(u)$. In the next section we specialize to the D5-brane extended along $AdS_4 \times S^2$ ($P=4$ and $Q=2$), and to the massless worldvolume fermion.

\section{Emergence of the p-wave Fermi surface}
\label{numerical}

\subsection{Properties of the Spectral Function}

For the probe D5-brane worldvolume fermions of the last section, we solved the linearized equations of motion, eqs. \eqref{eq:brokenphasesimplifiedeoms}, numerically, and used these solutions to extract the fermionic spectral functions. In this section we present a selection of our numerical results.

We work with two D5-branes extended along (when $T=0$) $AdS_4 \times S^2$ inside $AdS_5 \times S^5$. As shown in section \ref{eom1reduction}, we have many worldvolume fermions to choose from, with many different masses. In our numerical analysis we work exclusively with the massless worldvolume fermion. The dual operator is then the $l=0$ case of the mesino operator ${\cal F}_l$ written explicitly in eq. \eqref{eq:foperatorexplicit}. These mesinos are valued in the adjoint of the $SU(2)$ isospin symmetry, so we actually have three mesinos, ${\cal F}_0^+$, ${\cal F}^-_0$, and ${\cal F}_0^0$, where the superscript denotes the charge under $U(1)_3$. These are dual to the three fermions $\psi^{\pm}$ and $\psi^0$ in subsection \ref{sec:superfluidphase}.

In more detail, our procedure is as follows. We first choose the values of $T$ and $\mu$ that we want, and solve for the background $SU(2)$ gauge field functions $A_t^3(u)$ and $A_x^1(u)$, as in section \ref{probedpbranesinads}. Plugging the gauge field solution into eqs. \eqref{eq:brokenphasesimplifiedeoms}, we then solve for the bulk fermions. Near the horizon the fermions have the form of an in-going wave, eq. \eqref{eq:fermioningoingwave}, with the $\alpha$ in that equation being $\alpha = \frac{\omega}{4 \pi T}$ for our system. When $A_x^1(u)$ is nonzero, the fermions couple to one another, hence we employ the technique of section \ref{mixing} to compute the retarded Green's function, which then gives us the spectral function, as we explain below.

The normalization of our Green's functions is fixed by the normalization of the fermionic part of the D5-brane action, eq. \eqref{eq:dpbranefermionicaction}. The normalization includes various numerical factors, and in particular depends on the normalization of the $S^5$ spinor $\chi$ defined in eq. \eqref{typeIIBspinor}. We will omit the details, but we will mention that the normalization includes a factor of $N_f \, T_{D5} \propto \sqrt{\lambda} \, N_f N_c$. In what follows we will rescale our Green's functions by the overall normalization. In other words, we will divide the action by the normalization factor, so that we obtain an effective $AdS_4$ Dirac action with a Lagrangian of the form $i \bar{\Psi} \D \Psi$.

As explained in section \ref{mixing}, with three bulk fermions, the field theory retarded Green's function will be a $6 \times 6$ matrix. In the normal phase where the three bulk fermions decouple, the Green's function will be diagonal in isospin indices and in the subspaces defined by the $\Pi_{1,2}$ projectors defined in eq. \eqref{eq:defpionetwoprojectors} (see also eq. \eqref{eq:psiexplicit}). Explicitly, the retarded Green's function will have the form
\beq
\left [ G^R_{AB}(\omega,k_x,k_y) \right ] =  \text{diag} \left( G^R_{-2},G^R_{-1},G^R_{+2},G^R_{+1},G^R_{02},G^R_{01}\right),
\eeq
with $A,B = 1, \ldots, 6$, so that $A=1$ corresponds to the components of the ${\cal F}_0^-$ mesino in the $\Pi_2$ subspace, $A=2$ corresponds to ${\cal F}_0^-$ in the $\Pi_1$ subspace, $A=3$ corresponds to the components of the ${\cal F}_0^+$ mesino in the $\Pi_2$ subspace, and so on. In the superfluid phase, due to the bulk couplings, for generic momenta all the off-diagonal elements become nonzero.

The spectral function $\R_{ab}$ is defined as the anti-Hermitian part of the retarded Green's function,
\beq
\label{eq:defR}
\R_{AB}(\omega,k_x,k_y) \equiv i \, \left (G^R_{AB}(\omega,k_x,k_y)-G^{R \dagger}_{AB}(\omega,k_x,k_y) \right).
\eeq
We define the spectral measure $\R(\omega,k_x,k_y)$ as the trace over $\R_{AB}(\omega,k_x,k_y)$ (a trace over both flavor and spinor indices),
\beq
\R(\omega,k_x,k_y) \equiv \text{tr} \R_{AB}(\omega,k_x,k_y).
\eeq

Stability requires the eigenvalues of $\R_{AB}$, and hence both the diagonal elements of $\R_{AB}$ and the spectral measure (the sum of the eigenvalues), to be strictly non-negative. Otherwise, if we perturb the medium with one of the operators appearing in the spectral function, the resulting excitation would experience \textit{negative} energy dissipation into the medium, \textit{i.e.} the excitation would extract energy from the medium, signaling an instability. The eigenvalues of the spectral function are a direct measure of the states of the theory that have an overlap with the relevant operators. The off-diagonal elements of $\R_{AB}$, however, need not obey any positivity requirement. Indeed, the off-diagonal elements are more similar to interference effects than to a measure of a density of states. We have confirmed numerically that all of our spectral functions obey the correct positivity requirements.

We have confirmed that our numerical result for the spectral function obeys the following symmetries in both the normal and superfluid phases (in each case, any argument not shown is invariant):
\bea
\R_{11}(-\omega)=\R_{44}(\omega)\,, & \R_{22}(-\omega)=\R_{33}(\omega) ,\,& \R_{55}(-\omega)=\R_{66}(\omega)\, ,\hphantom{XXXXXXX}\\
\R_{11}(-k_x)=\R_{22}(k_x),\, & \R_{33}(-k_x)=\R_{44}(k_x),\, & \R_{55}(-k_x)=\R_{66}(k_x)\, ,\\
\R_{11}(-\omega,-k_y)=\R_{44}(\omega,k_y),\,& \R_{22}(-\omega,-k_y)=\R_{33}(\omega,k_y),\, & \R_{55}(-\omega,-k_y)=\R_{66}(\omega,k_y)\, .
\eea
An example for frequency and momentum symmetries in the off-diagonal elements is 
\beq
\R_{AB}(-k_y)=(-1)^{A+B}{\R_{BA}}^*(k_y) \, ,
\eeq
which is also true for all diagonal elements, stating their invariance under $k_y\to -k_y$. We have also confirmed that our numerical results obey
\beq
\R_{AB}={\R_{BA}}^* \, ,
\eeq
which follows directly from the definition of the spectral function in eq.~\eqref{eq:defR}.

Lastly, notice that because we study a fermionic operator of dimension $3/2$, the retarded Green's function, and hence the spectral function and measure, are dimensionless.

\subsection{Numerical Results}

\begin{figure}[ht]
\begin{center}
\psfrag{Rii}[c]{$\R_{11}$}
\psfrag{w}[c]{$\omega$}
\psfrag{kx}{$k_x/\pi T$}
\subfigure[][]{\includegraphics[width=0.48\textwidth]{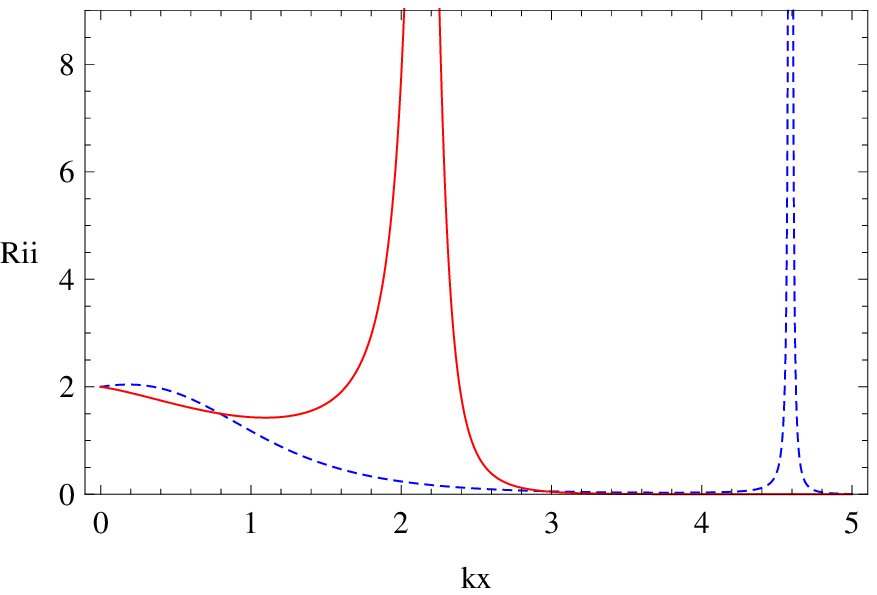}}
\hfill
\psfrag{Rii}[c]{$\R_{55}$}
\psfrag{kx}{$k_x/\pi T$}
\subfigure[][]{\includegraphics[width=0.48\textwidth]{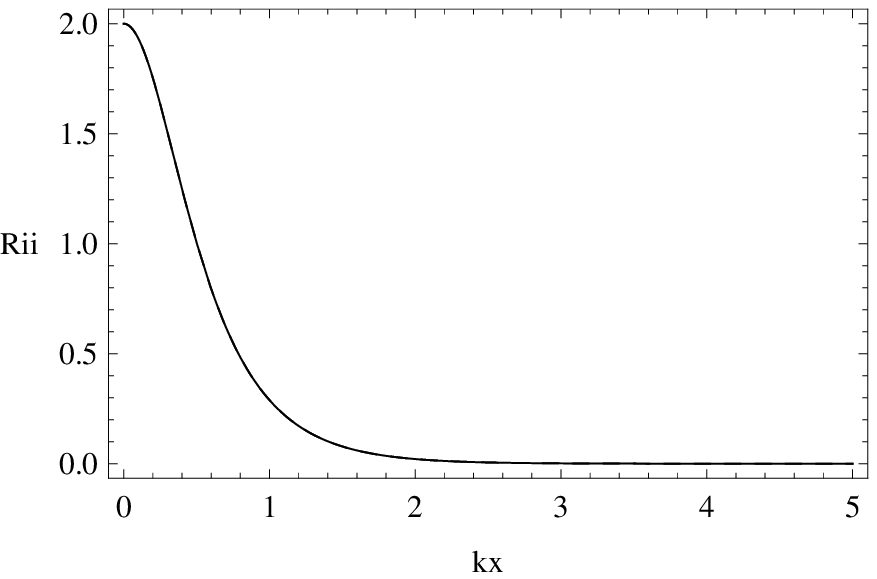}}
\caption{
\label{fig:normalphase1}
Two spectral function components $\R_{AA}(\omega, k_x,k_y)$, for $A=1,5$, plotted versus the rescaled momentum $k_x/\pi T$, with $\omega = k_y =0$, at temperatures $T \leq T_c$ but in the normal (non-superfluid) phase, \textit{i.e.} in the thermodynamically \textit{disfavored} phase. {\bf (a)} $\R_{11}$, corresponding to the fermionic operator  with charge $q=-1$, at $T=0.61\, T_c$. The two curves are for positive momentum $k_x$~(solid red curve) and negative momentum $-k_x$~(dashed blue curve), the latter case being equal to $\R_{22}(\omega,k_x,k_y)$ at positive momentum due to the symmetries of the spectral function. Multiple peaks are visible, just as in ref. \cite{Liu:2009dm}. {\bf (b)} $\R_{55}$, corresponding to a fermionic operator with charge $q=0$, at temperatures $T=T_c$~(solid curve), $T=0.75\, T_c,$~(grey dotted curve), $T=0.61\, T_c$~(dashed curve). The curves remain coincident since changing the chemical potential does not affect the uncharged operator.}
\end{center}
\end{figure}

First we compute the spectral function for temperatures below $T_c$ but in the normal (non-superfluid) phase, which we know is thermodynamically \textit{dis}favored. We do so for two reasons: first, to compare later to the superfluid phase, and second, to reproduce some of the finite-temperature results of ref. \cite{Liu:2009dm}, as a check of our methods. In practical terms, we use the solution for the gauge field with $A_t^3(u)$ (from eq. \eqref{gaugefieldansatznormal} with $P=4$) and zero $A_x^1(u)$. Figure \ref{fig:normalphase1} shows two diagonal spectral function components, $\R_{AA}(\omega,k_x,k_y)$ with $A=1,5$, as functions of $k_x/\pi T$, with $\omega = k_y = 0$, for $T\leq T_c$ in the normal phase.

Figure \ref{fig:normalphase1} (a) shows the component $\R_{11}$, which corresponds to the components of the mesino ${\cal F}_0^-$ in the $\Pi_2$ subspace, at temperature $T = 0.61 T_c$. For the moment, our main point is that figure \ref{fig:normalphase1} (a) is qualitatively similar to figure 4 of ref. \cite{Liu:2009dm}. As we lower the temperature, the peaks in the figure move to larger momenta and additional peaks appear near zero momentum. In fact, in figure \ref{fig:normalphase1} (a), $\R_{11}$ along negative momenta (dashed blue curve) already displays a small bump near $k_x / \pi T = 0$, which grows into a peak as we cool the system. Similar effects were observed in ref. \cite{Liu:2009dm}, and were interpreted as the emergence of multiple Fermi surfaces at different momenta. Additionally, the spectral functions for our other charged fermions in the low-temperature normal phase are similar to those in refs. \cite{Liu:2009dm,Faulkner:2009wj} (so we will not present them).

Figure \ref{fig:normalphase1} (b) shows the component $\R_{55}$, which corresponds to the neutral operator ${\cal F}^0_0$, in the $\Pi_2$ subspace, at temperatures $T = T_c$, $0.75 T_c$, and $0.61 T_c$. $\R_{55}$ is featureless here, but will not be so in the p-wave phase. Notice that $\R_{55}$ also does not change as the temperature decreases, or equivalently as the chemical potential increases, since when $q=0$ the chemical potential does not enter the relevant bulk fermion's equation of motion: see eq. \eqref{eq:K}.

\begin{figure}[htbp]
\begin{center}
\psfrag{R}[c]{$\R_{11}$}
\psfrag{w}[c]{$\omega$}
\psfrag{k}{$k_x/ \pi T$}
\subfigure[][]
{\includegraphics[width=0.48\textwidth]{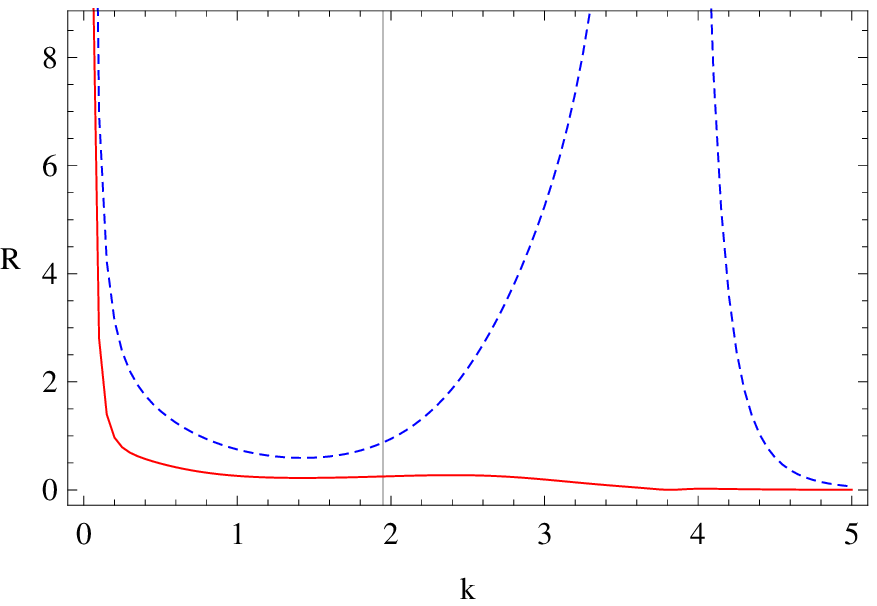}}
\hfill
\psfrag{R}[c]{$\R_{55}$}
\psfrag{k}{$k_x/ \pi T$}
\subfigure[][]
{\includegraphics[width=0.48\textwidth]{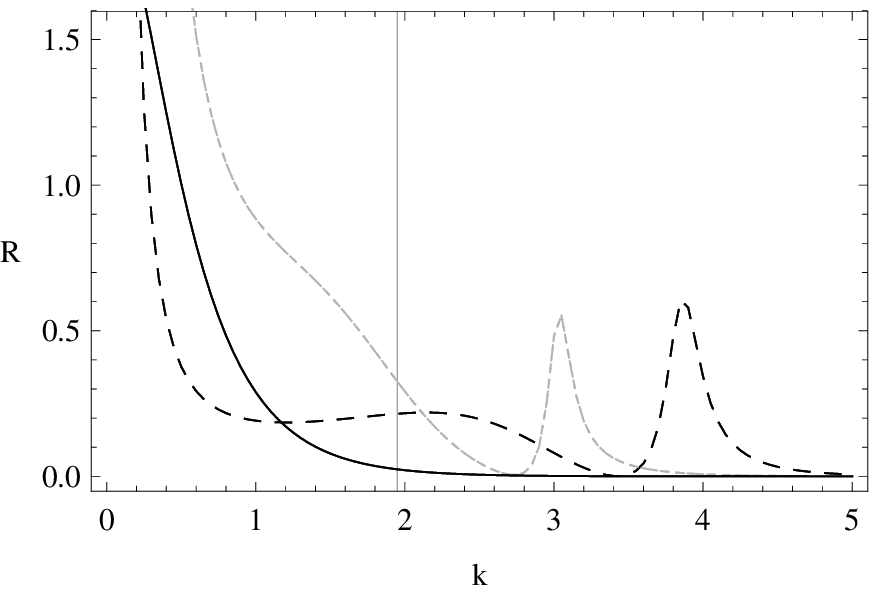}}
\caption{
\label{fig:pwavephase1}
Two spectral function components $\R_{AA}(\omega,k_x,k_y)$ plotted versus the rescaled momentum $k_x/\pi T$ with $\omega = k_y=0$ in the superfluid phase. {\bf (a)} Exactly the same spectral function as in figure \protect\ref{fig:normalphase1} (a), $\R_{11}$, at the same temperature $T=0.61\, T_c$, but now in the superfluid phase. The two curves correspond to positive momentum $k_x$~(solid red curve) or negative momentum $-k_x$~(dashed blue curve). {\bf (b)} Exactly the same spectral function component as in \protect\ref{fig:normalphase1} (b), $\R_{55}$, at the same temperatures $T_c$~(solid curve), $T=0.75\, T_c$~(grey dotted curve), $T=0.61\, T_c$~(dashed curve). Here we see operator mixing: a feature develops in the \textit{neutral} fermion's spectral function in the p-wave phase. A bump grows into a peak at the same momentum $k_x/\pi T \approx 3.87$ as the peak in $\R_{11}$.
}
\end{center}
\end{figure}

Next we plot the essentially the same thing as in figure \ref{fig:normalphase1}, but now in the thermodynamically favored superfluid phase. More precisely, figure \ref{fig:pwavephase1} shows two diagonal spectral function components, $\R_{AA}(\omega,k_x,k_y)$ with $A=1,5$, as a function of the rescaled momentum $k_x/ \pi T$, with $\omega = k_y = 0$, for $T\leq T_c$ in the superfluid phase, \textit{i.e.} now with nonzero $A_x^1(u)$. Figure \ref{fig:pwavephase1} (a) shows the same component of the spectral function as in figure \ref{fig:normalphase1} (a), $\R_{11}$, again with $T = 0.61 T_c$. Figure \ref{fig:pwavephase1} (b) shows the same component of the spectral function as in figure \ref{fig:normalphase1} (b), $\R_{55}$, at the same temperatures $T = T_c$, $0.75 T_c$, and $0.61 T_c$.

The operator mixing is obvious in figure \ref{fig:pwavephase1}: the spectral function for a \textit{neutral} fermion, $\R_{55}$, develops a bump that grows into a small peak located at the same momentum as the peak in $\R_{11}$, $k_x/\pi T = 3.87$. In bulk terms, the coupling between $\phi_3$ and $\phi_{\pm}$ in eq. \eqref{eq:brokenphasesimplifiedeoms} is allowing the peak in the charged fermions' spectral functions to ``leak'' into the spectral functions of the neutral fermions. That coupling is proportional to $A_x^1(u)$, hence the peak should grow as the temperature decreases and $A_x^1(u)$ grows, which is indeed what we see in figure \ref{fig:pwavephase1} (b). The method we developed in section \ref{mixing} for computing retarded Green's functions for coupled bulk fermions seems to work very well.

\begin{figure}[htbp]
\begin{center}
\psfrag{Rew}{$\text{Re}\,\frac{\omega}{\pi T}$}
\psfrag{Imw}[r]{$\text{Im}\,\frac{\omega}{\pi T}$}
\begin{minipage}{\textwidth}
\subfigure[][]{\includegraphics[width=0.49 \textwidth]{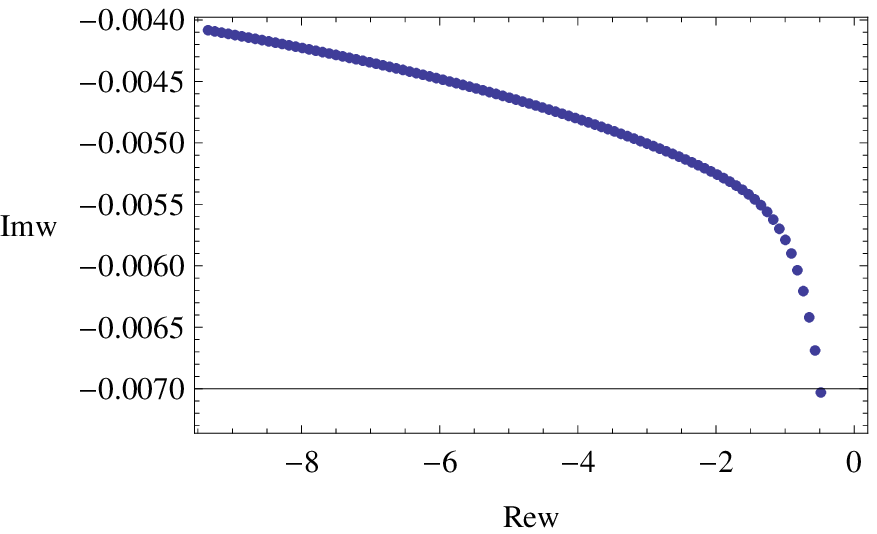}}
\hspace{10mm}
\subfigure[][]{\includegraphics[width=0.49 \textwidth]{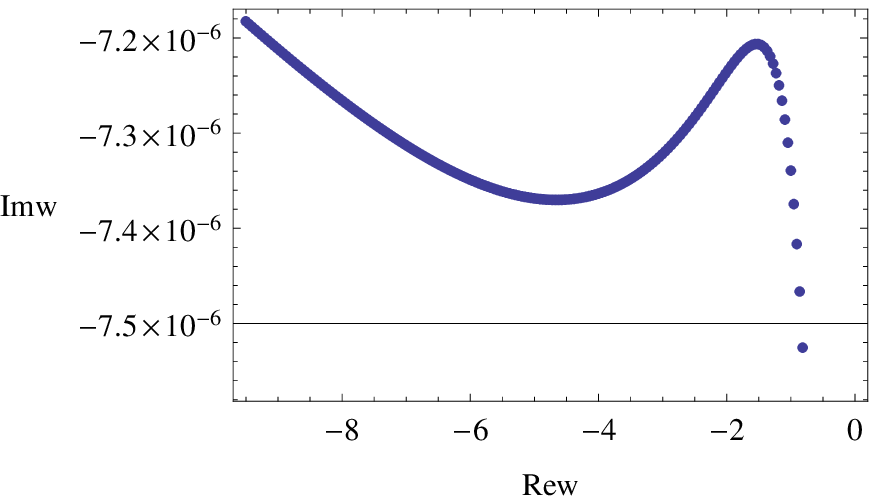}}
\vfill
\subfigure[][]{\includegraphics[width=0.49 \textwidth]{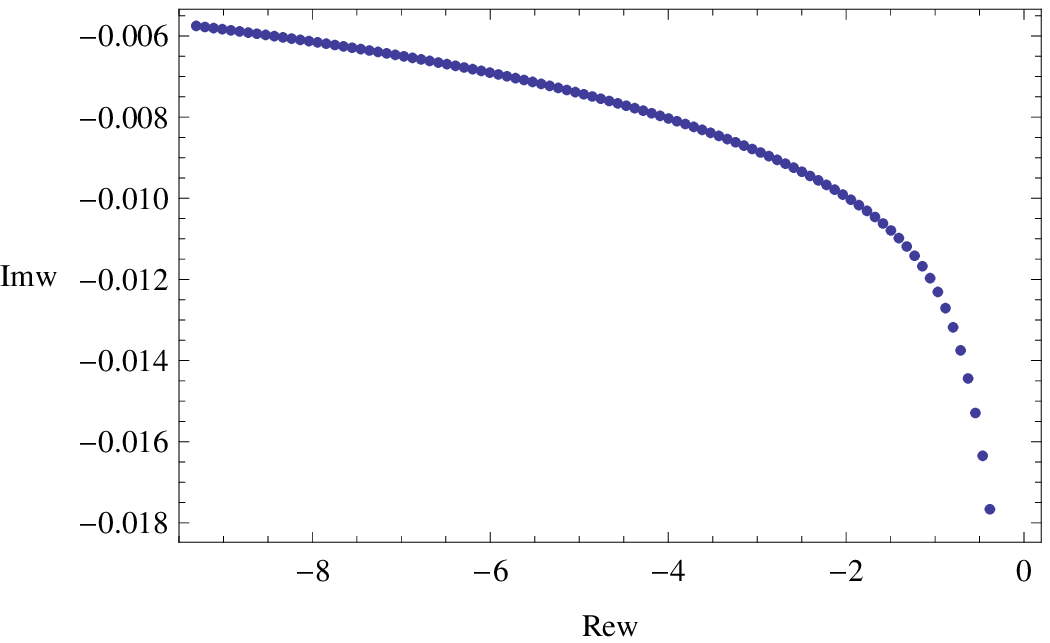}}
\hspace{10mm}
\subfigure[][]{\includegraphics[width=0.49 \textwidth]{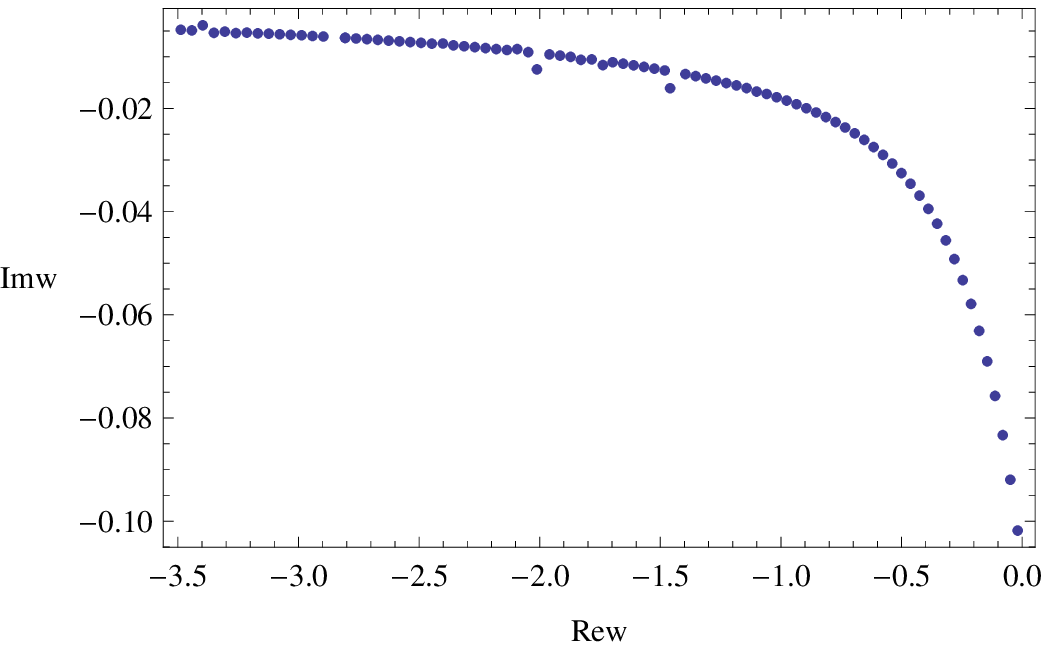}}
\end{minipage}
\caption{
\label{fig:qnms}
The movement of the pole in the retarded Green's function $G_{11}^R(\omega,k_x,k_y)$, for a mesino of charge $q=-1$, that is closest to $\omega=0$, as a function of momentum, for $T < T_c$, in both the normal (thermodynamically disfavored) phase and superfluid (thermodynamically favored) phase. \textbf{(a.)} The movement of the pole in the normal phase at $T = 0.91 T_c$, for $k_x/\pi T \in [2.58,12.58]$. As the momentum increases, the pole moves from the lower right, near $\text{Re} \, \omega/\pi T = 0$, toward the upper left. (The same applies for the following three figures.) The pole asymptotically approaches the real frequency axis, $\text{Im} \, \omega/\pi T = 0$, as $k_x$ increases. \textbf{(b)} The movement of the pole in the normal phase at $T = 0.48 T_c$, for $k_x/\pi T \in [6.15,16.15]$. Here we see that the distance to the real axis does not decrease monotonically, but rather a local minimum develops at $(\text{Re}\, \omega/ \pi T,\text{Im} \, \omega / \pi T) = (-1.79,-7.21\times 10^{-6})$ when $k_x /\pi T = 8.15$. \textbf{(c.)} The movement of the pole in the superfluid phase at $T = 0.91 T_c$, for $k_x/\pi T \in [2.49,12.29]$ and $k_y = 0$. The movement is qualitatively similar to (a). \textbf{(d)} The movement of the pole in the superfluid phase at $T = 0.48 T_c$, for $k_x/\pi T \in [4.83,8.98]$ and $k_y = 0$. The movement is again qualitatively similar to (a), in particular, the distance to the real frequency axis does not develop a local minimum, in contrast to the normal phase result in (b). We see qualitatively similar behavior when we set $k_x=0$ and increase $k_y$.
}
\end{center}
\end{figure}

As another comparison between the normal and broken phases when $T \leq T_c$, we focus on the pole in the retarded Green's function $G^R_{11}(\omega,k_x,k_y)$, for the mesino with charge $q=-1$ in the $\Pi_2$ subspace, that is closest to the origin of the complex frequency plane, $\omega = 0$, and follow the movement of the pole in the frequency plane as we change the momentum.\footnote{Poles in the retarded Green's function are holographically equivalent to the bulk fermion's quasi-normal modes \cite{Birmingham:2001pj,Son:2002sd,Kovtun:2005ev}. As explained in section \ref{mixing}, we can detect these quasi-normal modes from the vanishing of ${\det} P^-(\epsilon)$, where the matrix $P^-$ is defined in eqs. \eqref{eq:ptildedef} and \eqref{eq:pmatricesdef}.}

Figure \ref{fig:qnms} (a) shows the movement of the pole in the normal phase when $T = 0.91 T_c$ for values of $k_x/\pi T \in [2.579,12.580]$, which in the figure corresponds to starting at the point nearest $\text{Re} [\omega/\pi T] = 0$ and moving toward the upper left. The pole asymptotically approaches the real frequency axis $\text{Im} \, \omega / \pi T = 0$, as $k_x$ increases.

At a temperature $T^* \approx 0.6 T_c$, however, the distance to the real axis develops a local minimum. Figure \ref{fig:qnms} (b) shows the movement of the same pole as figure \ref{fig:qnms} (a) at a temperature $T = 0.48 T_c < T^*$, still in the normal phase. Here we see that the distance to the real frequency axis has a local minimum at $(\text{Re}\, \omega/ \pi T,\text{Im} \, \omega / \pi T) = (-1.79,-7.21\times 10^{-6})$ when $k_x /\pi T = 8.15$. Such behavior persists to lower temperatures, and indeed, the distance to the real frequency axis decreases. The lowest temperature we studied was $T = 0.19 \, T_c$, where the local minimum occurred at $(\text{Re} \, \omega/ \pi T,\text{Im} \, \omega / \pi T) = (5.32,2.64 \times 10^{-18})$ when $k_x /\pi T = 23.22$.

We seem to be seeing the emergence of a Fermi surface, which, as in ref. \cite{Liu:2009dm}, would occur at $T=0$ when the pole would reach the origin of the complex frequency plane at some finite momentum $k_F$, the Fermi momentum. Let us consider low temperature, and define $k^*$ as the value of momentum where the closest approach to the real frequency axis occurs. When $T=0$, $k^*$ would be the Fermi momentum, $k^* = k_F$. At our lowest temperature, $T = 0.19 \, T_c$, $k_x^* /\pi T \equiv 23.22$, and the closest approach to the real frequency axis occurs at an $\omega^*$ given by $\left(\text{Re} \, \omega^*, \text{Im} \, \omega^*\right) = (5.32,2.64 \times 10^{-18}) \pi T$. Letting $k^*$ play the role of $k_F$, we see behavior similar to the results of ref. \cite{Liu:2009dm}: for small but nonzero temperature, as we change the momentum the frequency of the pole behaves as
\beq
\omega - \omega^* \sim \left( k - k^*\right)^z,
\eeq
where our numerical results indicate that the exponent $z = 1.00 \pm 0.01$, and the spectral function behaves as
\beq
\R_{11} \sim \left( k - k^* \right)^{-\alpha},
\eeq
where our numerical results indicate that the exponent $\alpha = 2.0 \pm 0.1$. In fact, these results are rather robust: we find the same $z$ and $\alpha$ for many values of $T<T^*$, and for several other poles. These results suggest that the low-temperature normal phase may not be a Landau Fermi liquid, which would have $z=\alpha=1$.

Figures \ref{fig:qnms} (c) and (d) show the movement of the same pole, at the same temperatures, but in the superfluid phase.\footnote{Our independent caluclations in the normal and superfluid phases yield the same position for the pole at $T=T_c$ to within $0.1 \%$.} Figure \ref{fig:qnms} (c) shows the movement of the pole at $T = 0.91T_c$ for $k_x/\pi T \in [2.49,12.29]$ and $k_y=0$. Figure \ref{fig:qnms} (d) shows the movement of the pole at $T=0.48T_c$ for $k_x/\pi T \in [4.83,8.98]$ and $k_y = 0$. Unlike the normal phase result, here the distance to the real frequency axis does not develop a local minimum. In other words, here we do not see a Fermi surface emerge in the same fashion as in the normal phase.

To see the emergence of the p-wave superfluid Fermi surface, we study the spectral measure (largely following ref. \cite{Gubser:2010dm}), which as mentioned above, provides a direct measure of the density of states that have overlap with our fermionic operators.

Our main results concern the evolution of the spectral measure $\R(\omega,k_x,k_y)$ as we cool the system through the superfluid phase transition. In the spectral measure we will see the breaking of rotational symmetry as we take $T$ below $T_c$, and we will see the emergence of the Fermi surface as we approach $T=0$, although in the probe limit we will not reach $T=0$. Our results agree qualitatively with the $T=0$ results of ref. \cite{Gubser:2010dm}, in which the Fermi surface in the p-wave phase consists of isolated points.

\begin{figure}[htbp]
\begin{center}
\begin{minipage}{\textwidth}
\centering
\psfrag{kx}[r]{$\frac{k_x}{\pi T}$}
\psfrag{ky}{$\frac{k_y}{\pi T}$}
\psfrag{k}{$|k|/ \pi T$}
\psfrag{RHO}{$\R$}
\subfigure[][]{\includegraphics[width=0.49 \textwidth]{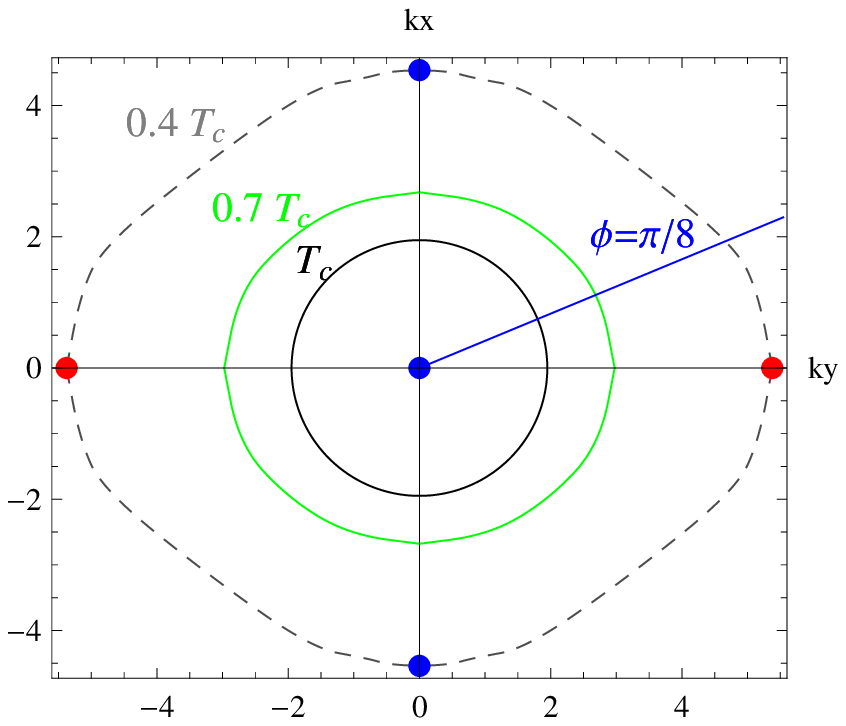}}
\hfill
\subfigure[][]{\includegraphics[width=0.49 \textwidth]{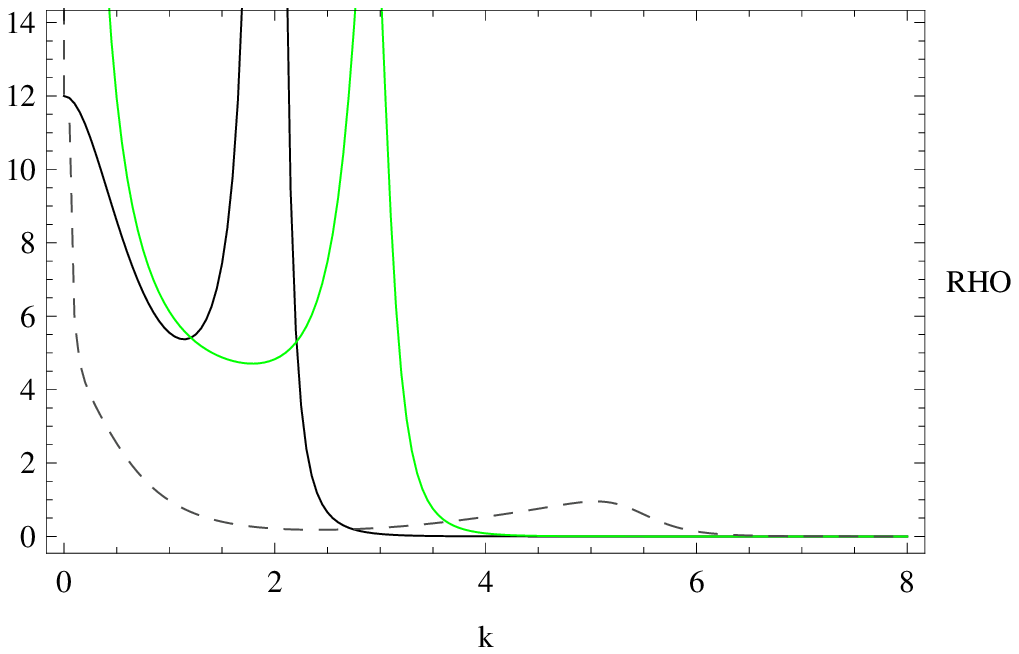}}
\end{minipage}
\caption{
\label{fig:fermiSurfaceCollapse}
The evolution of the spectral measure $\R(\omega,k_x,k_y) = \text{tr} \R_{AB}(\omega,k_x,k_y)$ in the $(k_x/ \pi T,k_y/ \pi T)$ plane at $\omega =0$, as we lower the temperature. \textbf{(a.)} The position of peaks in the spectral measure $\R$ are indicated by the curves as we lower the temperature from $T_c$~(black curve) through $0.7 T_c$~(green curve) to $0.43 T_c$~(dashed grey curve). The $T = 0.43 T_c$ case exhibits a small bump rather than a sharp peak, except for points on the $k_x$ and $k_y$ axes. We have indicated the bump with the grey dashed curve and the peaks with red and blue dots, including the blue dot at the origin. \textbf{(b.)} The spectral measure $\R$ plotted for a representative slice of the $(k_x/\pi T,k_y/ \pi T)$ plane (still with $\omega = 0$), namely along the line given by the polar angle $\phi = \pi/8$ drawn in (a.). We plot $\R$ versus the magnitude of the momentum $|k| = \sqrt{k_x^2 + k_y^2}$ divided by $\pi T$, at $T=T_c$~(black curve), $T = 0.7 T_c$~(green curve), and $T=0.43 T_c$~(dashed grey curve).
}
\end{center}
\end{figure}

Figure \ref{fig:fermiSurfaceCollapse} provides a road map for the evolution of the spectral measure as we lower the temperature. Here we set $\omega = 0$, so we are studying fluctuations with zero energy above the chemical potential. Figure \ref{fig:fermiSurfaceCollapse} (a) indicates the locations of peaks in the spectral measure, in the $(k_x/\pi T,k_y /\pi T)$ plane, with solid lines and the locations of small bumps as the dashed grey line. At $T=T_c$ we see rotational symmetry: the black solid line indicates peaks for any values of $(k_x/\pi T,k_y /\pi T)$ on the black circle. At $T=0.7T_c$ the rotational symmetry is mildly broken: the green line is not a perfect circle. At $T=0.43 T_c$, sharp peaks only appear at isolated points on the axes, denoted by the red and blue dots (and also at the blue dot at the origin), while the dashed line indicates a small bump, rather than a sharp peak. Here we are clearly seeing the emergence of the Fermi surface at isolated points.

To illustrate what the peaks and bumps look like, we choose a representative slice of the $(k_x/\pi T,k_y /\pi T)$ plane, namely the line given by the polar angle $\phi = \pi / 8$, which is drawn in figure \ref{fig:fermiSurfaceCollapse} (a), and plot the ($\omega = 0$) spectral measure versus the magnitude of the momentum $|k| = \sqrt{k_x^2 + k_y^2}$ divided by $\pi T$. In figure \ref{fig:fermiSurfaceCollapse} (b) a distinct peak is visible for both $T=T_c$ (black curve) and $T=0.7T_c$ (green curve), while for this generic (off-axis) value of $\phi$ the only feature of the spectral measure at $T=0.43 T_c$~(dashed grey curve) is a small bump. On the axes ($\phi = 0,\pi/2$), the picture is similar, except the bump becomes a sharp peak, corresponding to the red or blue dots in figure \ref{fig:fermiSurfaceCollapse} (a).

\begin{figure}[htbp]
\begin{center}
\begin{minipage}{\textwidth}
\subfigure[][$T=T_c,\,\omega=0$]{\includegraphics[width=0.45 \textwidth]{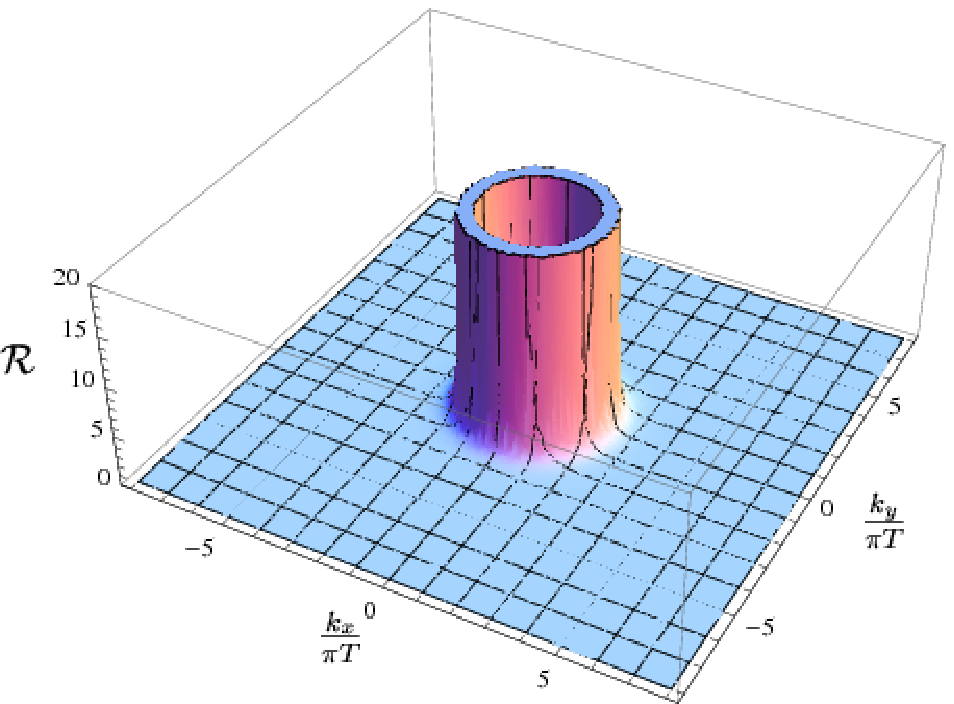}}
\hfill
\subfigure[][$T=0.91\,T_c,\,\omega=0$]{\includegraphics[width=0.45 \textwidth]{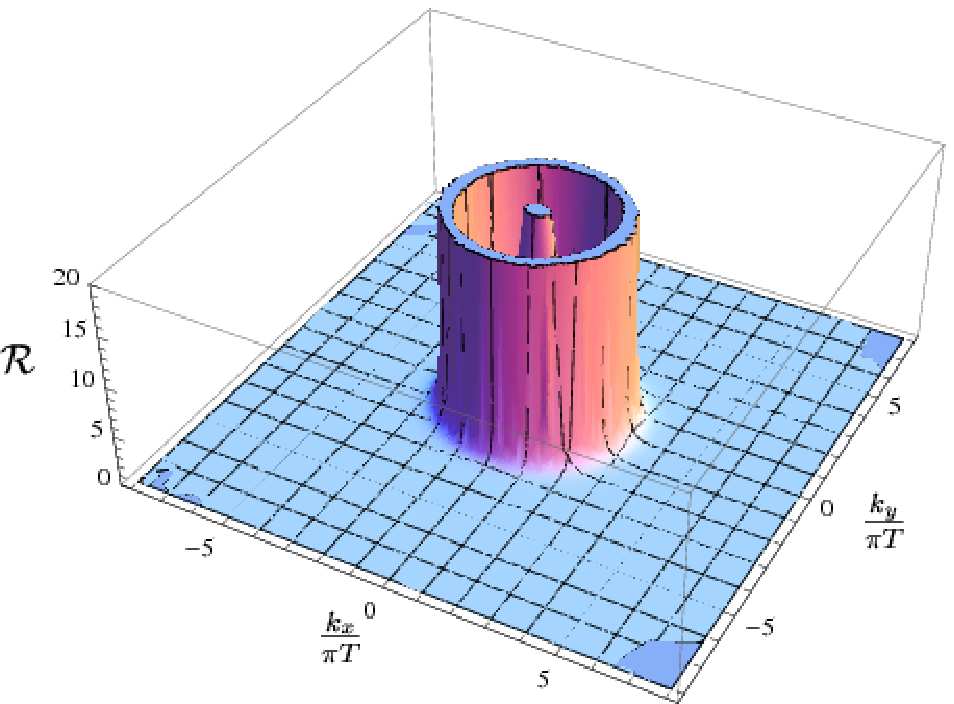}}
\vfill
\subfigure[][$T=0.69\,T_c,\,\omega=0$]{\includegraphics[width=0.45 \textwidth]{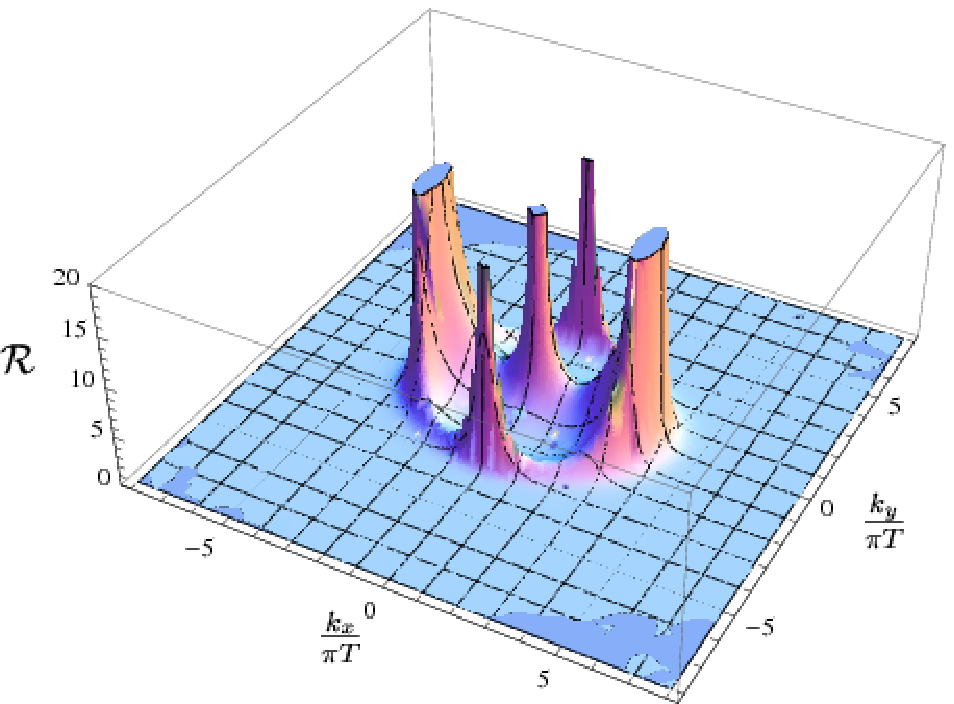}}
\hfill
\subfigure[][$T=0.43\,T_c,\,\omega=0$]{\includegraphics[width=0.45 \textwidth]{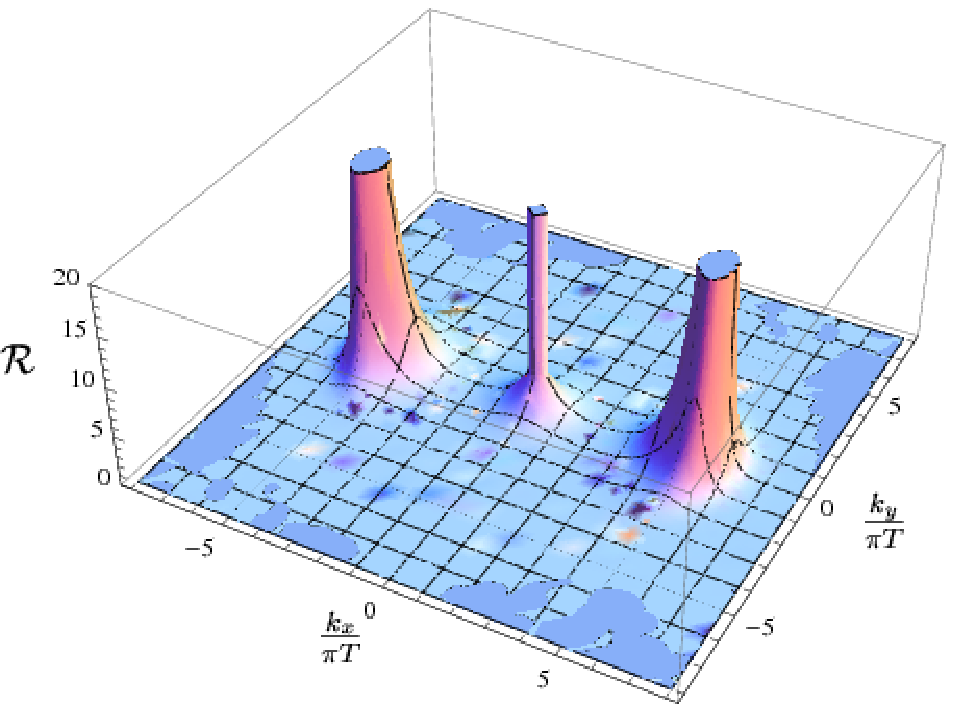}}
\end{minipage}
\caption{
\label{fig:3DRKxKyW0}
Three-dimensional plots of the spectral measure $\R(\omega,k_x,k_y)$ in the superfluid phase over the $(k_x/ \pi T,k_y/ \pi T)$ plane at zero frequency, $\omega=0$, and for distinct temperatures $T \leq T_c$. \textbf{(a)} The $T=T_c$ case, which is clearly rotationally invariant. The peaks correspond to the black circle in figure {\protect\ref{fig:fermiSurfaceCollapse}} (a). \textbf{(b)} The $T = 0.91 T_c$ case, where the spectral measure does not yet display any dramatic breaking of rotational symmetry. Notice the peak a the origin of the $(k_x/ \pi T,k_y/ \pi T)$ plane. \textbf{(c)} The $T=0.69 T_c$ case, where the breaking of rotational symmetry is obvious. We see that the ``cylinder'' of (a) breaks into five distinct peaks on the $k_x/ \pi T$ and $k_y/ \pi T$ axes. The peaks correspond to the green circle in figure {\protect\ref{fig:fermiSurfaceCollapse}} (a). \textbf{(d)} The $T = 0.4 T_c$ case, which still has five peaks, labeled by the red and blue dots in figure {\protect\ref{fig:fermiSurfaceCollapse}} (a). The peaks along the $k_y/\pi T$ axis are too narrow to appear in the three-dimensional plot with the resolution we use.
}
\end{center}
\end{figure}

To illustrate the evolution of the spectral measure in more detail, we present three-dimensional plots of $\R(\omega,k_x,k_y)$, for $\omega = 0$, over the $(k_x/\pi T,k_y /\pi T)$ plane, for temperatures $T = T_c$ (figure \ref{fig:3DRKxKyW0} (a)), $T= 0.91 T_c$ (figure \ref{fig:3DRKxKyW0} (b)), $T=0.69T_c$ (figure \ref{fig:3DRKxKyW0} (c)) and $T=0.4T_c$ (figure \ref{fig:3DRKxKyW0} (d)). In figure \ref{fig:3DRKxKyW0} (a) we see the peaks corresponding to the black circle in figure \ref{fig:fermiSurfaceCollapse} (a). Clearly here the spectral measure is rotationally symmetric. When we cool the system to $T=0.69 T_c$ (figure \ref{fig:3DRKxKyW0} (c)), we clearly see the emergence of five distinct peaks, two on the $k_x /\pi T$ axis, two on the $k_y /\pi T$ axis, and one at the origin. The circle of peaks corresponds to the green circle in figure \ref{fig:fermiSurfaceCollapse} (a). When we further cool the system to $T = 0.4T_c$, the five peaks are still present, although the resolution of our three-dimensional plot is insufficient to resolve the two on the $k_y /\pi T$ axis away from the origin.

Although these peaks have a much smaller footprint in the $(k_x/\pi T,k_y /\pi T)$ plane than the peaks on the $k_x /\pi T$ axis, they are much taller. The spectral measure $\R$ is of order $5 \times 10^5$ at the peaks on the $k_y /\pi T$ axis, but only order $10^2$ at the peaks on the $k_x /\pi T$ axis, and order $5 \times 10^4$ at the peak at the origin. Apparently a large number of states are piling up at two precise locations on the $k_y /\pi T$ axis.

For a bulk theory with $SU(2)$ gauge fields and fermions in the
fundamental representation, a combination of analytic and numerical
results reveal that the $\omega =0$ spectral measure at $T=0$ consists
of two isolated points on the $k_x$ axis, located symmetrically about
the origin \cite{Gubser:2010dm}. For the same bulk theory
but for fermions in the adjoint representation, as we consider here,
 the analytic arguments of ref. \cite{Gubser:2010dm} indicate that 
the $\omega = 0$ spectral measure at $T=0$ should consist of three isolated points, one at the origin and two on the $k_x$ axis, located symmetrically about the origin.

Therefore, what we see  appears to be 
consistent with the results of ref. \cite{Gubser:2010dm}
for the structure of the spectral measure in the superfluid phase. The main obstacle to a direct comparison is the probe limit, which restricts us to finite temperatures: we do not know which peaks in our spectral measure persist to $T=0$. Nevertheless, given that we see the three peaks on the $k_x/\pi T$ that we generically expect, and that the two peaks on the $k_y/\pi T$ axis have a shrinking footprint as we cool the system, we have good reason to believe that the results of ref. \cite{Gubser:2010dm} may apply to our system. To answer the question fully requires computing the back-reaction of the D5-branes in the bulk. Whether that produces a domain-wall geometry of the kind used in ref. \cite{Gubser:2010dm} is not guaranteed.

\begin{figure}[t!]
\begin{center}
\psfrag{kx}[rr]{\large$k_x/\pi T$}
\psfrag{ky}[bbb]{\large$\ \ \frac{k_y}{\pi T}$}
\psfrag{RHO}{$\R$}
\psfrag{w}{$\omega / \pi T$}
\psfrag{k}{$|k|/\pi T$}
\begin{minipage}[t]{\textwidth}
\begin{minipage}{0.5\textwidth}
\center
\subfigure[]{
\includegraphics[width=0.99 \textwidth]{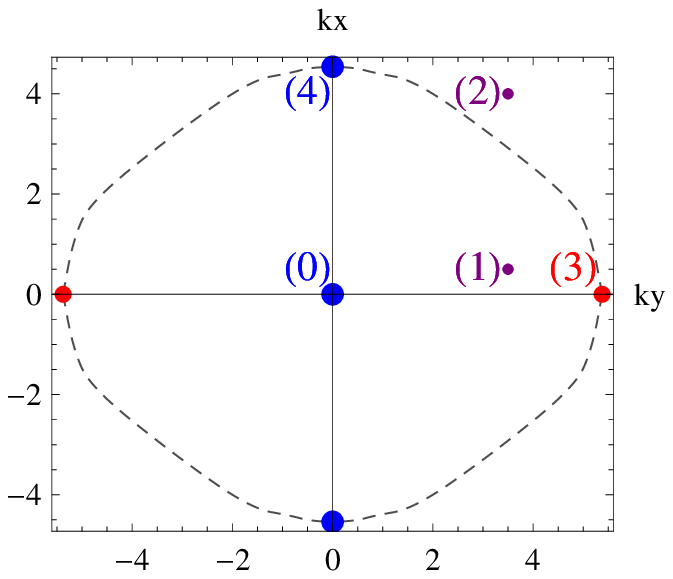}}
\vfill
\subfigure[]{\includegraphics[width=  \textwidth]{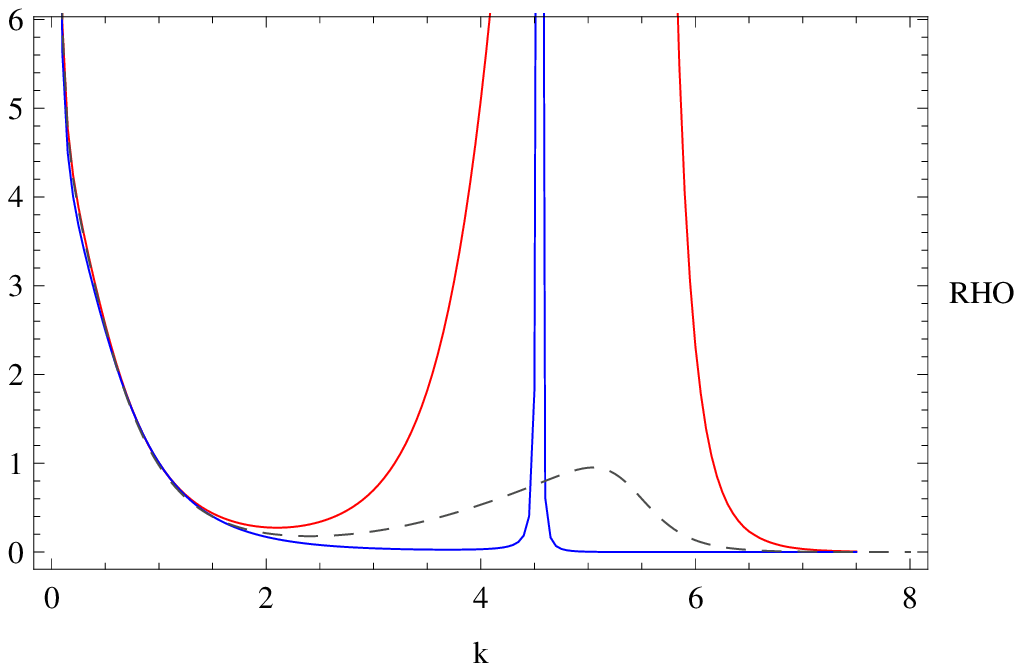}}
\end{minipage}
\renewcommand{\thesubfigure}{(\arabic{subfigure})}
\makeatletter
\makeatother
\setcounter{subfigure}{0}
\hfill
\begin{minipage}{0.44\textwidth}
\center
\vspace{-0.5cm}
\subfigure[][]{\includegraphics[width= 0.98\textwidth]{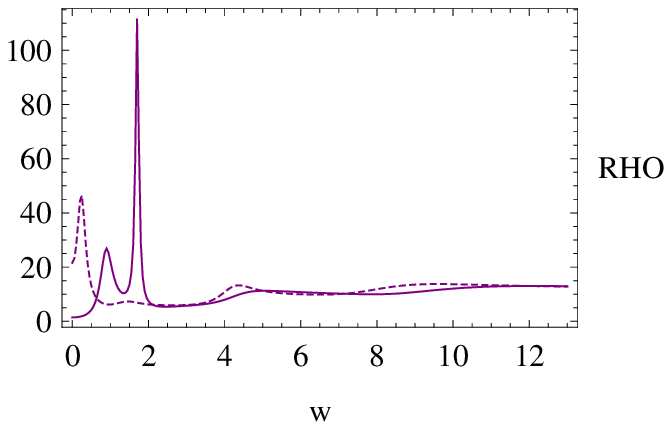}}
\vspace{-0.15cm}
\subfigure[(2)][]{\includegraphics[width=0.98\textwidth]{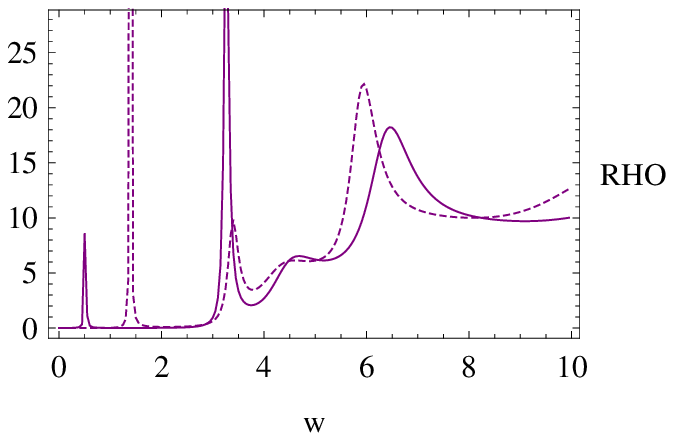}}
\vspace{-0.3cm}
\subfigure[][]{\includegraphics[width= 0.98\textwidth]{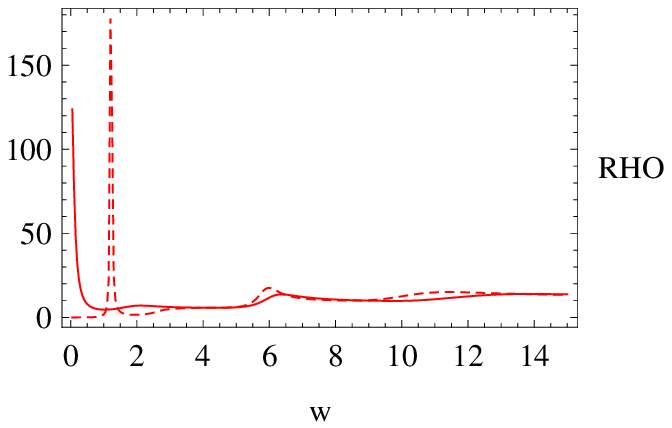}}
\end{minipage}
\end{minipage}
\caption{
\label{fig:lowTfermiPoints}
\textbf{(a)} Peaks in the $\omega=0$ spectral measure in the $(\frac{k_x}{\pi T},\frac{k_y}{\pi T})$ plane at $T=0.4 T_c$. The labeling is the same as figure {\protect\ref{fig:fermiSurfaceCollapse}} (a). \textbf{(b)} The $\omega = 0$ spectral measure along three lines in (a): along the $\frac{k_x}{\pi T}$ axis (red curve) and the $\frac{k_y}{\pi T}$ axis (blue curve), so the peaks at nonzero momenta correspond to (3) and (4) in (a), respectively, and along the polar angle $\phi = \pi /8$ (dashed grey curve), drawn in figure {\protect\ref{fig:fermiSurfaceCollapse}} (a), so the bump corresponds to where $\phi = \pi / 8$ intersects the dashed grey line in (a). In each case the peak at zero momentum corresponds to (0) in (a). \textbf{(1)} $\R$ as a function of $\omega/\pi T$ for the point labeled (1) in (a). The dotted and solid lines are for $T=0.55 T_c$ and $T=0.4T_c$, respectively. \textbf{(2)} and \textbf{(3)} show the same thing for the corresponding points (2) and (3) in (a).
}
\end{center}
\end{figure}

We begin to explore the $\omega$ dependence of the spectral measure $\R(\omega,k_x,k_y)$ in figure \ref{fig:lowTfermiPoints}. Figure \ref{fig:lowTfermiPoints} (a) simply reproduces the $T = 0.4 T_c$ part of figure \ref{fig:fermiSurfaceCollapse} (a), where $\omega =0$. Figure \ref{fig:lowTfermiPoints} (b) shows the spectral function along three lines in the $(k_x/ \pi T,k_y/ \pi T)$ plane: along the positive $k_x/\pi T$ axis (red curve), along the positive $k_y/\pi T$ axis (blue curve), and along the line given by the polar angle $\phi = \pi /8$ in figure \ref{fig:fermiSurfaceCollapse} (a). Here we see explicitly the difference in widths of the peaks on the $k_x /\pi T$ and $k_y / \pi T$ axes (the red and blue peaks).

To explore the $\omega$ dependence, we choose a few representative points in the $(k_x/ \pi T,k_y/ \pi T)$ plane and, for each point, plot the spectral measure versus $\omega$. For these points, we consider not only $T=0.4T_c$, as in figure \ref{fig:lowTfermiPoints} (a), but also the slightly higher temperature $T = 0.55T_c$, in order to study the behavior as we cool the system.

Our points are similar to those in figure 8 of ref. \cite{Gubser:2010dm}, where the same quantities were plotted (for the slightly different system of ref. \cite{Gubser:2010dm}): the spectral measure versus $\omega$ for fixed $k_x$ and $k_y$. We will thus compare our results to those of figure 8 of ref. \cite{Gubser:2010dm} along the way.

In figure \ref{fig:lowTfermiPoints} (1) we plot the $\omega$ dependence of $\R(\omega,k_x,k_y)$ for the point (1) labeled in figure \ref{fig:lowTfermiPoints} (a). The dotted line is for $T=0.55 T_c$ and the solid line is for $T = 0.40T_c$. Here we see that as we cool the system, a small gap (a depletion of states) opens near $\omega = 0$, while a sharp peak emerges near $\omega / \pi T \approx 1.7$. Such behavior at least appears to be approaching that of figure 8 (1) in ref. \cite{Gubser:2010dm}, where a genuine gap (zero states) appeared at $\omega =0$.

In figure \ref{fig:lowTfermiPoints} (2) we plot the $\omega$ dependence of $\R(\omega,k_x,k_y)$ for the point (2) labeled in figure \ref{fig:lowTfermiPoints} (a). The dotted line is for $T=0.55 T_c$ and the solid line is for $T = 0.40T_c$. Here we see that a sharp peak near $\omega / \pi T \approx 1.5$ when $T=0.55 T_c$ shrinks and begins moving toward $\omega =0$ as we lower the temperature to $T=0.4 T_c$. Moreover, the small peak near $\omega / \pi T \approx 3.5$ when $T=0.55 T_c$ grows much sharper at $T=0.4 T_c$. As in ref. \cite{Gubser:2010dm}, here we seem to see the emergence of the well-known ``peak-dip-hump'' shape, with the peak being at $\omega / \pi T \approx 3.5$, the dip being at $\omega / \pi T\approx 5.2$, and the hump being at $\omega / \pi T \approx 6.5$.

In figure 8 (2) of ref. \cite{Gubser:2010dm}, a gap was present in the spectral measure for small frequencies, except for a single genuine delta-function peak at finite frequency, and at larger frequency a continuum of states appears (the ``hump''). As argued in ref. \cite{Gubser:2010dm}, at finite temperature the delta-function peak will acquire a finite width and merge with the hump, producing the peak-dip-hump. Our spectral measure appears to be approaching the form of the spectral measure in figure 8 (2) of ref. \cite{Gubser:2010dm} (with the usual caveat that we cannot actually reach $T=0$).

In figure \ref{fig:lowTfermiPoints} (3) we plot the $\omega$ dependence of $\R(\omega,k_x,k_y)$ for the point (3) labeled in figure \ref{fig:lowTfermiPoints} (a), which is sitting right on top of the peak on the positive $k_x /\pi T$ axis. The dotted line is for $T=0.55T_c$ and the solid line is for $T=0.40T_c$. At the higher temperature (the dotted line), the primary feature is the peak near $\omega / \pi T \approx 1.2$, which moves toward $\omega =0$ and also shrinks (the peak is lower) as we lower the temperature, becoming the peak in the solid line. Assuming such a trend continues, our results would be consistent with figure 8 (3) of ref. \cite{Gubser:2010dm}, where, sitting right on top of the peak on the $k_x$ axis, the spectral measure went to a finite constant at $\omega =0$.

\begin{figure}[htpb]
\begin{center}
\psfrag{RHO}{\large$\R$}
\psfrag{w}{\large$\omega / \pi T$}
\begin{minipage}{\textwidth}
\subfigure[][]{\includegraphics[width=0.8 \textwidth]{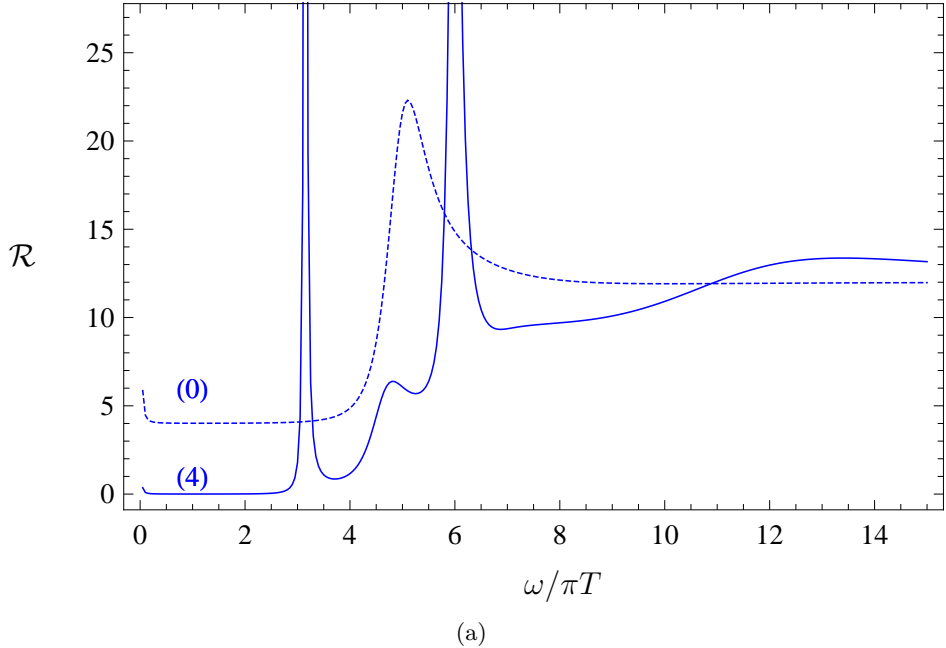}}
\end{minipage}
\caption{
\label{fig:plotOriginBluePoints}
The spectral measure $\R(\omega,k_x,k_y)$ as a function of $\omega / \pi T$ for $(k_x/\pi T, k_y/\pi T)$ values corresponding to points (0) (dotted blue curve) and (4) (solid blue curve) in figure {\protect\ref{fig:lowTfermiPoints}} (a). These points sit right on top of the peaks at the origin and on the positive $k_y /\pi T$ axis, respectively. Here $T=0.4T_c$.
}
\end{center}
\end{figure}

Figure \ref{fig:plotOriginBluePoints} shows $\R(\omega,k_x,k_y)$ versus $\omega / \pi T$ for the points (0) and (4) in figure \ref{fig:lowTfermiPoints} (a), sitting right on top of the peaks at the origin and on the positive $k_y / \pi T$ axis, respectively. Here we use only $T=0.4T_c$. We clearly see a gap developing at low frequency in both cases. Such behavior is similar to the gap that develops in the spectral function of vector fluctuations \cite{Ammon:2008fc,Ammon:2009fe}, which is immediately related (via a Kubo formula) to a gap in the conductivity. Whether these two gaps are related is unclear, but deserves further study.

\begin{figure}[htbp]
\begin{center}
\begin{minipage}{\textwidth}
\subfigure[][$T=T_c,\,\omega / \pi T=0.25$]{\includegraphics[width=0.45 \textwidth]{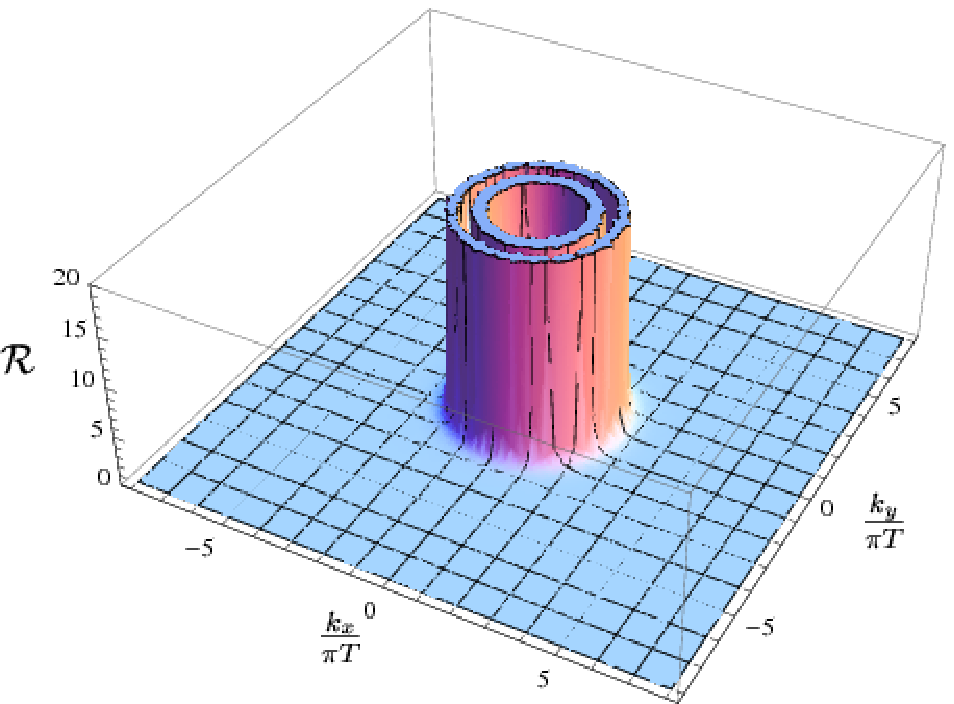}}
\hfill
\subfigure[][$T=0.91\,T_c,\,\omega / \pi T=0.25$]{\includegraphics[width=0.45 \textwidth]{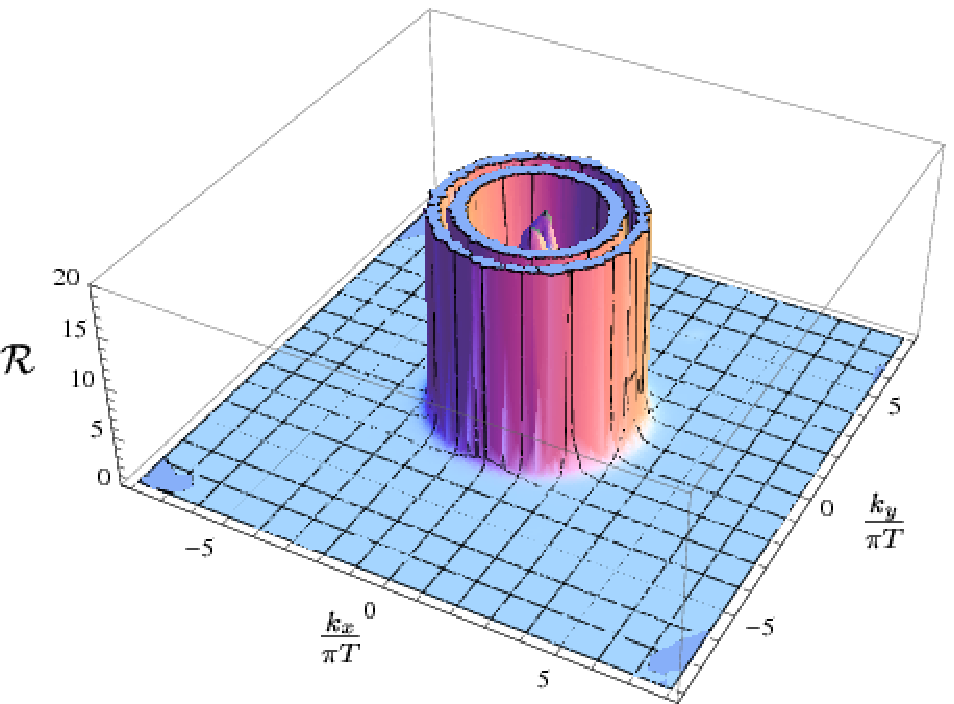}}
\vfill
\subfigure[][$T=0.54\,T_c,\,\omega / \pi T=0.25$]{\includegraphics[width=0.45 \textwidth]{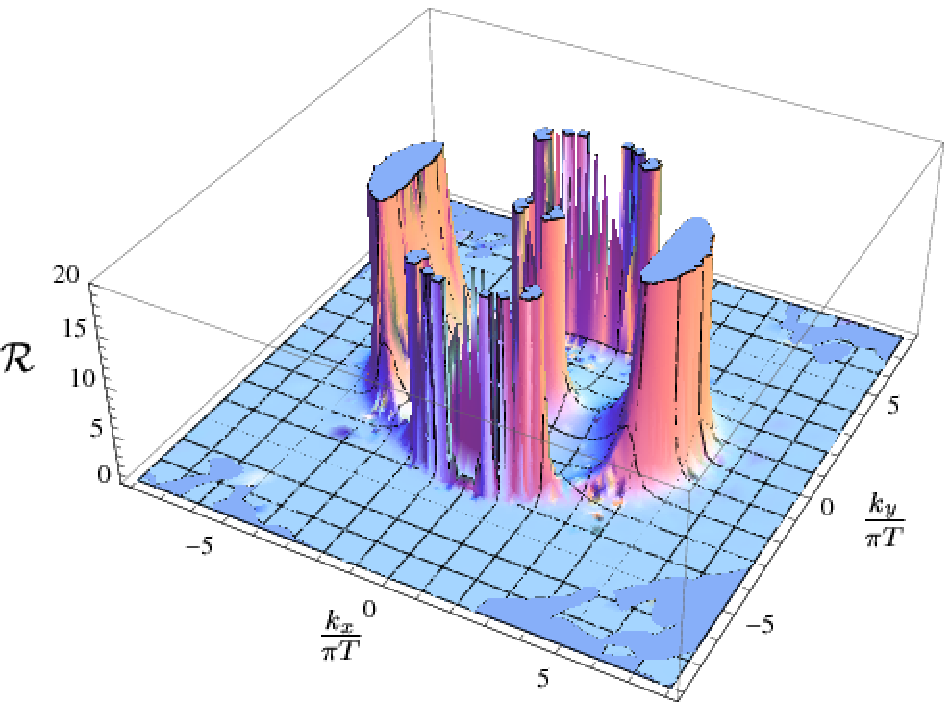}}
\hfill
\subfigure[][$T=0.43\,T_c,\,\omega / \pi T=0.25$]{\includegraphics[width=0.45 \textwidth]{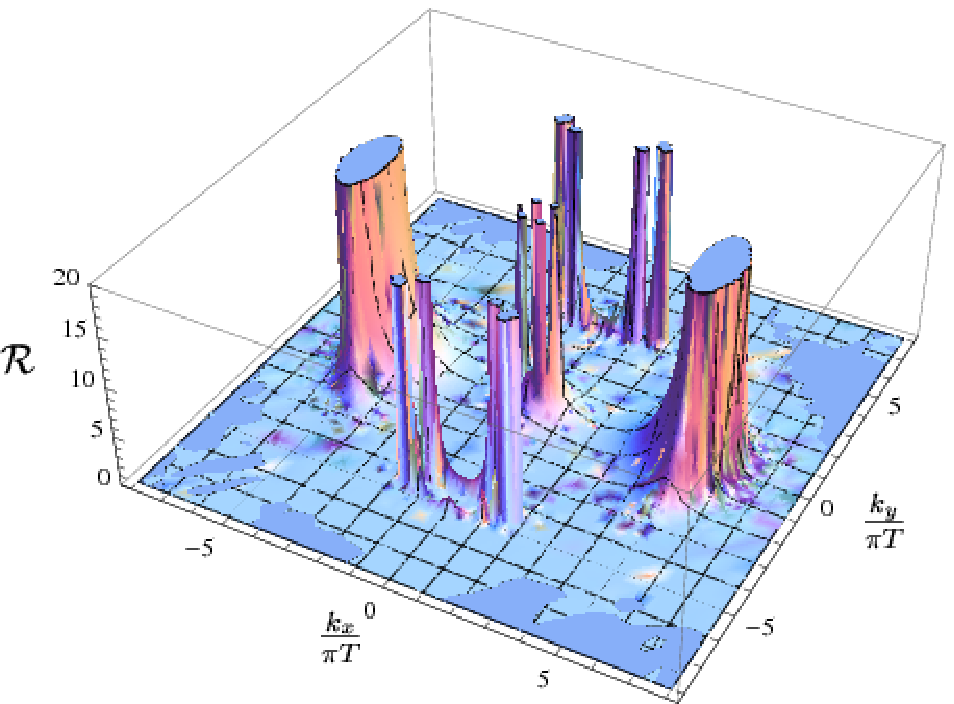}}
\end{minipage}
\caption{
\label{fig:3DRKxKyW025}
Three-dimensional plots of the spectral measure $\R(\omega,k_x,k_y)$ in the superfluid phase over the $(k_x/ \pi T,k_y/ \pi T)$ plane at $\omega / \pi T =0.25$, and for distinct temperatures $T \leq T_c$. The four plots are arranged in a similar fashion as in figure {\protect\ref{fig:3DRKxKyW0}}. \textbf{(a)} is the $T=T_c$ case, \textbf{(b)} is the $T=0.91 T_c$ case,  \textbf{(c)} is the $T=0.54 T_c$ case, and \textbf{(d)} is the $T=0.4 T_c$ case. The main differences from figure {\protect\ref{fig:3DRKxKyW0}} are that the ``cylinder'' we saw in figure {\protect\ref{fig:3DRKxKyW0}} (a) is now two concentric cylinders, and the peaks along the $k_y / \pi T$ axis each split into multiple peaks which then move away from one another along the $k_x / \pi T$ axis as we cool the system.
}
\end{center}
\end{figure}

Finally, to explore further the $\omega$ dependence of
$\R(\omega,k_x,k_y)$, we do not restrict to points in the $(k_x/ \pi
T,k_y/ \pi T)$ plane, but rather restrict to a single nonzero value of
frequency, $\omega / \pi T = 0.25$, and plot the spectral me asure over the entire $(k_x/ \pi T,k_y/ \pi T)$ plane. The result appears in figure \ref{fig:3DRKxKyW025}, where (a) is for $T=T_c$, (b) is for $T=0.91T_c$, (c) is for $T = 0.54 T_c$ and (d) is for $T=0.4T_c$. We see a number of differences from the $\omega = 0$ case of figure \ref{fig:3DRKxKyW0}. At the transition, $T=T_c$, the spectral measure is again rotationally symmetric, but now with two concentric ``cylinders.'' By the time we cool the system to $T=0.54 T_c$, we see a number of peaks clumped near the momentum axes. If we compare the spectral measure at $T=0.4T_c$ at $\omega =0$ and $\omega / \pi T = 0.25$, figures \ref{fig:3DRKxKyW0} (d) and \ref{fig:3DRKxKyW025} (d), respectively, then we see that the $\omega = 0$ peaks on the $k_y / \pi T$ axis each split into a number of peaks at $\omega / \pi T = 0.25$ which then move apart along the $k_x / \pi T$ axis.

To summarize, in the superfluid phase we observe isolated peaks in the spectral measure, whose locations appear to be consistent with the p-wave nature of the condensate, as well as with the fact that our fermions are in the adjoint representation of $SU(2)$. Such a structure suggests nodes in the energy gap on the normal-phase Fermi surface. We plan to investigate this further in the near future, for instance by identifying the appropriate Dirac cones  and by studying the spectral measure for fixed values of $\omega$, as well as for further values of fixed $k_x$, $k_y$.

\section{Conclusions}
\label{conclusions}

The main results of this paper are:
\begin{itemize}
\addtolength{\itemsep}{-0.1\baselineskip}

\item We described an efficient algorithm for computing retarded Green's functions from bulk fermions that couple to one another,

\item We performed holographic renormalization for fermions in AdS,

\item We determined a simple embedding into string theory of a charged fermion in AdS, using probe D-branes,

\item Using the above ingredients, we numerically computed fermionic spectral functions in p-wave superfluid states, with emphasis on the emergence of the isolated points of the Fermi surface as we cooled the system through the p-wave superfluid transition.

\end{itemize}

Using the machinery that we have developed, we can think of a number of directions for future research.

Introducing a magnetic field on the worldvolume of a probe Dp-brane is straightforward to do, and has many interesting consequences \cite{Ammon:2009fe,Wapler:2009rf,Wapler:2009tr,Filev:2009ai,Jensen:2010ga,O'Bannon:2008bz,Filev:2007gb,Albash:2007bk,Erdmenger:2007bn,Evans:2010iy,Jensen:2010vd}. Our embedding of p-wave superfluids, and of charged fermions, into string theory may be a useful arena in which to study the effects of magnetic fields both on holographic p-wave superfluids, along the lines of ref. \cite{Ammon:2009fe}, and on holographic Fermi surfaces, along the lines of refs. \cite{Albash:2009wz,Basu:2009qz,Denef:2009yy,Albash:2010yr}.

Typically the mesinos carry some R-charge, so another avenue to study holographic Fermi surfaces would be to embed probe Dp-branes into background geometries corresponding to field theory states with finite R-charge density. In those cases the $T=0$ limit is accessible within the probe approximation. These geometries will generically be charged dilatonic black hole solutions. The most attractive of such solutions, for condensed matter applications, may be the solution discussed in ref. \cite{Gubser:2009qt}, which produced in the field theory specific heats linear in temperature.

We focused on probe D5-branes extended along $AdS_4 \times S^2$ inside $AdS_5 \times S^5$, and worked only with the massless worldvolume fermion. Other Dp-branes, and worldvolume fermions of other masses, deserve further study.

Lastly, knowing the $T=0$, finite isospin density ground state of the D3/D5 theory would be extremely useful. In bulk terms, the problem is to find a solution of type IIB supergravity representing fully back-reacted D5-branes with non-trivial worldvolume gauge fields.

We plan to investigate these and related issues in the future.

\section*{Acknowledgements}
We thank Chris Herzog, Ingo Kirsch, Luca Martucci, Silviu Pufu, Fabio Rocha, Ronny Thomale, and Amos Yarom for useful discussions. We especially thank Carlos Fuertes for sharing his unpublished notes with us, Jegors Korovins for proof-reading the appendix, and Patrick Kerner for assistance with the figures. This work was supported in part by the Cluster of Excellence ``Origin and Structure of the Universe.'' The work of M. K. was supported by the \textit{Deutsche Forschungsgemeinschaft} (DFG). M. A. would like to thank the Studienstiftung des deutschen Volkes for financial support. 

\begin{appendix}
\section{Holographic Renormalization of Fermions in AdS}
\label{appa}

In this appendix we perform the holographic renormalization of a Dirac fermion in spaces that asymptotically approach $AdS_{d+1}$. We will work in Euclidean-signature $AdS_{d+1}$, unless stated otherwise. We assume the bulk metric asymptotically approaches the metric of eq. (\ref{eq:fgeucads}),
\beq
\label{eq:fgeucads2}
ds^2 = g_{AB} \, dx^A dx^B = \frac{du^2}{u^2} + \frac{1}{u^2} \delta_{ij} \, dx^i dx^j,
\eeq
but, unlike section \ref{mixing}, we leave the number of spatial dimensions $d$ unconstrained.

Our results will only be valid for spaces whose metrics asymptotically approach the $AdS_{d+1}$ metric. Various generalizations are possible. One generalization would involve spaces for which surfaces of fixed $u$ are not simply flat space, that is, spaces for which $\delta_{ij} \rightarrow g_{ij}(x)$ in eq. (\ref{eq:fgeucads2}), where $g_{ij}(x)$ has non-trivial curvature. Another generalization would be to study spaces that are not asymptotically $AdS_{d+1}$, for example, spaces that asymptotically approach the near-horizon geometry of Dp-branes, with $p\leq 5$ but $p\neq 3$. We also leave an analysis of the Lorentzian-signature case, along the lines of ref. \cite{Skenderis:2008dh, Skenderis:2008dg, Giecold:2009tt}, for the future.

As mentioned in section \ref{mixing}, the form of the counterterms will be fixed by symmetries (in particular Lorentz invariance on the $u=\epsilon$ surface) and by the requirement that they cancel divergences of the on-shell action. Moreover, for fermions we have an additional constraint. The counterterms can be built only from $\Psi_-$ (defined in eq. \ref{eq:psiexplicit}), because that is held fixed under variations, and hence does not spoil the stationarity of the action.

Otherwise, the procedure for fermions very closely parallels that for scalars, which was reviewed for example in ref. \cite{Skenderis:2002wp}. Our main results are the counterterms in eqs. (\ref{eq:countertermactionnothalfinteger}) and (\ref{eq:countertermactionhalfinteger}) and the renormalized on-shell actions in eqs. (\ref{eq:regulatedonshelldiracnothalfinteger}), (\ref{eq:regulatedonshelldirachalfintegernotonehalf}), and (\ref{eq:regulatedonshelldiraconehalf}).

\subsection{Solving the Equation of Motion}

We begin with the Dirac action (plus boundary terms),
\beq
\label{eq:diracactionappendix}
S = \int d^{d+1} x \, \sqrt{g} \, \left ( \bar{\Psi} \, \D \Psi - m \, \bar{\Psi} \Psi \right) + S_{bdy},
\eeq
We will work with a single Fourier mode, so we let $\Psi \rightarrow e^{ikx}\,\Psi$, where, without loss of generality, we have chosen the momentum to point in the $\hat{x}$ direction. The equation of motion is then eq. (\ref{eq:ddimsimplifieddiraceq}),
\beq
\label{eq:ddimsimplifieddiraceq2}
\left [u \g^u \partial_u + i k\, u \, \g^x - \frac{d}{2} \g^u - m \right] \Psi = 0.
\eeq
As in section \ref{mixing} we define $\Psi_{\pm} = \frac{1}{2} \left( 1 \pm \g^u\right) \Psi$ so that $\g^u \Psi_{\pm} = \pm \Psi_{\pm}$. We will refer to $\Psi_+$ and $\Psi_-$ as having opposite chirality (even when $d+1$ is odd). With $\Psi_{\pm}$, we obtain the same decoupled second-order equations as in eq. (\ref{eq:2ndorderdiraceq}), 
\beq
\label{eq:2ndorderdiraceq2}
\left [ \partial_u^2 - \frac{d}{u} \partial_u + \frac{1}{u^2} \left( - m^2 \pm m + \frac{d^2}{4} + \frac{d}{2} \right) - k^2 \right]\Psi_{\pm} = 0.
\eeq
We can solve eq. (\ref{eq:2ndorderdiraceq2}) exactly. The form of the solution depends on the value of $m$.

First, suppose $m$ is \textit{not} a half-integer. The solution of eq. (\ref{eq:2ndorderdiraceq2}) is then
\beq
\label{eq:soljj}
\Psi_{\pm} = u^{\frac{d+1}{2}} \left [ C^{\pm}_1(k) \, J_{m\mp\frac{1}{2}} (\sqrt{-k^2} \, u) + C^{\pm}_2(k) \, J_{-(m\mp\frac{1}{2})} (\sqrt{-k^2} \, u) \right],
\eeq
where the $J$ functions are Bessel functions of the first kind\footnote{ Notice also that since the arguments of the Bessel functions are imaginary, we could equally well write them using the modified Bessel functions $I_n(x)$ and $K_n(x)$. For our asymptotic analysis, the distinction is not important. If the space was pure $AdS_{d+1}$ (not just asymptotically $AdS_{d+1}$), however, then eq. (\ref{eq:soljj}) is the solution for all $u$, and regularity in the bulk would force us to discard the $I_n(x)$ solution. Indeed, we will discuss this later in the appendix.} and $C^{\pm}_1(k)$ and $C^{\pm}_2(k)$ are spinors of the same chirality\footnote{In the rest of the paper, $1$ and $2$ subscripts referred to the projectors $\Pi_{1,2}$. We never use these projectors in this appendix. Here, the subscripts $1$ and $2$ on $C^{\pm}_1(k)$ and $C^{\pm}_2(k)$ simply distinguish the two linearly independent solutions of eq. (\ref{eq:2ndorderdiraceq2}).} as $\Psi_{\pm}$, which may depend on $k$, as indicated.

When $m$ is half-integer, $m\pm\frac{1}{2}$ is an integer and hence the order of the Bessel functions is an integer. In that case the two Bessel functions are not linearly independent: if the order $n$ of $J_n(x)$ is an integer, then $J_n(x)$ obeys the special identity $J_{-n}(x) = (-1)^n J_n(x)$, so in that case $J_n(x)$ and $J_{-n}(x)$ are obviously linearly related.\footnote{The Wronskian \beq W(J_n(x),J_{-n}(x)) = - \frac{\sin n \pi}{\pi x}, \nonumber \eeq clearly vanishes when $n$ is integer, indicating linear dependence. The Wronskian \beq W(J_n(x),Y_{n}(x)) =  \frac{2}{\pi x}, \nonumber \eeq for any $n$.} When $m$ is half-integer, we must introduce a Bessel function of the second kind, $Y_n(x)$, which is linearly independent from $J_n(x)$. When $m$ is half-integer, the solution thus becomes
\beq
\label{eq:soljy}
\Psi_{\pm} = u^{\frac{d+1}{2}} \left [ C^{\pm}_1(k) \, J_{m\mp\frac{1}{2}} (\sqrt{-k^2} \, u) + C^{\pm}_2(k) \, Y_{m\mp\frac{1}{2}} (\sqrt{-k^2} \, u) \right].
\eeq
Notice that if the order $n$ of $Y_n(x)$ is an integer, then $Y_n(x)$ obeys a relation similar to that for $J_n(x)$, namely $Y_{-n}(x) = (-1)^n Y_n(x)$.

The key difference between the Bessel functions of the first and second kinds that will be important for us is that the small-$x$ expansion of $J_n(x)$ involves only powers of $x$, whereas the expansion of $Y_n(x)$ involves both powers of $x$ and logarithms of $x$. Indeed, $J_n(x)$ has a series expression (for any $n$),
\beq
\label{eq:jnseriesform}
J_n(x) = \left( \frac{x}{2} \right)^n \, \sum_{k=0}^{\infty} \frac{(-1)^k}{k! \, \G(k+n+1)} \, \left(\frac{x}{2}\right)^{2k},
\eeq
from which we can immediately read the expansion for small $x$. $Y_n(x)$ has a series expression, when $n$ is a non-negative integer,
\bea
\label{eq:ynseriesform}
Y_n(x) & = & \frac{2}{\pi} J_n(x) \log \frac{x}{2} - \frac{1}{\pi} \left( \frac{x}{2} \right)^{-n} \, \sum_{k=0}^{n-1} \frac{(n-k-1)!}{k!} \left( \frac{x}{2} \right)^{2k} \nonumber \\ & & - \frac{1}{\pi} \left( \frac{x}{2} \right)^n \, \sum_{k=0}^{\infty} \frac{(-1)^k}{k!(n+k)!} \left [ \psi(n+k+1) + \psi(k+1) \right] \left( \frac{x}{2} \right)^{2k},
\eea
where $\psi(x) = \frac{\G'(x)}{\G(x)}$ is the digamma function. We can obtain the series form of $Y_{-n}(x)$ simply from $Y_{-n}(x) = (-1)^n Y_n(x)$. The two features to notice are that the logarithmic terms in $Y_n(x)$ are simply of the form $J_n(x) \log x$, and $Y_n(x)$ also includes terms with inverse powers of $x$, from the $x^{-n}$ multiplying the first sum.

Using the series expressions for the Bessel functions, we can rewrite the solutions in a form useful for our purposes (and introduce some notation to keep expressions simple and compact). First, consider $m$ not-half-integer. We write
\bea
\label{eq:psisolnotationdef}
\Psi_{\pm} & = & u^{\frac{d+1}{2}} \left [ C^{\pm}_1(k) \,  J_{m\mp\frac{1}{2}} (\sqrt{-k^2} \, u)+ C^{\pm}_2(k) \, J_{-(m\mp\frac{1}{2})} (\sqrt{-k^2} \, u) \right], \nonumber \\ & = & u^{\frac{d+1}{2}} \left [ c^{\pm}_1(k) \, u^{m\mp\frac{1}{2}} \left( 1+s_a^{\pm}(u,k) \right)  + c^{\pm}_2(k) \, u^{-\left(m\mp\frac{1}{2}\right)} \, \left(1+s_b^{\pm}(u,k)\right) \right],\nonumber \\ & = & c_1^{\pm}(k) \, u^{\frac{d}{2} + m \mp \frac{1}{2} + \frac{1}{2}} \, \left( 1+ s_a^{\pm}(u,k) \right) + c_2^{\pm}(k) \, u^{\frac{d}{2} - m \pm \frac{1}{2} + \frac{1}{2}} \, \left( 1+ s_b^{\pm}(u,k) \right),
\eea
where in the second line we have absorbed various factors into $C^{\pm}_1(k)$ and $C^{\pm}_2(k)$, which we then relabeled as $c^{\pm}_1(k)$ and $c^{\pm}_2(k)$, and we have defined the series
\beq
\label{eq:asumdef}
s_a^{\pm}(u,k) \equiv \sum_{j=1}^{\infty} a_{j}^{\pm}(m) \left( -k^2 \right)^{j} u^{2j}, \qquad a_{j}^{\pm}(m) \equiv \frac{\left(-1\right)^j}{j! \,2^{2j}} \frac{\G\left(1+\left(m\mp\frac{1}{2}\right)\right)}{\G\left(j+1+\left(m\mp\frac{1}{2}\right)\right)},
\eeq
and $s_b^{\pm}(u,k)$ and $b_j^{\pm}(m)$ are defined similarly, but with $\left(m\mp\frac{1}{2}\right) \rightarrow - \left(m\mp\frac{1}{2}\right)$. We have isolated the leading powers of $u$ near the boundary $u\ra0$: the sums $s_a^{\pm}(u,k)$ and $s_b^{\pm}(u,k)$ involve sub-leading powers of $u$  (each sum starts at order $u^2$). Written separately (to facilitate counting powers of $u$), $\Psi_{\pm}$ are
\bea
\label{eq:psisolseparate}
\Psi_+ & = & c_1^+(k) \, u^{\frac{d}{2} + m} \left(1+s_a^+(u,k) \right) + c_2^+(k) \, u^{\frac{d}{2} - m+1} \left(1+s_b^+(u,k)\right) \nonumber \\ \Psi_- & = & c_1^-(k) \, u^{\frac{d}{2} + m+1} \left(1+s_a^-(u,k) \right) + c_2^-(k) \, u^{\frac{d}{2} - m} \left(1+s_b^-(u,k)\right).
\eea
The coefficients in eq. (\ref{eq:psisolseparate}) are actually not independent. If we return to eq. (\ref{eq:ddimsimplifieddiraceq2}) and insert $\Psi = \Psi_+ + \Psi_-$, using our solutions for $\Psi_{\pm}$, then we find, upon generalizing $k \g^x \ra \ks$ and collecting powers\footnote{When $m$ is an integer, another term of order $u^{\frac{d}{2}+m+1}$ appears in eq. (\ref{eq:mnothalfintegerasymptoticdiracleadingtwoterms}), with a coefficient proportional to $b_m^+(m) c_2^+(k) + b_m^-(m) i \ks \, c_2^-(k) $. This term has opposite chirality from the term shown in eq. (\ref{eq:mnothalfintegerasymptoticdiracleadingtwoterms}), hence we set them to zero independently. Using the definition of the $b_j^{\pm}(m)$, we then recover exactly the same relation between $c_2^+(k)$ and $c_2^-(k)$ as in eq. (\ref{eq:onshellcoeffsrelations}), so we obtain no new information.} of $u$,
\bea
\label{eq:mnothalfintegerasymptoticdiracleadingtwoterms}
0 & = & \quad \left [ (-2m+1) \, c_2^+(k) + i \ks \, c_2^-(k) \right] u^{\frac{d}{2}-m+1} \nn \\ & & + \left [ - (2m+1) \, c_1^-(k) + i \ks \, c_1^+(k) \right] u^{\frac{d}{2}+m+1} + O\left(u^{\frac{d}{2}-m+2}\right).
\eea
We thus conclude that
\beq
\label{eq:onshellcoeffsrelations}
c_1^-(k) = \frac{1}{2m+1} \, i \ks \, c_1^+, \qquad c_2^+(k) = \frac{1}{2m-1} \, i \ks \, c_2^-(k).
\eeq

When $m$ is half-integer, we can plug the series expressions for the Bessel functions in eqs. (\ref{eq:jnseriesform}) and (\ref{eq:ynseriesform}) into the solutions for $\Psi_{\pm}$, and rearranging various terms, we can write
\bea
\Psi_{\pm} & = & u^{\frac{d+1}{2}} \left [ C^{\pm}_1(k) \, J_{m\mp\frac{1}{2}} (\sqrt{-k^2} \, u) + C^{\pm}_2(k) \, Y_{m\mp\frac{1}{2}} (\sqrt{-k^2} \, u) \right], \nn \\ & = & u^{\frac{d+1}{2}} \left [ c^{\pm}_1(k) \, u^{m\mp\frac{1}{2}} \, \left(\ln u \right)\, \left( 1+s_a^{\pm}(u,k) \right)  + c^{\pm}_2(k) \, u^{-\left(m\mp\frac{1}{2}\right)} \, \left(1+s_d^{\pm}(u,k)\right) \right],\nonumber \\ & = & c_1^{\pm}(k) \, u^{\frac{d}{2} + m \mp \frac{1}{2} + \frac{1}{2}} \, \left( \ln u \right) \, \left( 1+ s_a^{\pm}(u,k) \right) + c_2^{\pm}(k) \, u^{\frac{d}{2} - m \pm \frac{1}{2} + \frac{1}{2}} \, \left( 1+ s_d^{\pm}(u,k) \right), \nonumber
\eea
where in the second line we have absorbed various factors into $C_1^{\pm}(k)$ and $C_2^{\pm}(k)$, which we then relabeled as $c_2^{\pm}(k)$ and $c_1^{\pm}(k)$. Notice in particular that we exchanged the indices $1$ and $2$. Recall that we are using units in which the radius of $AdS_{d+1}$ is equal to one. The arguments of the logarithms include factors of the $AdS_{d+1}$ radius to render them dimensionless. The sums $s_a^{\pm}(u,k)$ are defined in eq. (\ref{eq:asumdef}), while the sums $s_d^{\pm}(u,k)$ are
\beq
\label{eq:dsumdef}
s_d^{\pm}(u,k) = \sum_{j=1}^{\infty} d_j^{\pm}(m,k) (-k^2)^j u^{2j},
\eeq
where the coefficients $d_j^{\pm}(m,k)$ are not particularly illuminating to see, so we will not write them. They can be derived straightforwardly from eqs. (\ref{eq:jnseriesform}) and (\ref{eq:ynseriesform}). That derivation also shows that they can depend not only on $m$ but also on $k$, as indicated. We have defined our notation to isolate the leading powers of $u$ near the boundary $u\ra0$, and also to show that the solutions are identical in form to those for the $m$-not-half-integer case in eq. (\ref{eq:psisolnotationdef}), except for an extra logarithmic factor in the $c_1^{\pm}(k)$ terms. Written separately (to facilitate counting powers of $u$), $\Psi_{\pm}$ are
\bea
\label{eq:psisolseparate2}
\Psi_+ & = & c_1^+(k) \, u^{\frac{d}{2} + m} \, \left (\ln u \right) \, \left(1+s_a^+(u,k) \right) + c_2^+(k) \, u^{\frac{d}{2} - m+1} \left(1+s_d^+(u,k)\right) \nonumber \\ \Psi_- & = & c_1^-(k) \, u^{\frac{d}{2} + m+1} \, \left (\ln u \right) \, \left(1+s_a^-(u,k) \right) + c_2^-(k) \, u^{\frac{d}{2} - m} \left(1+s_d^-(u,k)\right).
\eea
Here again, the coefficients in eq. (\ref{eq:psisolseparate2}) are not independent. Inserting the solutions for $\Psi_{\pm}$ into the Dirac equation, eq. (\ref{eq:ddimsimplifieddiraceq2}), and generalizing $k \g^x \ra \ks$, produces 
\bea
\label{eq:dirachalfinteger}
0 & = & \quad \left [ - (2m+1) \, c_1^-(k) + i \ks \, c_1^+(k) \right] u^{\frac{d}{2}+m+1} \ln u + O\left(u^{\frac{d}{2}+m+2} \ln u\right) \nn \\ & & + (-2m+1) \, c_2^+(k) \, \, u^{\frac{d}{2}-m+1} \left(1+s_d^+(u,k) \right) \, + i \ks \, c_2^-(k) \, u^{\frac{d}{2}-m+1} \, \left( 1+s_d^-(u,k) \right)\nn \\ & & + c_1^+(k) \, u^{\frac{d}{2} + m} + c_2^+(k) \, u^{\frac{d}{2}-m+1} \, \left[ u\partial_u s_d^+(u,k)\right] + O\left(u^{\frac{d}{2}-m+2}\right),
\eea
where in the first line we have indicated the sub-leading order of terms involving powers and the logarithm of $u$, while in the third line we have indicated the sub-leading order of terms involving just powers of $u$. When $m\neq1/2$, the vanishing of the $u^{\frac{d}{2}+m+1} \ln u$ and $u^{\frac{d}{2}-m+1}$ terms in eq. (\ref{eq:dirachalfinteger}) requires
\beq
\label{eq:halfmonshellrelations}
c_1^-(k) = \frac{1}{2m+1} \, i \ks \, c_1^+, \qquad c_2^+(k) = \frac{1}{2m-1} \, i \ks \, c_2^-(k),
\eeq
which are identical to what we found in the $m$ non-half-integer case, eq. (\ref{eq:onshellcoeffsrelations}). In what follows, we will also need to write $c_1^+(k)$ in terms of $c_2^-(k)$. We thus turn to higher-order terms in eq. (\ref{eq:dirachalfinteger}). When $m\neq1/2$, the $u^{\frac{d}{2}+m}$ term will always be the same power of $u$ as $u^{\frac{d}{2}-m+1}$ times a particular term in the summations $s_d^+(u,k)$, $s_d^-(u,k)$ and $u\partial_u s_d^+(u,k)$. Recall that these summations involve powers of $u^2$. Some term in the summations, say the $j$th term, will have $j=m-\frac{1}{2}$. Recalling the definition of the summations in eq. (\ref{eq:dsumdef}), we find from the second and third lines in eq. (\ref{eq:dirachalfinteger})
\bea
\label{eq:halfintegermc1c2}
c_1^+(k) & = & -\left [ c_2^+(k) (-2m+1)d_j^+(m,k) + i\ks \, c_2^-(k) d_j^-(m,k) + c_2^+(k) d_j^+(m,k) \, 2j \right] \left(-k^2\right)^j, \nn \\ & = & -i\ks \, c_2^-(k) \, d_{m-\frac{1}{2}}^-(m,k) \left(-k^2\right)^{m-\frac{1}{2}},
\eea
where we plugged $j=m-\frac{1}{2}$ into the first line, so that the two $c_2^+(k)$ terms canceled, producing the second line. The factor of $2j=2m-1$ in the first line is crucial for the cancelation: it comes from the $u\partial_u$ acting on  $s_d^+(m,k)$ in the third line of eq. (\ref{eq:dirachalfinteger}). The $m=1/2$ story is similar: $u^{\frac{d}{2}-m+1}$ and $u^{\frac{d}{2}+m}$ become the same power $u^{\frac{d}{2} + \frac{1}{2}}$, so that, from the second and third lines in eq. (\ref{eq:dirachalfinteger}), we immediately find
\beq
c_1^+(k) = - i \ks \, c_2^-(k).
\eeq

In the following subsections we will restrict to positive values of $m$ unless stated otherwise. We can recover results for negative $m$ as follows. For $m$ not-half-integer, if $m<0$, then we can insert $m=-|m|$ in eq. (\ref{eq:soljj}) to obtain
\bea
\Psi_{\pm} & = & u^{\frac{d+1}{2}} \left [ C^{\pm}_1(k) \,  J_{m\mp\frac{1}{2}} (\sqrt{-k^2} \, u)+ C^{\pm}_2(k) \, J_{-(m\mp\frac{1}{2})} (\sqrt{-k^2} \, u) \right], \nonumber \\ & = & u^{\frac{d+1}{2}} \left [ C^{\pm}_1(k) \,  J_{-\left(|m|\pm\frac{1}{2}\right)} (\sqrt{-k^2} \, u)+ C^{\pm}_2(k) \, J_{|m|\pm\frac{1}{2}} (\sqrt{-k^2} \, u) \right].
\eea
To obtain results for negative $m$, we can work with positive $m$ and then in all formulas take $m\ra|m|$ and exchange $C_1^{\pm}(k) \leftrightarrow C_2^{\mp}(k)$, which means
\beq
c_1^{\pm}(k) \ra c_2^{\mp}(k), \qquad c_2^{\pm}(k) \ra c_1^{\mp}(k), \qquad \mbox{$m$ not-half-integer.}
\eeq
For $m$ half-integer, if $m<0$, then we can insert $m=-|m|$ in eq. (\ref{eq:soljy}) to obtain
\bea
\Psi_{\pm} & = & u^{\frac{d+1}{2}} \left [ C^{\pm}_1(k) \, J_{m\mp\frac{1}{2}} (\sqrt{-k^2} \, u) + C^{\pm}_2(k) \, Y_{m\mp\frac{1}{2}} (\sqrt{-k^2} \, u) \right], \nn \\ & = & u^{\frac{d+1}{2}} \left [ C^{\pm}_1(k) \, J_{-(|m|\pm\frac{1}{2})} (\sqrt{-k^2} \, u) + C^{\pm}_2(k) \, Y_{-(|m|\pm\frac{1}{2})} (\sqrt{-k^2} \, u) \right], \nn \\ & = & \left(-1\right)^{|m|\pm\frac{1}{2}} u^{\frac{d+1}{2}} \left [ C^{\pm}_1(k) \, J_{|m|\pm\frac{1}{2}} (\sqrt{-k^2} \, u) + C^{\pm}_2(k) \, Y_{|m|\pm\frac{1}{2}} (\sqrt{-k^2} \, u) \right], \nn
\eea
where in the third equality we have used $J_{-n}(x) = (-1)^n J_n(x)$ and $Y_{-n}(x)=(-1)^nY_n(x)$. To recover results for negative $m$, we can work with positive $m$ and then in all formulas take $m\ra|m|$, $C_1^{\pm}(k) \ra (-1)^{|m|\mp\frac{1}{2}} C_1^{\mp}(k)$, and $C_2^{\pm}(k) \ra (-1)^{|m|\mp\frac{1}{2}} C_2^{\mp}(k)$, which means
\beq
c_1^{\pm}(k) \ra (-1)^{|m|\mp\frac{1}{2}} c_1^{\mp}(k), \qquad c_2^{\pm}(k) \ra (-1)^{|m|\mp\frac{1}{2}} c_2^{\mp}(k), \qquad \mbox{$m$ half-integer.}
\eeq

\subsection{Determining the Counterterms}

As reviewed in section \ref{mixing}, the AdS/CFT dictionary equates the exponential of (minus) the on-shell supergravity action with the generating functional of field theory correlation functions. For the action in eq. (\ref{eq:diracactionappendix}), clearly the bulk term vanishes when evaluated on a solution. The only nonzero contribution to the on-shell action comes from the boundary terms. We will split $S_{bdy}$ into two terms,
\beq
S_{bdy} = S_{var} + S_{CT},
\eeq
where $S_{var}$ are terms required for the variational principle to be well-posed \cite{Contino:2004vy,Henneaux:1998ch} and $S_{CT}$ includes the counterterms that will cancel any divergences. As shown in refs. \cite{Contino:2004vy,Henneaux:1998ch},
\beq
\label{eq:svardef}
S_{var} = \int d^dx \, \sqrt{\g} \, \bar{\Psi}_+ \Psi_-,
\eeq
where the integration is over only the $u=\epsilon$ surface, $\sqrt{\g}$ is the determinant of the induced metric on the $u=\epsilon$ surface, which for us is simply $\sqrt{\g} = \epsilon^{-d}$, and $\Psi_{\pm}$ are evaluated at $u=\epsilon$. We will plug the solutions for $\Psi_{\pm}$ into $S_{var}$ and isolate any terms that diverge as $\epsilon \rightarrow 0$. We will then introduce into $S_{CT}$ any terms necessary to cancel the divergences and hence render the on-shell action finite. The terms in $S_{CT}$ must respect the symmetries of the on-shell $S_{var}$ and must be built only from $\Psi_-$, to preserve stationarity of the action, as explained above.

For $m$ not half-integer (and positive), if we plug the solutions for $\Psi_{\pm}$ from eq. (\ref{eq:psisolseparate}) into eq. (\ref{eq:svardef}), we find\footnote{Starting now, we drop the $k$ dependence in the coefficients: $c_1^{\pm}(k) \rightarrow c_1^{\pm}$ and $c_2^{\pm}(k) \rightarrow c_2^{\pm}$.}
\bea
\label{eq:onshellsvar}
S_{var} & = & \int d^dx \, \frac{1}{\epsilon^d} \, \left [ \bar{c}_1^+ c_2^- \, \epsilon^{d} \, \left( 1 + f_{a^+b^-}(\epsilon,k) \right) + \bar{c}_2^+ c_1^- \, \epsilon^{d + 2} \, \left( 1 + f_{a^-b^+}(\epsilon,k)\right)\right . \\  & & \qquad + \left . \bar{c}_1^+ c_1^- \, \epsilon^{d + 2m +1} \, \left( 1 + f_{a^+a^-}\left(\epsilon,k\right)\right) + \bar{c}_2^+ c_2^- \, \epsilon^{d - 2m +1} \,  \left( 1 + f_{b^+b^-}\left(\epsilon,k\right)\right) \right], \nn
\eea
where we have introduced one more piece of notation: we have defined
\beq
f_{a^+a^-}(\epsilon,k) = s_a^+(\e,k) + s_a^-(\e,k) + s_a^+(\e,k) \, s_a^-(\e,k), \nn
\eeq
and similarly for $f_{b^+b^-}(\epsilon,k)$, $f_{a^+b^-}(\epsilon,k)$, and $f_{a^-b^+}(\epsilon,k)$, all of which are summations in powers of $\e^2$ starting with $\e^2$. We now ask what happens to each of the terms in the brackets in eq. (\ref{eq:onshellsvar}) as $\e\ra0$. The first term clearly remains finite (the $\epsilon^d$'s cancel). The second and third terms vanish as $\e^2$ and $\e^{2m+1}$, respectively. The fate of the fourth term depends on $m$. If $m<1/2$, then the fourth term vanishes as $\e^{-2m+1}$, and so we are done: no divergences appear and no counterterms are necessary. If $m>1/2$, however, then the fourth term may have one or more divergent terms, with the leading divergence going as $\e^{-2m+1}$, so we must add counterterms. Starting now, we will assume $m>1/2$. Using eq. (\ref{eq:onshellcoeffsrelations}), we can write the on-shell $S_{var}$ as
\beq
\label{eq:onshellsvar2}
S_{var} = \int d^dx \, \frac{1}{\e^d} \left [ \bar{c}_1^+ c_2^- \, \epsilon^{d} +\frac{1}{2m-1} \, \bar{c}_2^- \, i \ks \, c_2^- \epsilon^{d - 2m +1} \,  \left( 1 + f_{b^+b^-}\left(\epsilon,k\right)\right) + O\left( \e^{d+2}\right)\right].
\eeq

Having isolated the divergences in $S_{var}$, we now must write an $S_{CT}$ that obeys the constraints mentioned in section \ref{mixing}: it must cancel all divergences of, while respecting all symmetries of, the on-shell $S_{var}$, and must be built only from the boundary value of $\Psi_-$. The $S_{CT}$ that does the job is
\beq
\label{eq:countertermactionnothalfinteger}
S_{CT} = \int d^dx \sqrt{\g} \, \sum_{j=0}^{\infty} \, \a_j(m) \, \bar{\Psi}_- \ds_{\e} \, \Box^j_{\e} \Psi_- = \int d^dx \frac{1}{\e^d} \, \sum_{j=0}^{\infty} \e^{1+2j} \, \a_j(m) \, \bar{\Psi}_- \ds \, \Box^j \Psi_-,
\eeq
where $\ds_{\e} = \epsilon \, \ds$ (the power of $\epsilon$ comes from the inverse vielbein evaluated at $u=\epsilon$) and $\Box^j_{\e}$ is some power $j$ of the scalar Laplacian $\Box_{\e}$ on the $u=\epsilon$ surface (so the derivatives act only in field theory directions), which in our case is simply $\Box_{\e} = \epsilon^2 \, \partial^2$. When we take $\Psi \ra e^{ikx} \Psi$, the counterterms become
\beq
S_{CT} = \int d^dx \frac{1}{\e^d} \, \sum_{j=0}^{\infty} \e^{1+2j} \, \a_j(m) \, \bar{\Psi}_- \, i \ks \, \left(-k^2\right)^j \Psi_-,
\eeq
so that, plugging in the solutions for $\Psi_{\pm}$, we find
\bea
\label{eq:onshellct}
S_{CT} & = & + \int d^dx \, \frac{1}{\e^d} \, \sum_{j=0}^{\infty} \e^{1+2j} \, \a_j(m) \nn \\ & & \left [ + \e^{d+2m+2} \,\, \bar{c}_1^- i \ks \, (-k^2)^j \, c_1^- \left( 1+f_{a^-a^-}(\e,k) \right) \right . \nn \\ & & \,\,\, + \e^{d+1} \qquad \bar{c}_1^- i \ks \, (-k^2)^j \, c_2^- \left( 1+f_{a^-b^-}(\e,k) \right) \nn \\ & &\,\,\, + \e^{d+1} \qquad \bar{c}_2^- i \ks \, (-k^2)^j \, c_1^- \left( 1+f_{b^-a^-}(\e,k) \right)\nn \\ & & \,\, \left . + \e^{d-2m} \quad \,\,\, \bar{c}_2^- i \ks \, (-k^2)^j \, c_2^- \left( 1+f_{b^-b^-}(\e,k) \right) \right ].
\eea
Of the terms in brackets, the term that is potentially divergent when $\e\ra0$ goes as $\e^{d-2m}$. All of the other terms vanish when $\e\ra0$. We fix the coefficients $\a_j(m)$ by demanding that the divergent terms in eqs. (\ref{eq:onshellsvar2}) and (\ref{eq:onshellct}) cancel each other, which means that the quantity
\beq
\frac{1}{2m-1}\left(1+f_{b^+b^-}(\e,k)\right) + \sum_{j=0}^{\infty} \a_j(m) (-\e^2 k^2)^j \left ( 1+f_{b^-b^-}(\e,k) \right)
\eeq
must vanish order-by-order in $-\e^2k^2$, up to order $\e^{2m-1}$. We immediately see that $\a_0(m) = -\frac{1}{2m-1}$, and we can write a formal recursive solution for all the other $\a_j(m)$ (here we drop the dependence on $m$ for notational clarity, so $\a_j(m)\ra\a_j$, etc.),
\beq
\label{eq:formalajsolution}
\a_j = - \frac{1}{(2m-1)} \left [ b_j^+ + b_j^- + \sum_{i=1} b_i^+ b_{j-i}^-\right] -\sum_{i<j} \a_i \, \left [ 2 \, b_{j-i}^- + \sum_{k=1} b_k^- \, b_{j-i-k}^- \right],
\eeq
where we define $b_j^{\pm}(m) \equiv 0$ if $j\leq0$. The first four $\a_j(m)$ are
\bea
\a_0 & = & - \frac{1}{2m-1}, \qquad \a_1 = -\frac{b_1^-+b_1^+}{(2m-1)} - \a_0 \, 2b_1^-, \nn \\ \a_2 & = & -\frac{ b_2^- + b_2^+ + b_1^- b_1^+}{(2m-1)} - \a_1 \, 2b_1^- -\a_0 \left(2 b_2^- + b_1^{-\, 2} \right), \nn \\ \a_3 & = & -\frac{b_3^+ + b_3^- + b_2^+ b_1^- + b_1^+ b_2^-}{(2m-1)} - \a_2 \, 2b_1^- -\a_1 \left(2b_2^- + b_1^{-\, 2} \right) -\a_0\left(2b_3^- + 2b_1^-b_2^- \right). \nn
\eea
Plugging in the explicit forms for the $b_j^{\pm}(m)$ (which are just the $a_j^{\pm}(m)$ from eq. (\ref{eq:asumdef}) with $m\mp\frac{1}{2} \ra - \left(m\mp\frac{1}{2}\right)$), we can write these explicitly:
\bea
\a_0(m) & = & -\frac{1}{2m-1}, \nn \\ \a_1(m) & = & -\frac{1}{(2m-1)^2(2m-3)}, \nn \\ \a_2(m) & = &\frac{2}{(2m-1)^3(2m-3)(2m-5)} \nn \\ \a_3(m) & = & \frac{17-10m}{(2m-1)^4(2m-3)^2(2m-5)(2m-7)}. \nn
\eea
We have thus determined the counterterms when $m$ is not half-integer.

We now consider half-integer $m$, in which case the solutions for $\Psi_{\pm}$ appear in eq. (\ref{eq:psisolseparate2}). Oncer again, to determine the divergences, we plug the solutions for $\Psi_{\pm}$ into $S_{var}$, with the result
\bea
\label{eq:onshellsvar3}
S_{var} & = &  \int d^dx \, \sqrt{\g} \, \bar{\Psi}_+ \Psi_-\nn \\ & = & \int d^dx \, \frac{1}{\epsilon^d} \left [  \bar{c}_1^+ c_2^- \, \epsilon^{d} \, \left( \ln \e \right) \, \left( 1 + f_{a^+d^-}(\epsilon,k) \right) + \bar{c}_2^+ c_1^- \, \epsilon^{d + 2} \, \left( \ln \e \right) \, \left( 1 + f_{a^-d^+}(\epsilon,k)\right) \right . \nonumber \\ & & + \left . \bar{c}_1^+ c_1^- \, \epsilon^{d + 2m +1} \, \left( \ln \e\right)^2 \, \left( 1 + f_{a^+a^-}\left(\epsilon,k\right)\right) + \bar{c}_2^+ c_2^- \, \epsilon^{d - 2m +1} \,  \left( 1 + f_{d^+d^-}\left(\epsilon,k\right)\right) \right]
\eea
where $f_{a^+d^-}(\epsilon,k)$, $f_{a^-d^+}(\epsilon,k)$ and $f_{d^+d^-}\left(\epsilon,k\right)$ are defined similarly to $f_{a^+a^-}\left(\epsilon,k\right)$. We now ask what happens to each of the terms in brackets in eq. (\ref{eq:onshellsvar3}) when $\e\ra0$. The first term diverges as $\ln\e$. The second and third terms vanish as $\epsilon^{2} \, \ln\e$ and $\epsilon^{2m +1} \, \left(\ln\e\right)^2$, respectively. The fate of the fourth term depends on $m$. If $m=1/2$, then the fourth term is finite. The first term still diverges, however, so in this case we need counterterms (in contrast to the $m$-not-half-integer cases). If $m>1/2$, then the fourth term may have one or more power-law divergent terms, with the leading divergence going as $\e^{-2m+1}$. Notice, however, that the fourth term will also always produce something finite, since the sum $f_{d^+d^-}(\e,k)$ will always have a term that goes as $\e^{2m-1}$ which will cancel the $\e^{-2m+1}$. When $m\neq1/2$, we can use eqs. (\ref{eq:halfmonshellrelations}) and (\ref{eq:halfintegermc1c2}) to rewrite the $S_{var}$ in eq. (\ref{eq:onshellsvar3}) as
\bea
\label{eq:onshellsvar4}
S_{var} & = & \int d^dx \, \frac{1}{\epsilon^d} \left [ \bar{c}_1^+ c_2^- \, \epsilon^{d} \, \left( \ln \e \right) \, \left( 1 + f_{a^+d^-}(\epsilon,k) \right) \right . \nn \\ & & \left . \qquad \qquad \,\,\,\, + \bar{c}_2^+ c_2^- \, \epsilon^{d - 2m +1} \,  \left( 1 + f_{d^+d^-}\left(\epsilon,k\right)\right) + O\left(\e^{d+2} \ln \e\right) \right ] \nn \\ & = & \int d^dx \, \frac{1}{\epsilon^d} \, \bar{c}_2^- i \ks \, c_2^- \left [ -d^-_{m-\frac{1}{2}}(m,k) \left(-k^2\right)^{m-\frac{1}{2}} \, \e^d \, \ln \e \, \left( 1+f_{a^+d^-}(\e,k)\right) \right. \\ &  & \left .  \qquad \qquad \qquad \qquad + \frac{1}{2m-1} \, \e^{d-2m+1} \, \left( 1+ f_{d^+d^-}(\e,k) \right) + O\left(\e^{d+2} \, \ln \e \right) \right ]. \nn
\eea
The $m=1/2$ case is similar, except in the third line $d^-_{m-\frac{1}{2}}(m,k) \left(-k^2\right)^{m-\frac{1}{2}} \rightarrow 1$, and the fourth line remains identical to the first line.

We now include logarithmic terms in $S_{CT}$,
\beq
\label{eq:countertermactionhalfinteger}
S_{CT} = \int d^dx \, \sum_{j=0}^{\infty} \, \left( \a_j(m,k) + \b_j(m,k) \ln\e\right)\, \sqrt{\g} \,  \bar{\Psi}_- \, \ds_{\e} \, \Box^j_{\e} \Psi_-.
\eeq
Plugging the solutions for $\Psi_{\pm}$ into $S_{CT}$, we find
\bea
\label{eq:onshellsct}
S_{CT} & = & + \int d^dx \, \frac{1}{\e^d} \, \sum_{j=0}^{\infty} \e^{1+2j} \, \left ( \a_j(m,k) + \b_j(m,k) \ln \e \right) \nonumber \\ & & \left[ + \e^{d+2m+2} \, \left(\ln \e\right)^2 \, \bar{c}_1^- i \ks \, (-k^2)^j \, c_1^- \left( 1+f_{a^-a^-}(\e,k) \right) \right . \nn \\ & & + \e^{d+1} \qquad \left(\ln \e\right)^1 \, \bar{c}_1^- i \ks \, (-k^2)^j \, c_2^- \left( 1+f_{a^-d^-}(\e,k) \right) \nn \\ & &+ \e^{d+1} \qquad \left(\ln \e\right)^1 \, \bar{c}_2^- i \ks \, (-k^2)^j \, c_1^- \left( 1+f_{d^-a^-}(\e,k) \right)\nn \\ & & \left . + \e^{d-2m} \quad \left(\ln \e\right)^0 \, \bar{c}_2^- i \ks \, (-k^2)^j \, c_2^- \left( 1+f_{d^-d^-}(\e,k) \right) \right ] \nn \\ & = & \int d^dx \, \frac{1}{\e^d} \, \sum_{j=0}^{\infty} \, \left(\a_j(m,k)+\b_j(m,k) \ln \e \right) \, \e^{2j} \, \e^{d-2m+1} \nn \\ & & \qquad \left[  \bar{c}_2^- i \ks \left(-k^2\right)^j c_2^- \, \left( 1+f_{d^-d^-}(\e,k)\right) + O\left(\e^{2m+1} \ln \e \right) \right].
\eea
Of the terms in brackets in the first equality, only the fourth term is potentially divergent when $\e\ra0$. All the other terms vanish when $\e\ra0$. In the second equality we have thus isolated the potentially divergent terms. We now compare eq. (\ref{eq:onshellsvar4}) and eq. (\ref{eq:onshellsct}), and adjust the coefficients $\a_j(m,k)$ and $\b_j(m,k)$ such that all divergences cancel.

We first consider the logarithmically-divergent terms. For these we must adjust the coefficients $\b_j(m,k)$ such that
\beq
-d^-_{m-\frac{1}{2}}(m,k) \left(-k^2\right)^{m-\frac{1}{2}} + \sum_{j=0}^{\infty} \b_j(m,k) (-k^2)^j \e^{2j-2m+1} = 0.
\eeq
We thus conclude that all $\b_j(m,k)$ are zero except for one, $\b_{m-\frac{1}{2}}(m,k) = d^-_{m-\frac{1}{2}}(m,k)$. If $m=1/2$, the result is that only $\b_0(m,k) = 1$ is nonzero.

Next we consider the power-law divergences, which appear when $m\neq1/2$. For these we adjust the coefficients $\a_j(m,k)$ such that the quantity
\beq
\frac{1}{2m-1}\left(1+f_{d^+d^-}(\e,k)\right) + \sum_{j=0}^{\infty} \a_j(m,k) (-\e^2 k^2)^j \left ( 1+f_{d^-d^-}(\e,k) \right)
\eeq
vanishes order-by-order in $-\e^2k^2$, up to order $\e^{2m-3}$, to guarantee that all divergences cancel. The $\alpha_j(m,k)$ are thus straightforward to obtain, indeed, they are identical in form to the $\a_j(m,k)$ in eq. (\ref{eq:formalajsolution}), but with $b_j^{\pm}(m) \ra d_j^{\pm}(m,k)$. We have thus determined the counterterms when $m$ is half-integer.

Notice that when $m$ is half-integer, a finite counterterm is also possible: we may introduce a nonzero $\alpha_{m-1/2}$, producing a finite counterterm proportional to $\bar{c}_2^- i \ks \, (-k^2)^{m-1/2} \, c_2^-$. Something similar happens for scalars \cite{Skenderis:2002wp}. The scalar equation of motion has Bessel-function solutions, and for certain values of the scalar mass $m$ the linearly-independent solutions are Bessel functions of the first and second kind, which means the asymptotic expansion includes logarithmic terms. In those cases, a finite counterterm also appears, and is proportional to the matter conformal anomaly of the dual CFT \cite{Skenderis:2002wp} (and references therein). Here we are simply seeing the fermionic version of the scalar story.

To summarize: for the action in eq. (\ref{eq:diracaction}), clearly the bulk term vanishes when evaluated on a solution. The only nonzero contribution to the on-shell action comes from the boundary terms, hence $S=S_{bdy}$ on-shell. Generically, $S_{bdy}$ diverges and we must add counterterms. When $m$ is not half-integer, the resulting on-shell action is\footnote{Recall once again that we restricted to positive $m$. At the end of the last subsection we explained how to recover results for negative $m$ from the results for positive $m$.}
\beq
\label{eq:regulatedonshelldiracnothalfinteger}
S = \int d^dx \, \bar{c}_1^+ c_2^- + O\left( \e^2\right)
\eeq
while when $m$ is half-integer but $m\neq1/2$ (here we take $d_j^{\pm}(m,k) \ra d_j^{\pm}$),
\beq
\label{eq:regulatedonshelldirachalfintegernotonehalf}
S = \int d^dx \, \bar{c}_2^+ c_2^- \left( d_{m-\frac{1}{2}}^+ + d_{m-\frac{1}{2}}^- + \sum_{i=1} d_i^+ d_{m-\frac{1}{2}-i}^- \right) \left ( -k^2 \right)^{m-\frac{1}{2}} + O\left( \e^2 \ln \e\right),
\eeq
and for $m=1/2$,
\beq
\label{eq:regulatedonshelldiraconehalf}
S = \int d^dx \, \bar{c}_2^+ c_2^- + O\left( \e^2 \ln \e \right).
\eeq
These on-shell actions remain finite as $\e\ra0$, and hence, upon functional differentiation, will produce renormalized field theory correlators, as we next discuss.

\subsection{Computing Renormalized Correlators}

The field $\Psi$ in the bulk is dual to some fermionic operator $\Op$. As reviewed in section \ref{mixing}, the renormalized on-shell bulk action, $S_{ren} = \lim_{\epsilon \rightarrow 0} S$, acts as the generating functional for correlators involving $\Op$. In other words, to compute renormalized correlators of $\Op$, we take functional derivatives of $S$ with respect to some source. We identify the source for $\Op$ as the coefficient of the dominant term in $\Psi$'s near-boundary expansion. In eqs. (\ref{eq:psisolseparate}) and (\ref{eq:psisolseparate2}), the dominant term is the $u^{\frac{d}{2}-m}$ term, hence we 
identify $c_2^-(k)$ as the source for $\Op$. More formally, we equate (see eq. \eqref{eq:generatingdef})
\beq
e^{-S_{ren}[c_2^-,\bar{c}_2^-]} = \left \< \mbox{exp} \left [ \int d^dx \, \left(\bar{c}_2^- \, \Op + \bar{\Op} \, c_2^- \right)\right]  \right\>,
\eeq
where the left-hand-side in the exponential of minus the on-shell bulk action in eq. (\ref{eq:diracaction}), and the right-hand-side in the generating functional of the dual field theory, with $c_2^-(k)$ acting as the source for the operator $\Op$. Upon taking minus the logarithm of both sides, we find that the on-shell bulk action is the generator of connected correlators.

For any value of the bulk fermion's mass $m$, a quantization exists for which $\Op$ has dimension $\Delta = \frac{d}{2} + |m|$ \cite{Faulkner:2009wj,Iqbal:2009fd}. When $|m| \in [0,1/2)$, a second quantization exists for which $\Delta = \frac{d}{2} - |m|$. These two quantizations correspond to two different field theories, one of which (the one with $\Delta = \frac{d}{2} - |m|$) is ``unstable'' against a relevant deformation by a double-trace operator of $\Op$, and flows to the other, ``stable,'' theory, as in the scalar case \cite{Klebanov:1999tb} \footnote{For the Dp-branes we study in section \ref{eom1reduction}, $m$ is always integer or half-integer, so the second quantization is only possible when $m=0$. Indeed, in the bulk of the paper we use $m=0$. In that case, however, the two quantizations are equivalent (see appendix A of ref. \cite{Liu:2009dm} and ref. \cite{Iqbal:2009fd}).}. When $|m|>d/2$, the dual operator is irrelevant: $\Delta > d$.

When $d$ is even, $\Op$ will be an operator of definite chirality, since in that case $\g^u$ is the chirality operator of the field theory and $c_2^-(k)$ has definite chirality. Notice that taking $m \rightarrow -m$ takes $c_2^-(k)\ra c_1^+(k)$ when $m$ is not half-integer and $c_2^-(k)\ra \left(-1\right)^{|m|+\frac{1}{2}} \, c_2^+(k)$ when $m$ is half-integer, and hence switches the chirality of $\Op$.

We reviewed in section \ref{mixing} how to compute the Euclidean Green's function from the renormalized on-shell action. In particular, from eqs. \eqref{eq:secondderivativeofonshellaction} and \eqref{eq:euclideangreensfunctionfromonshellaction}, we have
\beq
\left \< \Op \, \bar{\Op} \right \>_{ren} = - \frac{\delta^2 S_{ren}}{\delta c_2^- \delta \bar{c_2^-}} = - \frac{\delta c_1^+}{\delta c_2^-} = G(k) \, \g^t,
\eeq
where $G(k)$ is defined by $c_1^+ = - G(k) \, \g^t c_2^-$. As we mentioned below eq. \eqref{eq:euclideangreensfunctionfromonshellaction}, in general, we extract $G(k)\, \g^t$ from from a solution by imposing some regularity condition in the bulk of the spacetime. As an example, let us consider the simplest case: pure $AdS_{d+1}$ with $m$ positive and not half-integer. The solution in eq. (\ref{eq:soljj}) is then the solution for all $u$, not just the asymptotic solution as $u \rightarrow 0$. We need to write $c_1^+$ in terms of $c_2^-$. To do so, we impose a regularity condition in the bulk of $AdS_{d+1}$. Deep in the interior of $AdS_{d+1}$, where $u\rightarrow \infty$, the solution in eq. (\ref{eq:soljj}) diverges unless $C_1^+ = \left(-1\right)^{m+1/2} C_2^+$. Translating that condition into a condition on $c_1^+$ and $c_2^+$ is trivial, once we recall the definition of $c_1^+$ and $c_2^+$ in eq. (\ref{eq:psisolnotationdef}). We find
\beq
c_1^+ = - 2^{-2m} \left( 2m-1\right) \, \frac{\G\left( \frac{1}{2} - m\right)}{\G\left(\frac{1}{2} + m\right)} \, k^{2m-1} \, c_2^+.
\eeq
We then use eq. (\ref{eq:onshellcoeffsrelations}) to write $c_2^+$ in terms of $i \ks \, c_2^-$. We finally obtain for the two-point function,
\beq
\left \< \Op \, \bar{\Op} \right \>_{ren} = 2^{-2m} \, \frac{\G\left( \frac{1}{2} - m\right)}{\G\left(\frac{1}{2} + m\right)} \, k^{2m-1} \, i\ks,
\eeq
which agrees with the result in refs. \cite{Henningson:1998cd,Henneaux:1998ch,Mueck:1998iz} and, up to normalization, is the correct momentum-space form for the two-point function of a quasi-primary operator of dimension $\Delta = d/2 + m$. Similar arguments work when $m$ is half-integer, where again we find the correct momentum-space form for a quasi-primary operator of dimension $\Delta = d/2+m$. In that case, we can also adjust the normalization to any value we like using the finite counterterm.

\end{appendix}

\bibliographystyle{JHEP}
\bibliography{d5fermi_v2}

\end{document}